\documentclass[11pt,a4paper]{article}
\usepackage[utf8]{inputenc}
\usepackage{appendix}
\usepackage{amsthm}
\usepackage{bbm}
\usepackage{tensor}
\usepackage[dvipsnames]{xcolor}
\usepackage{tikz}
\usepackage{easybmat}
\usepackage{mathdots}
\usepackage{youngtab}
\usepackage{lmodern}
\usepackage[T1]{fontenc}
\usepackage{microtype}
\usepackage[margin=2.5cm]{geometry}

\usepackage{hyperref}

\hypersetup{
    colorlinks=true,
    linkcolor=black,
    filecolor=magenta,      
    urlcolor=cyan,
    citecolor=SeaGreen,
}

\usepackage{rotating}

\usepackage{pdflscape}
\usepackage{mathtools,amsmath,amssymb}

\usepackage{cite}

\numberwithin{equation}{section}
\newcommand{\Q}{q}
\newcommand{\bS}{s}

\newcommand{\ID}{\mathrm{I}}
\newcommand{\IDb}{\mathrm{J}}

\DeclareMathOperator{\tr}{tr}

\newcommand{\Qop}{\mathrm{Q}}
\newcommand{\Pop}{\mathrm{P}}

\newcommand{\oa}{\mathbf{a}}
\newcommand{\oad}{\mathbf{\bar{a}}}
\newcommand{\wpp}{\bar{\mathbf{w}}}
\newcommand{\wmm}{\mathbf{w}}

\newcommand{\ap}{\bar{\mathbf{A}}}
\newcommand{\am}{{\mathbf{A}}}

\newcommand{\F}{J}

\newcommand{\W}{W}

\usetikzlibrary{fit, arrows, calc, positioning}

\tikzstyle{yellowRectangle} = [
    circle,
    node distance=2 cm,
    text width=2 em,
    text centered,
    rounded corners,
    minimum height=0.4 cm,
    minimum width=1 cm,
    thick
]
\tikzstyle{yellowRectangle2} = [
    circle,
    node distance=2 cm,
    text width=6 em,
    text centered,
    rounded corners,
    minimum height=0.4 cm,
    minimum width=1 cm,
    thick
]

\tikzstyle{container} = [
    rectangle,
    draw,
    inner sep=1 cm,
    dashed
]

\tikzstyle{empty} = [
]

\tikzstyle{line} = [
    draw,
->,
    thick, SeaGreen!60,
]

\newcommand{\h}{\ell}
\newcommand{\m}{f}
\newcommand{\NN}{\mathbf{N}}

\newcommand{\QQ}{\mathcal{Q}}
\newcommand{\WW}{\widetilde{W}}

\begin{document}

\addtocontents{toc}{\protect\setcounter{tocdepth}{2}}

\begingroup\parindent0pt
\begin{flushright}\footnotesize
% \texttt{HU-MATH-2017-04}\\
% \texttt{HU-EP-17/14}\\
% \texttt{TCDMATH-17-13}\\
\end{flushright}
%\vspace*{1em}
\centering
\begingroup\LARGE
\bf
QQ-system and  Weyl-type  transfer matrices in integrable $SO(2r)$ spin chains
\par\endgroup
\vspace{3.5em}
\begingroup\large
{\bf Gwena\"el Ferrando}$\,^{a,b}$, 
{\bf Rouven Frassek}$\,^a$, 
{\bf Vladimir Kazakov}$\,^a$ 
\par\endgroup
\vspace{2em}
\begingroup\sffamily\footnotesize
$^a\,$Laboratoire de Physique de l'\'Ecole Normale Sup\'erieure,\\ CNRS,
Universit\'e PSL, Sorbonne Universit\'es,\\
24 rue Lhomond, 75005 Paris, France\\
\vspace{1em}
$^b\,$Institut de Physique Th\'eorique,\\
Université Paris-Saclay, CNRS, CEA Saclay,\\
91191 Gif-sur-Yvette, France
\par\endgroup
\vspace{2em}
%\today
% \vfill
\begin{abstract}
\noindent
We propose the full system of Baxter Q-functions (QQ-system) for the integrable spin chains with the symmetry of the $D_r$ Lie algebra. We use this QQ-system to derive new Weyl-type formulas expressing  transfer matrices in all symmetric and antisymmetric (fundamental) representations through $r+1$ basic Q-functions. 
Our functional relations are consistent with the Q-operators proposed recently by one of the authors and verified explicitly on the level of operators at small finite length. 
\end{abstract}

\endgroup

\thispagestyle{empty}

% \newpage
\tableofcontents
% \newpage

\section{Introduction}

Baxter Q-operators play an important role in the theory of integrable spin chains~\cite{Baxter:1982zz}, in 2D integrable quantum field theory and sigma models~\cite{Bazhanov:1996dr}, in integrable examples of higher dimensional CFTs, such as QCD in BFKL limit~\cite{Faddeev:1994zg,Derkachov:2001yn,Lipatov:1993qn}, $\mathcal{N}=4$ super Yang-Mills theory and ABJM theory~\cite{Beisert:2010jr} where Q-functional approach has led to the elegant description of spectrum of the systems in terms of the quantum spectral curve (QSC)~\cite{Gromov:2013pga,Gromov:2014caa}, the ODE/IM correspondence \cite{Dorey:2007zx}, the fermionic basis \cite{Boos:2008rh}, stochastic processes \cite{Kuniba:2016fpi,Lazarescu_2014} and pure mathematics \cite{Frenkel:2013uda}. In particular, they provide a natural formulation for the  Bethe ansatz equations~(BAE) whose solutions (Bethe roots) yield the spectrum of energy  for the Heisenberg-type spin chains and are at the heart of  Sklyanin's separation of variables (SoV) construction~\cite{Sklyanin:1984sb}. They also allow for  natural representations of transfer matrices (T-operators), encoding all quantum conserved charges of the system.

All these operators, T and Q, commute due to the underlying integrable structure, so that on a given eigenstate we can operate with their eigenvalues --  the functions of a spectral parameter: $T(x)$ and  $Q(x)$. For $A$-type spin chains all these operators can be built within the framework of the quantum inverse scattering method \cite{Faddeev:1996iy} from solutions of the Yang-Baxter equation. The transfer matrices are built from Lax matrices of finite dimension while, as noted in \cite{Bazhanov:1996dr,Bazhanov:1994ft,Bazhanov:1998dq}, the construction of Q-operators is related to an infinite-dimensional Hilbert space. These methods were further developed in \cite{Antonov:1996ag,Bazhanov:2001xm,Rossi:2002ed,Korff_2006,Bazhanov:2008yc,Kojima:2008zza,Boos:2010ss,Boos:2013noa,Frassek:2010ga,Tsuboi:2019vvv}. For us the most relevant articles are \cite{Bazhanov:2010ts,Bazhanov:2010jq,Frassek:2011aa} for Q-operators of $A$-type spin chains and the recent generalisation to some Q-operators of $D$-type spin chains \cite{Frassek:2020nki}. An alternative approach, based on the formalism of co-derivatives~\cite{Kazakov:2007na}, was proposed in \cite{Kazakov:2010iu} and further developed in \cite{Alexandrov:2011aa} in relation to the interplay between quantum and classical integrability of $A$-type spin chains.

\begin{figure}[htb]
\begin{center}
\begin{tikzpicture}

    \node [empty](origin){};
    
    \node [yellowRectangle, above=0cm of origin] (q1) {$Q_{\{1,2\}}$};
    \node [yellowRectangle, below=0.5cm of q1] (qsc1) {$Q_{\{1\}}$};
    \node [yellowRectangle, right=1cm of q1](q2){$Q_{\{1,3\}}$};
    \node [yellowRectangle, right=1cm of qsc1] (qsc2) {$Q_{\{2\}}$};
    \node [yellowRectangle, right=1cm of q2](q3){$Q_{\{2,3\}}$};
    \node [yellowRectangle, right=1cm of qsc2](qsc3){$Q_{\{3\}}$};
    \node [yellowRectangle, below=0.5cm of qsc2](q0){$Q_{\varnothing}$};
    \node [yellowRectangle, above=0.5cm of q2](qf){$Q_{\{1,2,3\}}$};
    \path [line] (qsc1) --  (q2);
    \path [line] (qsc1) --  (q1);
    \path [line] (qsc2) --  (q1);
    \path [line] (qsc2) --  (q3);
    \path [line] (qsc3) -- (q2);
    \path [line] (qsc3) -- (q3);
    \path [line] (q0) --  (qsc1);
    \path [line] (q0) -- (qsc2);
    \path [line] (q0) -- (qsc3);
    \path [line] (q1) --  (qf);
    \path [line] (q2) -- (qf);
    \path [line] (q3) -- (qf);
\end{tikzpicture}
\caption{Hasse diagram for $A_2$}\label{fig:HasseA2}
\end{center}
\end{figure}
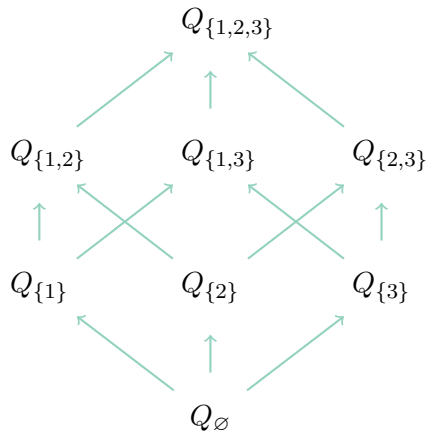

T-functions represent a quantum generalisation of the characters for the symmetry algebra of the spin chain. They depend on the  representation $f$ in the auxiliary space and, generically, on the twist $\tau$ -- a group element introduced into the spin chain in the form of twisted, quasi-periodic boundary conditions or, alternatively, as generalized "magnetic" fields. That is why we will denote the T-functions as $T_{f}^{(\tau)}$.~\footnote{Though the explicit $\tau$ dependence will often be omitted.} Generally, one has an infinite number of different T-functions since there exists an infinite number of inequivalent representations. However, most of them are not independent quantities. The most constructive way to see that is to represent T-functions in terms of Baxter Q-functions since the latter always form a finite variety. Say, for $A_{r}$ algebra, the $2^{r+1}$ Q-functions are usually  labeled by subsets of integers $I\subset \{1,2,\dots,r,r+1\}$ where $r$ is the rank of the algebra (for instance $I=\{1,3,4\}\subset \{1,2,3,4,5\}$). This QQ-system~\footnote{We decided to call it QQ-system, to avoid the confusion with the ``Q-system" established in the mathematical literature to denote the quadratic,  Hirota-type relations for characters of ``rectangular" representations. This hints on Pl\"ucker QQ-relations or on  ``Quantum Q"-relations.} can be conveniently depicted as a Hasse diagram in the shape of an $r+1$-dimensional hypercube with the vertices labeling the corresponding Q-functions~\cite{Tsuboi:2009ud}, see Figure~\ref{fig:HasseA2} for the example of $A_2$. 

As we will see, in $D_{r}$ algebra the labeling is similar but slightly different. Moreover, only $r+1$ Q-functions are algebraically independent as in the $A_r$ case~\cite{Pronko:1999gh,Bazhanov:2001xm,Krichever:1996qd,Dorey:2000ma}, see also the supersymmetric generalisation \cite{Tsuboi:2009ud,Kazakov:2007fy,Gromov:2014caa,Kazakov:2015efa}. The system of all Q-functions, which we will call here QQ-system, is endowed with a Gra\ss mannian structure. The remaining Q-functions can thus be expressed through a chosen basis of $r+1$ $Q$'s by various Pl\"ucker QQ-relations, often in the form of Wronskian determinants (Casoratians).

For the Heisenberg spin chains with $A_{r}$ symmetry, the most traditional representation of T-functions is given in terms of the basis of Q-functions of the type $Q_{\{1\}},Q_{\{1,2\}},Q_{\{1,2,3\}},\dots,Q_{\{1,2,\dots, r+1\}}$ (or different re-labelings of the same basis), cf.~Figure~\ref{fig:HasseA2}. The same functions enter into the formulation of the standard nested system of BAE's. The T-functions in this basis are usually represented by the so-called tableaux formulas which are direct generalizations of Schur polynomials for  characters~\cite{Krichever:1996qd,Tsuboi:1997iq,Kuniba:2010ir}.

The other well-known, so-called Cherednik-Bazhanov-Reshetikhin (CBR), formulas for T-functions in an arbitrary finite dimensional representation $\h$, are given in terms of determinants of T-functions in the simplest symmetric or antisymmetric representations~\cite{cherednik1989special,Bazhanov:1989yk}. They have been proven in \cite{Kazakov:2007na}, including the supersymmetric $A_{r|s}$ algebra. They represent the quantum generalization of the Jacobi-Trudi formulas for characters.  For the $D_{r}$ algebra the corresponding determinant representations have been found in \cite{Tsuboi:1996sh}. For both algebras, the CBR type formulas appear to be  solutions~\cite{Kuniba:1994na}, with appropriate boundary conditions, of Hirota finite difference equations for T-functions~\cite{Reshetikhin:1983vw,Klumper:1992vt,Kuniba:2010ir} (TT-system).

The most natural representation of T-functions in terms of Q-functions, using the basis of the single-index $Q$'s, $\{Q_{\{1\}},Q_{\{2\}},\dots,Q_{\{r+1\}}\}$, was constructed only for the $A_r$ algebra~\cite{Pronko:1999gh,Bazhanov:2001xm,Tsuboi:2009ud,Bazhanov:2010jq,Gromov:2010km,Derkachov:2006fw,Derkachov:2010qe,Kazakov:2015efa,Gromov:2014caa,Nepomechie:2020ixi}. Irreducible representations of $A_r$ are labelled by highest weights $(\lambda_1,\cdots,\lambda_{r})\in\mathbb{N}^{r}$ such that $\lambda_{1}\geqslant \lambda_{2}\geqslant \ldots\geqslant \lambda_{r}$ and, assuming $Q_\emptyset(x) = 1$, the T-functions read
\begin{equation}\label{TQAr}
T_{\lambda}^{(\tau)}(x)=Q_{\{1,\dots,r+1\}}(x)\frac{\underset{1\leqslant i,j\leqslant r+1}{\det} Q_{\{i\}}(x+\mu_j)}{\underset{1\leqslant i,j\leqslant r+1}{\det}Q_{\{i\}}(x+r+1-j)}\equiv Q_{\{1,\dots,r+1\}}(x) \frac{\left|Q_{\{i\}}^{[2\mu_j]}\right|_{r+1}}{\left|Q_{\{i\}}^{[2(r+1-j)]}\right|_{r+1}}\, .
\end{equation}
Here we introduced the shifted weights $\mu_j=\lambda_j+r-j+1$ for $j=1,\ldots,r$ and $\mu_{r+1}=0$ as well as the twist matrix ${\rm diag}(\tau_1,\tau_2,\dots,\tau_{r+1})$. We set $\prod_i\tau_i=1$ to restrict to  $SL(r+1)$.  In order to shorten the formulas we shall use the following notations throughout the article: $|M_{i,j}|_p \equiv \underset{1\leqslant i,j\leqslant p}{\det}M_{i,j}$ and $M^{[k]}\equiv M(x+\frac{k}{2})$, where $x$ is the spectral parameter. The single-index Q-functions in \eqref{TQAr} are  polynomials up to an exponential prefactor:
\begin{equation}\label{QofA}
    Q_{\{i\}}(x)=(\tau_{i})^{x}\left(x^{m_{i}}+C_{i,m_{i}-1}\,x^{m_{i}-1}+\dots+C_{i,0}\right)\, .
\end{equation}
The representation \eqref{TQAr} is the direct generalization of Weyl's formula for characters:
\begin{equation}\label{chiAr}
\chi_{\lambda}^{SL(r+1)}(\tau)=\frac{|\tau_i^{\mu_j}|_{r+1}}{|\tau_i^{r+1-j}|_{r+1}}\, .
\end{equation}
It is clear that \eqref{TQAr} behaves as $Q_{\{1,\dots,r+1\}}(x) \chi_{\lambda}^{SL(r+1)}(\tau) $ in the "classical" limit $x\to\infty$.

The goal of this article is to construct  a similar  QQ-system, together with a similar Weyl-type representation for T-matrices, for the $D_{r}$ algebra. The standard Weyl formula for $D_{r}$ characters is 
\begin{equation}\label{Weyl_ch_D}
\chi_{f}^{SO(2r)}(\tau)=\frac{|\tau_i^{\h_j}+\tau_i^{-\h_j}|_{r} + |\tau_i^{\h_j}-\tau_i^{-\h_j}|_{r}}{|\tau_i^{r-j}+\tau_i^{-r+j}|_{r}\,}\,,
\end{equation}
see e.g.~\cite{fulton2013representation,okounkov1996shifted,Campoleoni:2015qrh,Hatayama:1998zp}, with $\h_j=\m_j+r-j$ and the highest weights $\m_1\geqslant \m_2\geqslant\dots\geqslant \m_{r-1}\geqslant |\m_{r}|$ are all integers or all half-integers (the last one can also be negative).

However, in general, the situation for $D_{r}$  is more complicated than for $A_{r}$ algebra. The representations of the Lie algebra do not  "quantize" trivially, i.e. cannot be lifted to the Yangian algebra (apart from the symmetric and spinorial representations), see \cite{Ogievetsky:1986hu} and \cite{MacKay:1990mp} for an instructive example. Instead, in order to construct the T-functions, one has to introduce the representations acting in the so-called Kirillov-Reshetikhin modules~\cite{kirillov1990representations}. Such modules are known only for rectangular representations  $(a,s)$, see Appendix~\ref{app:Characters}. These representations have highest weights $\m_{1}=\m_{2}=\dots=\m_{a}=s$ and $f_{a+1}=\ldots=f_r=0$ for $a\leqslant r-2$ and $f_1=\ldots=f_{r-1}=\pm f_r = s/2$ for $a=\pm$.
The Kirillov-Reshetikhin characters are linear combinations of the above mentioned Weyl characters. The symmetric and spinorial  characters in Kirillov-Reshetikhin representation are not different from the Weyl characters \eqref{Weyl_ch_D} but in other representations they do differ. 

The generating function for characters in symmetric representations 
reads
\begin{align}\label{ChiD}   K_s(t,\{\tau_i\})=\frac{1-t^2}{\prod _{i=1}^r \left(1-t (\tau_i+1/\tau_i)+t^2\right)}
%=\frac{t^{-1}-t}{t^{r-1}\prod _{a=1}^r %\left(t^{-1}+t- u_a\right)}
=\sum_{k=0}^{\infty} t^k\, \chi_{k}(\tau)
\end{align}
so that they coincide with standard Weyl  characters \eqref{Weyl_ch_D}.
On the contrary, the KR-characters for totally antisymmetric representations already differ from usual $D_r$ Weyl characters \eqref{Weyl_ch_D}. 
The generating function of KR~characters for these representations   reads
\begin{align}\label{KaD}
K_a(t,\{\tau_i\})=\frac{\prod _{i=1}^r \left(1-t (\tau_i+1/\tau_i)+t^2\right)}{1-t^2}\,=\sum_{k=0}^{r-2} t^k\, \Psi_{k}(\tau)+\dots
%=\frac{t^{r-1}\prod _{a=1}^r \left(y- u_a\right)}{t^{-1}-t}\,.
\end{align}
where only the coefficients of $t^k$,  up to $t^{r-2}$ term, give the KR type antisymmetric characters $\Psi_{a=k}(\tau)$.
       
In this work, we propose a  QQ-system, appropriate for the $D_{r}$~algebra, and discuss the corresponding Hasse diagram. We will also introduce new  QQ$'$-type conditions. From either the QQ-system or these QQ$'$-type conditions, one can derive new Weyl-type formulas for T-functions in the symmetric and antisymmetric representations of Kirillov-Reshetikhin modules: T-functions are then given in terms of ratios of determinants involving a basic set of $r+1$  Q-functions, generalizing the classical Weyl-type formulas \eqref{ChiD} and \eqref{KaD}. In particular, we will show that these Weyl-type formulas are consistent with the tableau sum formulas for T-functions. The QQ-relations and QQ$'$-type conditions were checked using explicit expressions for T and Q-operators found in~\cite{Frassek:2020nki} at small lengths of the spin chain.

 \paragraph{Note added in version 3:} Shortly after this preprint appeared on the arXiv, the preprint \cite{Ekhammar:2020enr} with  partially overlapping results on QQ-systems for $D_r$ algebra was posted on the arXiv.

\section{Lax matrix construction and eigenvalues of T-operators}
\label{sec:Lax}
We start by introducing the the fundamental R-matrix of $\mathfrak{so}(2r)$ which was  written down in  \cite{Zamolodchikov:1978xm}. It is a matrix is of size $(2r)^2\times (2r)^2$ and it reads
\begin{equation}\label{eq:normalR}
\begin{split}
 R(x)=x(x+\kappa)\ID +(x+\kappa)\Pop -x\Qop \,.
% %  
%  \frac
%  {z^2}{(z+\frac{\kappa}{2})(z-\frac{\kappa}{2}) } \mathfrak{L}_{1}(z)&=\ID+(z-\frac{\kappa}{2})^{-1}\Pop -(z+\frac{\kappa}{2})^{-1}\Qop \,.\\
%  &=z^2+z(\Pop-\Qop)+\frac{\kappa}{2}(\Pop+\Qop-\frac{\kappa}{2})\,.\\ 
 \end{split}
\end{equation}  
Here $\kappa=r-1$, the letter $\ID$ denotes the identity matrix and the permutation and trace operator ($\Pop$, $\Qop$) are  defined as the tensor products
\begin{equation}
 \Pop =\sum_{i,j=1}^{2r}E_{ij}\otimes E_{ji}\,,\qquad  \Qop =\sum_{i,j=1}^{2r}E_{i,j}\otimes E_{i',j'}\,.
\end{equation} 
The elementary $2r\times 2r$ matrices $E_{ij}$ obey the standard relations $E_{ij}E_{kl}=\delta_{jk}E_{il}$.
We use the notation $i'=2r-i+1$. The R-matrix in \eqref{eq:normalR} is related by a similarity transformation to the one originally obtained in \cite{Zamolodchikov:1978xm}, cf.~\cite{Frassek:2020nki}, and generates the extended Yangian $X(\mathfrak{so}(2r))$ \cite{Arnaudon2006}.
% \begin{equation}
%  i'=2r-i+1\,.
% \end{equation} 
It is invariant under transformations 
\begin{equation}\label{eq:comR}
 [R(x),B\otimes B]=0\,,
\end{equation} 
if $B$ satisfies the orthogonality condition $BB'=\theta \ID$ with $\theta \in \mathbb{C}$ and  $B'_{ij}\equiv B_{j'i'}$.

\subsection{Transfer matrix construction for first fundamental}
In the following we focus on spin chains of length $N$ with the defining representation at each site. The quantum space of the spin chain is 
\begin{equation}
 V=\mathbb{C}^{2r}\otimes \ldots \otimes \mathbb{C}^{2r}\,.
\end{equation} 
The R-matrix \eqref{eq:normalR} allows to construct the fundamental transfer matrix $T=T_{1,1}$, i.e. with the defining representation in auxiliary space, which contains the Hamiltonian of the spin chain. It is also convenient to introduce the symmetric generalisations $T_{1,s}$ at this point. The required Lax matrix was given in  \cite{Reshetikhin:1986vd}. It reads
\begin{equation}\label{eq:fundlax}
\mathcal{L}(x)=x^2\ID+x\sum_{i,j=1}^{2r} \F_{ij}\otimes E_{ji}+\sum_{i,j=1}^{2r}G_{ij}\otimes E_{ji}\,.
\end{equation}
with
\begin{equation}
G_{ij}=\frac{1}{2}\sum_{k=1}^{2r}\F_{kj}\F_{ik}+\frac{\kappa}{2}\F_{ij}-\frac{1}{4}\left((\kappa-1)^2+2\kappa s+s^2\right)\delta_{ij}\,.
\end{equation} 
Here we introduce the generators $J_{ij}$ of $\mathfrak{so}(2r)$ obeying the commutation relations
\begin{equation}
 [\F_{ij},\F_{kl}]=\delta_{jk}\F_{il}-\delta_{i'k}\F_{j'l}-\delta_{jl'}\F_{ik'}+\delta_{il}\F_{j'k'}\,,
\end{equation} 
with $\F_{ij}=-\F_{j'i'}$. We stress that the formula for the Lax matrix only holds for symmetric representations with generators acting on the highest weight state $|{\rm hws}\rangle$  as follows
\begin{equation}
 \F_{ij}|{\rm hws}\rangle=0\,,\quad \text{for}\quad i<j\,,\qquad \F_{ii}|{\rm hws}\rangle=s\delta_{1i}|{\rm hws}\rangle
\end{equation} 
where $s\in\mathbb{N}$ for finite dimensional representations.
The generators in such representation satisfy the characteristic identity
\begin{equation}
\sum_{j,k=1}^{2r}\left(J_{ij} - \delta_{ij}\right) (J_{jk} + s\delta_{jk}) (J_{kl} -( s+ 2 \kappa)\delta_{kl})=0\,,
\end{equation} 
which is needed in order to satisfy the Yang-Baxter equation, see also \cite{Isaev:2015hak} for a recent discussion of such constraints.
A realisation of the generators $\F_{ij}$ for general $s$ in terms of oscillators can be found in \cite{Frassek:2020nki}.
The defining representation  $s=1$ can be realised via
\begin{equation}
 \F_{ij}=E_{ij}-E_{j'i'}\,.
\end{equation} 
We recover the R-matrix $\mathcal{L}(x)=R(x-\frac{\kappa}{2})$.

The first space in \eqref{eq:fundlax} with generators $J_{ij}$ serves as our auxiliary space and the quantum space is built from $N$ copies of the second one with matrix elements $E_{ij}$.
The  transfer matrix constructed from this monodromy is defined via
\begin{equation}\label{eq:transferm}
 T_{1,s}(x)=\tr\mathcal{D} \mathcal{L}_1(x)\mathcal{L}_2(x)\cdots \mathcal{L}_N(x)
\end{equation} 
where $\mathcal{L}_i(x)$ denotes the Lax matrix acting non-trivially on the $i$th spin chain site and the trace is taken over the representation with generators $J_{ij}$. We further introduced a diagonal twist
\begin{equation}\label{eq:ttwist}
 \mathcal{D}= \prod_{k=1}^r\tau_k^{J_{kk}}\,,
\end{equation} 
with the parameters $\tau\in \mathbb{C}^r$ that we already encountered in the definition of characters. Some symmetries of the transfer matrix constructed via \eqref{eq:transferm} can be found in  Appendix~\ref{app:sym}.

The Hamiltonian of the spin chain is obtained from the fundamental transfer matrix $T$ by taking the logarithmic derivative at the permutation point
\begin{equation}\label{eq:ham}
H= \left.\frac{\partial}{\partial x}\ln T(x)\right|_{x=\frac{\kappa}{2}}=\sum_{i=1}^N \mathcal{H}_{i,i+1}\,.
\end{equation} 
The Hamiltonian density is obtained from the logarithmic derivative of the R-matrix at the permutation point and it reads
\begin{equation}
 \mathcal{H}_{i,i+1}=\kappa^{-1}\left(\ID-\Qop+\kappa\Pop\right)_{i,i+1}
\end{equation} 
and $\mathcal{D}_N$ the twist \eqref{eq:ttwist} at site $N$ enters via $\mathcal{H}_{N,N+1}=\mathcal{D}_N\mathcal{H}_{N,1}\mathcal{D}_N^{-1}$. We also remind the reader that $\kappa=r-1$.

\subsection{Diagonalisation of fundamental transfer matrix}\label{sec:diag}

As discussed at the end of the previous section, the fundamental transfer matrix $T=T_{1,1}$ with $s=1$ contains the nearest-neighbour Hamiltonian and higher local charges. It has been diagonalised in \cite{Reshetikhin:1986vd,deVega:1986xj} using the algebraic Bethe ansatz, see also \cite{Martins:1997wb} for a different nesting procedure and \cite{Gerrard:2019dtc} for the trigonometric case.
One of the key observations is that the transfer matrix can be written as
\begin{equation}\label{eq:t}
 T(x)=T_+(x)+T_-(x)
\end{equation} 
where the two terms are related via
\begin{equation}
 T_\pm^t(-x)|_{\tau_i\to\tau_i^{-1}}=T_\mp(x)\,.
\end{equation} 
We note that the twist only slightly modifies the derivation of the spectrum of the transfer matrix in \cite{Reshetikhin:1986vd,deVega:1986xj}.
Following the same logic as in the references above we find the contributions of $T_\pm$ to the eigenvalues of the transfer matrix
\begin{equation}\label{eq:abaT}
 T_\pm(x)=q_0^{[1-r]}q_0^{[r-1]}\sum_{k=1}^{r}\tau_k^{\mp1}\frac{\Q_{k-1}^{[\pm(k-r+2)]}}{\Q_{k-1}^{[\pm(k-r)]}}\frac{\Q_{k}^{[\pm(k-r-1)]}}{\Q_{k}^{[\pm(k-r+1)]}}\,.
%  =\sum_{k=1}^{r}\frac{\Q_{k-1}^{[k-r+2]}(x)}{\Q_{k-1}^{[k-r]}(x)}\frac{\Q_{k}^{[k-r-1]}(x)}{\Q_{k}^{[k-r+1]}(x)}
\end{equation} 
with the notation $q^{[k]}\equiv q(x+\frac{k}{2})$. In \eqref{eq:abaT} above we introduced the Q-functions along the tail of the Dynkin diagram, cf. Figure~\ref{Dynkin}. 
This equation is valid on the level of operators. In the diagonal form the Q-functions are written in terms of the Bethe roots $x_i^{(j)}$ at level $j\in\{1,2,\ldots, r-2,+,-\}$ corresponding to the nodes of the Dynkin diagram as given in Figure~\ref{Dynkin}. The index $i$ takes values $i\in\{1,2,\ldots,m_j\}$. Here $m_j$ denotes the magnon numbers $
 \vec m=(m_1,\ldots,m_{r-2},m_+,m_{-})$. They are determined for a given state labelled by weight vector $\vec n$
%  through the  Cartan matrix $C$ of $D_r$ 
 via
% \begin{equation}\label{eq:weightmagnon}
%  \vec w = \vec m\cdot C-(N,0,\ldots,0)
% \end{equation} 
% \begin{equation}
% \small
%  \vec w = \left(2m_1-m_0-m_2,\ldots,2m_{r-3}-m_{r-4}-m_{r-2},2m_{r-2}-m_{r-3}-m_+-m_-, 2m_+-m_{r-2}, 2m_--m_{r-2}\right)
% \end{equation} 
\begin{equation}\label{pols}
\small
 \vec n = \left(\begin{array}{c} 2m_1-m_0-m_2\\ \vdots\\2m_{r-3}-m_{r-4}-m_{r-2}\\2m_{r-2}-m_{r-3}-m_+-m_-\\ 2m_+-m_{r-2}\\ 2m_--m_{r-2}\end{array}\right)
\end{equation} 
where $m_0=N$ is the length of the spin chain, see \cite{Reshetikhin:1986vd} and 
 $n_i=f_i-f_{i+1}$ for $1\leqslant i<r$ and $n_r=f_{r-1}+f_{r}$.
The first Q-functions along the tail of the Dynkin diagram  are then given by 
% \begin{equation}\label{eq:abaT}
%  \T_+(x)=\sum_{k=1}^{r}\frac{\Q_{k-1}^{[k-r+2]}(x)}{\Q_{k-1}^{[k-r]}(x)}\frac{\Q_{k}^{[k-r-1]}(x)}{\Q_{k}^{[k-r+1]}(x)}
% %  =\sum_{k=1}^{r}\frac{\Q_{k-1}^{[k-r+2]}(x)}{\Q_{k-1}^{[k-r]}(x)}\frac{\Q_{k}^{[k-r-1]}(x)}{\Q_{k}^{[k-r+1]}(x)}
% \end{equation}
\begin{equation}
 \Q_0(x)=x^N\,,\qquad
 \Q_i(x)=\prod_{j=1}^{m_i}(x-x_j^{(i)})\,,\quad 1\leqslant i\leqslant r-2\,.\\
\end{equation} 
Here $q_0$ does not depend on any Bethe roots and plays a role similar to that of the Q-functions for the full sets in $A$-type.
The last two Q-functions factorise:
\begin{equation}
 \Q_{r-1}=s_{+}s_{-}\,,\qquad 
 \Q_{r}=s_{+}^{[+1]}s_{+}^{[-1]}\,,
\end{equation} 
where $s_\pm$ are the Q-functions that correspond to the spinorial nodes. They are  polynomials of degree $m_\pm$ in the spectral parameter
\begin{equation}
s_\pm(x)=\prod_{i=1}^{m_\pm}(x-x_i^{(\pm)})\,.
\end{equation} 
It immediately follows that the last term in \eqref{eq:abaT} reduces to the more familiar form
\begin{equation}
 \frac{\Q_{r-1}^{[\pm 2]}}{\Q_{r-1}^{[0]}}\frac{\Q_{r}^{[\mp 1]}}{\Q_{r}^{[\pm1]}}=\frac{s_{-}^{[\pm2]} s_{+}^{[\mp2]}}{s_{-} s_{+}}\,.
\end{equation} 

From the definition of the Hamiltonian \eqref{eq:ham} and the eigenvalue equation \eqref{eq:t} of the transfer matrix we obtain the energy formula. The eigenvalues of the Hamiltonian are parametrised by the Bethe roots and read
\begin{equation}\label{Energy}
 E= \frac{r}{r-1}N+\frac{q'_{1}\left(-\frac{1}{2}\right)}{q_{1}\left(-\frac{1}{2}\right)}-\frac{q'_{1}\left(\frac{1}{2}\right)}{q_{1}\left(\frac{1}{2}\right)}=
 \frac{r}{r-1}N-\sum_{k=1}^{m_1}\left(\frac{1}{x_k^{(1)}+\frac{1}{2}}-\frac{1}{x_k^{(1)}-\frac{1}{2}}\right)\,,
\end{equation} 
cf.~\cite{Reshetikhin:1986vd}. As for the first fundamental representation of $A$-type, the energy eigenvalues only depend on the Bethe roots at the first nesting level.

\section{QQ-relations from Bethe ansatz equations }
\label{sec:QQfrom BAE}

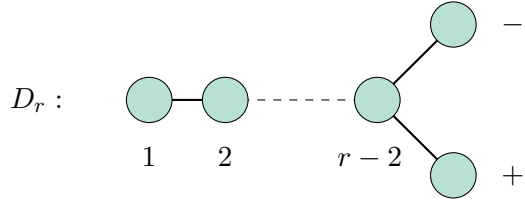
\begin{figure}[htb]
\begin{center}
 \begin{tikzpicture}
\foreach \a in {1} {
    \begin{scope}[shift={(0.7*\a,0)}]
      \draw[fill=SeaGreen!40] (0.3*\a,0) circle (0.3cm);
      \draw[black,thick] (0.3*\a+0.3,0)--(0.3*\a+0.7,0);
    \end{scope}
  }
      \draw[black,dashed] (2,0)--(4,0);
      \draw[black,thick] (4,0)--(5,1);
      \draw[black,thick] (4,0)--(5,-1);
      \draw[fill=SeaGreen!40] (2,0) circle (0.3cm);
      \draw[fill=SeaGreen!40] (4,0) circle (0.3cm);
      \draw[fill=SeaGreen!40] (5,1) circle (0.3cm);
      \draw[fill=SeaGreen!40] (5,-1) circle (0.3cm);
       \node [left] at (0,0) {$D_r:$};
       \node [below] at (1,-0.5) {$1$};
       \node [below] at (2,-0.5) {$2$};
       \node [below] at (3.9,-0.5) {$r-2$};
       \node [right] at (5.5,-1) {$+$};
       \node [right] at (5.5,1) {$-$};
\end{tikzpicture}
\end{center}
\caption{Dynkin diagram for $D_r$ Lie algebra.}
\label{Dynkin}
\end{figure}
The Bethe equations can be read off from the eigenvalue equation of the transfer matrix
\begin{equation}\label{eq:abaT2}
 T(x)=q_0^{[1-r]}q_0^{[r-1]}\sum_{k=1}^{r}\left[\tau_k^{-1}\frac{\Q_{k-1}^{[k-r+2]}}{\Q_{k-1}^{[k-r]}}\frac{\Q_{k}^{[k-r-1]}}{\Q_{k}^{[k-r+1]}}+\tau_k\frac{\Q_{k-1}^{[r-k-2]}}{\Q_{k-1}^{[r-k]}}\frac{\Q_{k}^{[r-k+1]}}{\Q_{k}^{[r-k-1]}}\right]\,,
\end{equation} 
which is obtained by combining \eqref{eq:t} and \eqref{eq:abaT}. 
When demanding that the transfer matrix is regular and Bethe roots are distinct the Bethe equations arise as pole cancellation conditions. They are conveniently written in terms of Q-functions as
\begin{align}\label{BAE}
-\frac{\tau_{k+1}}{\tau_{k}}&=\left(\frac{\Q_{k-1}^{[-1]}}{\Q_{k-1}^{[+1]}}\,\,
\frac{\Q_{k}^{[+2]}}{\Q_{k}^{[-2]}}\,\,\frac{\Q_{k+1}^{[-1]}}{\Q_{k+1}^{[+1]}}\right)_{k}\,,\qquad (k=1,2,\dots,r-3)\notag\\
-\frac{\tau_{r-1}}{\tau_{r-2}}&=\left(\frac{\Q_{r-3}^{[-1]}}{\Q_{r-3}^{[+1]}}\,\,
\frac{\Q_{r-2}^{[+2]}}{\Q_{r-2}^{[-2]}}\,\,\frac{\bS_+^{[-1]}}{\bS_+^{[+1]}}\,\,\frac{\bS_-^{[-1]}}{\bS_-^{[+1]}}\right)_{r-2}, \notag\\
-\frac{1}{\tau_{r-1}\tau_{r}}&=\left(\frac{\Q_{r-2}^{[-1]}}{\Q_{r-2}^{[+1]}}\,\,\frac{\bS_+^{[+2]}}{\bS_+^{[-2]}}\right)_{+}, \notag \\
-\frac{\tau_{r}}{\tau_{r-1}}&=\left(\frac{\Q_{r-2}^{[-1]}}{\Q_{r-2}^{[+1]}}\,\,\frac{\bS_-^{[+2]}}{\bS_-^{[-2]}}\right)_{-}\,,\notag\\
\end{align}   
where \((\dots)_k\) with $1\leqslant k\leqslant r-2$ indicates that the expression is taken at a root of \(\Q_k\) and \((\dots)_\pm\) at a root of $s_\pm$.

Along the tail of the Dynkin diagram, cf.~Figure~\ref{Dynkin}, we induce the standard \(A_n\) type Pl\"ucker  QQ-relation
\begin{equation} 
\frac{\tau_k-\tau_{k+1}}{\sqrt{\tau_k\tau_{k+1}}}\Q_{k-1}\Q_{k+1}=\sqrt{\frac{\tau_{k}}{\tau_{k+1}}}\Q_{k}^{+}\widetilde\Q_{k}^{\,-}-\sqrt{\frac{\tau_{k+1}}{\tau_{k}}}\Q_{k}^{-}\widetilde\Q_{k}^{\,+}    
\end{equation}  
where \(\Q_{k}\) and \(\widetilde\Q_{k}\) are two different Q-functions at the same level of the Hasse diagram, see Section~\ref{sec:qq-system} for that details. The form of the eigenvalue equation \eqref{eq:abaT2} is unchanged by such transformation. The Bethe ansatz equations can be restored by shifting
its argument  \(x\to x\pm 1\), taking each of relations at \(\Q_{k}=0\) and divide one by another.  At the fork of the Dynkin diagram, \((r-2)th \) node, the QQ-relation takes the form
\begin{equation}  \frac{\tau_{r-2}-\tau_{r-1}}{\sqrt{\tau_{r-2}\tau_{r-1}}}\Q_{r-3}\,\bS_+\,\bS_-=\sqrt{\frac{\tau_{r-2}}{\tau_{r-1}}}\Q_{r-2}^{+}\widetilde\Q_{r-2}^{\,-}
-\sqrt{\frac{\tau_{r-1}}{\tau_{r-2}}}\Q_{r-2}^{-}\widetilde\Q_{r-2}^{\,+}   \,.
\end{equation}
At the spinorial nodes $\pm$, the QQ-relations are
\begin{equation} 
\frac{\tau_{{r-1}}\tau_{r}-1}{\sqrt{\tau_{{r-1}}\tau_{r}}} \Q_{r-2}=\sqrt{\tau_{{r-1}}\tau_{r}}\bS_+^{+}\,\widetilde\bS_+^{\,-}-\frac{1}{\sqrt{\tau_{{r-1}}\tau_{r}}}\bS_+^{-}\,\widetilde\bS_+^{+}\,,
\end{equation}
\begin{equation}
\frac{\tau_{{r-1}}-\tau_{r}}{\sqrt{\tau_{{r-1}}\tau_{r}}}\Q_{r-2}=\sqrt{\frac{\tau_{r-1}}{\tau_{r}}}\bS_-^{+}\,\widetilde\bS_-^{\,-}-\sqrt{\frac{\tau_{r}}{\tau_{r-1}}}\bS_-^{-}\,\widetilde\bS_-^{\,+}\,.
\end{equation} 
These QQ-relations for spinorial nodes have  appeared in 
\cite{Masoero:2015lga} in relation to the ODE/IM correspodence \cite{Dorey:2007zx} and recently in \cite{Frenkel:2020iqq}. In Section~\ref{sec:qq-system} we propose a more general version of the QQ-relations.

\section{Basic (extremal) Q-functions}\label{sec:extremalQ}
A construction of the Q-operators corresponding to the extremal nodes of the Dynkin diagram, cf.~Figure~\ref{Dynkin}, was recently proposed in \cite{Frassek:2020nki}. The latter construction was inspired by the isomorphism $A_3\simeq D_3$, admits the expected asympotic behavior \eqref{pols} and has been checked by showing some functional relations of $r=4$ in some examples of finite length. All  functional relations in the following sections are consistent with the proposed Q-operators and have been verified  explicitly for several examples of finite length.

\subsection{Q-operator construction for first fundamental}\label{sec:firstfun}

We construct $2r$ Q-operators $Q_i$ with $1\leqslant i\leqslant2r$ corresponding to the first fundamental node. 
The Lax matrix needed is of the size $2r\times 2r$ with oscillators as entries and its leading order in the spectral parameter is quadratic. It reads
\begin{equation}\label{eq:quadlax}
L(z)=
 \left(\begin{BMAT}[5pt]{c|c|c}{c|c|c}
z^2+z(2-r-\wpp\wmm)+\frac{1}{4}\wpp\IDb\wpp^t\wmm^t\IDb\wmm&z\wpp-\frac{1}{2}\wpp\IDb\wpp^t\wmm^t\IDb&-\frac{1}{2}\wpp\IDb\wpp^t\\\
-z\wmm+\frac{1}{2}\IDb\wpp^t\wmm^t\IDb\wmm&z\ID-\IDb\wpp^t\wmm^t\IDb&-\IDb\wpp^t\\
-\frac{1}{2}\wmm^t\IDb\wmm&\wmm^t\IDb&1\\
      \end{BMAT}
\right)\,.
\end{equation}
The Lax matrix above contains $2(r-1)$ oscillators arranged into the vectors $\wpp$ and $\wmm$ as follows 
\begin{equation}
 \wpp=(\oad_{2},\ldots,\oad_{r},\oad_{r'},\ldots,\oad_{2'})\,,\qquad \wmm=(\oa_{2},\ldots,\oa_{r},\oa_{r'},\ldots,\oa_{2'})^t\,.
\end{equation}
They obey the standard commutation relations
\begin{equation}
 [\oa_i,\oad_j]=\delta_{ij}\,.
\end{equation} 
The matrix $\IDb$ is given in \eqref{eq:invar}.
The Q-operator $Q_1$ is defined as the regularised trace over the monodromy of the Lax matrices \eqref{eq:quadlax} which is constructed by taking the $N$-fold tensor product in the matrix space and multiplying in the auxiliary oscillator space:
\begin{equation}
 Q_1(x)=\tau_1^x\,\hat \tr\left[ D L^{[-1]}\otimes L^{[-1]}\otimes\ldots\otimes  L^{[-1]}\right]\,.
\end{equation} 
The twist matrix $D$ in the auxiliary space depends on the parameters $\tau_i$, cf.~\eqref{eq:ttwist} for the transfer matrix. In the case of the Q-operator $Q_1$ it reads
\begin{equation}
 D=\prod_{i=2}^r\left(\tau_i\tau_1^{-1}\right)^{\NN_i}\left(\tau^{-1}_i\tau_1^{-1}\right)^{\NN_{i'}}\,,
\end{equation} 
with the number operator $\NN_i=\oad_i\oa_i$. The trace is defined as
\begin{equation}
 \hat \tr (D X)=\frac{\tr(D X)}{\tr(D)}\,.
\end{equation} 
By construction of the Q-operators $Q_1$ belongs to the family of commuting operators. The Q-operator for the case $N=1$ is spelled out explicitly in Appendix~\ref{app:explicitQ}. 

From $Q_1$ we define the remaining $2r-1$ Q-operators at the first fundamental node.
For that we introduce the  transformation
\begin{equation}\label{eq:invtrans2}
\tilde B_{ij}=\sum_{\substack{k=1\\k\neq i,j}}^r (E_{k',k'}+E_{k,k})+E_{i',j'}+E_{j',i'}+E_{i,j}+E_{j,i}\,,
\end{equation} 
with $1\leqslant i\neq j\leqslant r$. It  belongs to the class of transformations discussed in \eqref{eq:comR} and commutes with the R-matrix. It follows that the Q-operators defined via
\begin{equation}
 Q_i(x)=\left. (\tilde B_{1,i}\otimes\ldots\otimes \tilde B_{1,i})\,Q_1(x)(\tilde B_{1,i}\otimes\ldots\otimes \tilde B_{1,i})\right|_{\tau_1\leftrightarrow\tau_i}\,,\qquad i=2,\ldots,r
\end{equation} 
and  
\begin{equation}\label{eq:Qrefl}
 Q_{i} (x)=\left.\left(\IDb\otimes\ldots\otimes\IDb\right)Q_{i'} (x)\left(\IDb\otimes\ldots\otimes\IDb\right)\right|_{\tau_i\to\tau_i^{-1}}\,,\qquad i=r+1,\ldots,2r
\end{equation} 
also belong to the family of commuting operators.
This defines us $2r$ Q-operators
\begin{equation}
\{ Q_1,Q_2,\ldots, Q_{2r}\}\,.
\end{equation} 
Up to the exponential prefactor, we identify the q-function $q_1$ with the eigenvalues of the Q-operator $Q_1$. Here we could have chosen any other single-index $Q$.
\subsection{Q-operator construction for spinor representations}\label{sec:spinq}

\newcommand{\LL}{\check L}
Similarly we proceed for the Q-operators corresponding to the spinorial nodes $\pm$ of the Dynkin diagram in Figure~\ref{Dynkin}. Here the Lax matrix is a $2\times 2$ block matrix with block size $r\times r$. It reads
\begin{equation}\label{eq:laxlin}
\LL(x)=
\left(\begin{BMAT}[5pt]{c:c}{c:c}
      x\,\ID+\ap\am&\ap\\
    \am&\ID
      \end{BMAT}
\right)\,,
\end{equation} 
and contains $\frac{r(r-1)}{2}$ pairs of oscillators $[\oa_{i,j},\oad_{k,l}]=\delta_{il}\delta_{jk}$. The submatrices $\ap$ and $\am$ are of the form
\begin{equation}\label{eq:ApAm}
\ap=\left(\begin{array}{cccc}
               \oad_{1,r'}&\cdots&\oad_{1,2'}&0\\
               \vdots  &\iddots &0&-\oad_{1,2'}\\
             \oad_{r-1,r'}&0&\iddots&\vdots\\
               0&-\oad_{r-1,r'}&\cdots&-\oad_{1,r'}
              \end{array}\right),\quad
              \am=\left(\begin{array}{cccc}
              - \oa_{r',1}&\cdots&-\oa_{r',r-1}&0\\
              \vdots&\iddots &0&\oa_{r',r-1}\\
               -\oa_{2',1}&0&\iddots&\vdots\\
               0&\oa_{2',1}&\cdots&\oa_{r',1}
              \end{array}\right).
\end{equation} 
Similar as before we define the Q-operator as the trace of the monodromy built out of the Lax matrix $\LL$ above as
\begin{equation}
 S(x)=\left(\tau_1\cdots\tau_r\right)^{\frac{x}{2}} \,\hat \tr\left[  \check D\, \LL^{[1-r]}\otimes \LL^{[1-r]}\otimes\ldots\otimes \LL^{[1-r]}\right]\,.
\end{equation} 
Here we introduced the twist in the auxiliary space  via
\begin{equation}
 \check D=\prod_{1\leqslant i<j\leqslant r}\left(\tau_{i}\tau_j\right)^{\oad_{i,j'}\oa_{j',i}}\,.
\end{equation} 
The remaining Q-operators at the spinorial nodes are obtained through the similarity transformation
\begin{equation}\label{eq:invtrans}
 B(\vec\alpha)=\frac{1}{2}\sum_{i=1}^r\left((1+\alpha_i)(E_{i',i'}+E_{i,i})+(1-\alpha_i)(E_{i',i}+E_{i,i'})\right)\,,
\end{equation} 
with $\alpha_i=\pm 1$, that commutes with the R-matrix, cf.~\eqref{eq:comR}, and subsequently inverting the twist parameters. For $\alpha_i=1$ the matrix $B(\vec\alpha)$ reduces to the identity. We define
\begin{equation}\label{eq:spinQ}
 S_{\vec\alpha}(x)=(B(\vec\alpha)\otimes\ldots\otimes B(\vec\alpha))S(x)(B(\vec\alpha)\otimes\ldots\otimes B(\vec\alpha))|_{\tau_i\to\tau_i^{\alpha_i}}\, ,
\end{equation} 
labelled by $\vec\alpha=(\alpha_1,\ldots,\alpha_r)$ with $\alpha_i=\pm1$. By construction the $2^r$ operators $S_{\vec\alpha}$ commute with one another. We choose to identify $s_\pm$ with $S_{(+1,\ldots,+1,\pm1)}$ up to the exponential prefactor.

\section{The QQ-system for $D_r$}\label{sec:qq-system}
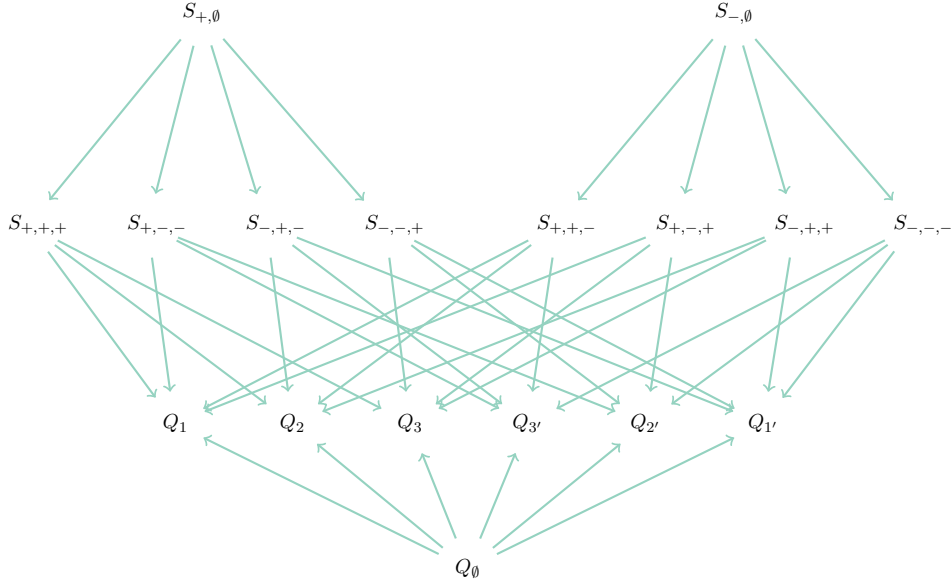
\begin{figure}[htb]
 \begin{center}
  \begin{tikzpicture}[scale=0.7, transform shape]

    \node [empty](origin){};
    
    \node [yellowRectangle, above=3cm of origin] (ppp) {$S_{+,+,+}$};
    \node [yellowRectangle, right=1cm of ppp] (pmm) {$S_{+,-,-}$};
    \node [yellowRectangle, right=1cm of pmm] (mpm) {$S_{-,+,-}$};
    \node [yellowRectangle, right=1cm of mpm] (mmp)  {$S_{-,-,+}$};
    \node [yellowRectangle, right=2cm of mmp] (ppm) {$S_{+,+,-}$};
    \node [yellowRectangle, right=1cm of ppm] (pmp) {$S_{+,-,+}$};
    \node [yellowRectangle, right=1cm of pmp] (mpp) {$S_{-,+,+}$};
    \node [yellowRectangle, right=1cm of mpp] (mmm) {$S_{-,-,-}$};
%     \node [yellowRectangle, above=1.5cm of origin] (s111) {$Q_{\{a_1,a_2\}}$};

    \node [yellowRectangle, right=2cm of origin] (q1) {$Q_{1}$};
    \node [yellowRectangle, right=1cm of q1] (q2) {$Q_{2}$};
    \node [yellowRectangle, right=1cm of q2] (q3) {$Q_{3}$};
    \node [yellowRectangle, right=1cm of q3] (q4) {$Q_{3'}$};
    \node [yellowRectangle, right=1cm of q4] (q5) {$Q_{2'}$};
    \node [yellowRectangle, right=1cm of q5] (q6) {$Q_{1'}$};

    \node [yellowRectangle, above=7cm  of origin,xshift=3.25cm] (one) {$S_{+,\emptyset}$};
    \node [yellowRectangle, above=7cm  of origin,xshift=13.25cm] (two) {$S_{-,\emptyset}$};

    \node [yellowRectangle, below=2cm  of origin,xshift=8.25cm] (zet) {$Q_\emptyset$};

    \path [line] (ppp) --  (q1);
    \path [line] (ppp) --  (q2);
    \path [line] (ppp) --  (q3);
    \path [line] (pmm) --  (q1);
    \path [line] (pmm) --  (q4);
    \path [line] (pmm) --  (q5);
    \path [line] (mpm) --  (q2);
    \path [line] (mpm) --  (q4);
    \path [line] (mpm) --  (q6);
    \path [line] (mmp) --  (q3);
    \path [line] (mmp) --  (q5);
    \path [line] (mmp) --  (q6);
    \path [line] (one) --  (ppp);
    \path [line] (one) --  (pmm);
    \path [line] (one) --  (mpm);
    \path [line] (one) --  (mmp);
    
%     \tikzstyle{line} = [
%     draw,
% %     -latex'->,
% ->,
%     thick, Red,
% ]
    \path [line] (two) --  (mmm);
    \path [line] (two) --  (ppm);
    \path [line] (two) --  (mpp);
    \path [line] (two) --  (pmp);
    
       \path [line] (mmm) --  (q6);
         \path [line] (mmm) --  (q5);
           \path [line] (mmm) --  (q4);
       \path [line] (mpp) --  (q2);
         \path [line] (mpp) --  (q3);
           \path [line] (mpp) --  (q6);
       \path [line] (pmp) --  (q1);
         \path [line] (pmp) --  (q3);
           \path [line] (pmp) --  (q5);
       \path [line] (ppm) --  (q1);
         \path [line] (ppm) --  (q2);
           \path [line] (ppm) --  (q4);

 \path [line] (zet) --  (q1);
    \path [line] (zet) --  (q2);
    \path [line] (zet) --  (q3);
    \path [line] (zet) --  (q4);
    \path [line] (zet) --  (q5);
    \path [line] (zet) --  (q6);
\end{tikzpicture}
 \end{center}
\caption{Hasse diagram of mixed orientation for $D_3$. In a particular gauge, the functions at the first and last level nodes can be chosen as in  \eqref{eq:polfix}.  }
 \label{fig:Hasse2}
\end{figure}

In this section we introduce the QQ-system. It has been verified at small finite length using the construction \cite{Frassek:2020nki} that was reviewed in Section~\ref{sec:extremalQ}. In total we have $3^r-2^{r-1}r+2$  Q-functions, see Figure~\ref{fig:Hasse2} and Figure~\ref{fig:hasse4} for $r=3,4$ examples.

The QQ-relations along the tail of the Dynkin diagram have a structure similar to those for $A_r$ but the labeling of single-index functions is different. We shall say that a subset $I$ of $\{1,\dots,2r\}$ is acceptable if for all $1\leqslant k\leqslant r$, the integers $k$ and $k'=2r-k+1$ do not both belong to $I$. In particular, an acceptable set cannot have more than $r$ elements: $|I|\leqslant r$. A Q-function $Q_I$ is associated to each acceptable $I$ and these functions satisfy the relations
\begin{equation} \label{eq:QQrel}
 Q_{J\cup \{i\}}^{[+1]}Q_{J\cup\{j\}}^{[-1]}- Q_{J\cup\{i\}}^{[-1]}Q_{J\cup\{j\}}^{[+1]}= \frac{\tau_i-\tau_j}{\sqrt{\tau_i\tau_j}}Q_{J} Q_{J\cup\{i,j\}}
\end{equation} 
where $\tau_i=\tau_{i'}^{-1}$ for $i>r$, $\{i,i'\}\cap\{j,j'\} = \emptyset$, $J$ is acceptable of order at most $r-2$ and does not contain $i,i',j$ or $j'$. We have excluded here the case where $k$ and $k'$ are contained in the same set as the Q-functions defined this way would  not have the  expected asymptotic behavior.
For the $D_r$ spin chains under consideration, the Q-operator of the empty set can be conveniently  fixed as 
\begin{equation}
 Q_\emptyset(x)=x^N\,,
\end{equation} 
though such a choice for a generic $D_r$ QQ-system can be changed by a gauge transformation, see below in this section.

As discussed at the end of the Section~\ref{sec:Lax}, the Q-operators $Q_I$ with $|I|=r-1$ or $|I|=r$ factorise into spinorial Q-functions. More precisely,
\begin{equation}\label{eq:fac1}
 Q_{\{i_1,\ldots,i_{r-1}\}}=S_{\{i_1,\ldots,i_{r-1},i_r\}}S_{\{i_1,\ldots,i_{r-1},i_r'\}}\,,
\end{equation} 
and 
\begin{equation}\label{eq:fac2}
 Q_{\{i_1,\ldots,i_r\}}=S^{[+1]}_{\{i_1,\ldots,i_r\}}S_{\{i_1,\ldots,i_r\}}^{[-1]}\,.
\end{equation} 
The set notation for the Q-operators $S_I$ can be mapped to the notation $S_{\vec \alpha}$ in the previous subsection using $\vec\alpha$ as follows: to an acceptable set $I$ of order $r$ we associate $\vec\alpha$ such that, for $1\leqslant i\leqslant r$,
\begin{equation}\label{Itoalpha}
\alpha_i =  \left\{
\begin{array}{ll}
+1\quad \text{if}\quad   i\in I\\
-1\quad \text{if}\quad   i'\in I
\end{array}
\right.\, .
\end{equation} 
We thus obtain a one-to-one correspondence between $S_{\{i_1,\ldots,i_r\}}$ and $S_{\vec \alpha}$ as defined in \eqref{eq:spinQ}. We further remark that the polynomial structure of the spinorial Q-functions allows to determine them from the quadratic relations \eqref{eq:fac1} and \eqref{eq:fac2}.

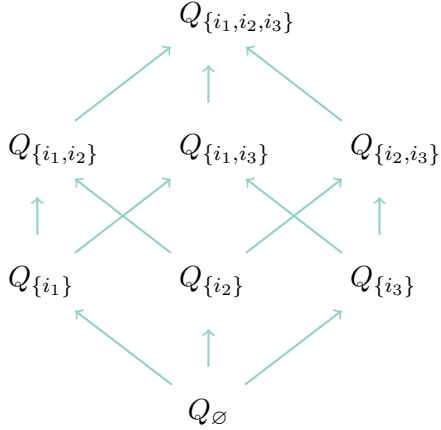
\begin{figure}[htb]
\begin{center}
\begin{tikzpicture}

    \node [empty](origin){};
    
    \node [yellowRectangle, above=0cm of origin] (q1) {$Q_{\{i_1,i_2\}}$};
    \node [yellowRectangle, below=0.5cm of q1] (qsc1) {$Q_{\{i_1\}}$};
    \node [yellowRectangle, right=1cm of q1](q2){$Q_{\{i_1,i_3\}}$};
    \node [yellowRectangle, right=1cm of qsc1] (qsc2) {$Q_{\{i_2\}}$};
    \node [yellowRectangle, right=1cm of q2](q3){$Q_{\{i_2,i_3\}}$};
    \node [yellowRectangle, right=1cm of qsc2](qsc3){$Q_{\{i_3\}}$};
    \node [yellowRectangle, below=0.5cm of qsc2](q0){$Q_{\varnothing}$};
    \node [yellowRectangle, above=0.5cm of q2](qf){$Q_{\{i_1,i_2,i_3\}}$};
    \path [line] (qsc1) --  (q2);
    \path [line] (qsc1) --  (q1);
    \path [line] (qsc2) --  (q1);
    \path [line] (qsc2) --  (q3);
    \path [line] (qsc3) -- (q2);
    \path [line] (qsc3) -- (q3);
    \path [line] (q0) --  (qsc1);
    \path [line] (q0) -- (qsc2);
    \path [line] (q0) -- (qsc3);
    \path [line] (q1) --  (qf);
    \path [line] (q2) -- (qf);
    \path [line] (q3) -- (qf);
\end{tikzpicture}
\caption{Directed Hasse diagram for $D_3$}\label{fig:Hasse1}
\end{center}
\end{figure}

The QQ-relation \eqref{eq:QQrel} can be summarised in a Hasse diagram which reminds that of the $A_r$ case. The latter is exemplified for $D_3$ in Figure~\ref{fig:Hasse1} and for $D_4$ in Figure~\ref{fig:hasse4}. However, the last two levels are nontrivial:  the level $|I|=(r-1)$ factorises according to \eqref{eq:fac1} and the level  $|I|=r$, cf.~\eqref{eq:fac2}. In total, there are 
\begin{equation}\label{eq:number}
 2^k\binom{r}{k}
\end{equation}
Q-funtions $Q_{I}$ at level $k$. At the last two levels the Q-functions split according to \eqref{eq:fac1} and \eqref{eq:fac2} such that \eqref{eq:number} remains valid for $1\leqslant k\leqslant r-2$ and $2\cdot2^{r-1}$ spinorial Q-functions $S_{\vec\alpha}$ distinguished by  $\prod_{i=1}^r\alpha_i=\pm1$ are assigned to $(r-1)$'th and $r$'th spinor node, respectively.

Let $S_I$ and $S_{J}$ denote two Q-functions labelled by some acceptable sets $I$ and $J$ verifying $|I\cap J|=r-2$, i.e.
\begin{equation}
 I=\{i_1,\ldots,i_{r-2},i_{r-1},i_{r}\} \qquad\text{and}\qquad
 J=\{i_1,\ldots,i_{r-2},i_{r-1}',i_{r}'\} \, .
\end{equation}
It follows that they must belong to the same node of the Dynkin diagram.
 Among them we have the QQ-relations
\begin{equation}\label{SS relation}
 S^{[+1]}_{I} S^{[-1]}_{J}- S^{[-1]}_{I} S^{[+1]}_{J}=\frac{\tau_{i_{r-1}}\tau_{i_r}-1}{\sqrt{\tau_{i_{r-1}}\tau_{i_r}}}
 Q_{I\cap J}
\end{equation} 
which relate the spinorial Q-functions to the last Q-functions on the tail of the Dynkin diagram, i.e. at the $r-2$'th node. Notice that for each level $r-2$ Q-function there are two ways to obtain them from spinorial Q-functions, e.g.: when $r=4$, $I\cap J= \{1,3\}$ can come from $I=\{1,3,2,4\}$ and $J=\{1,3,7,5\}$ or from $I=\{1,3,2,5\}$ and $J=\{1,3,7,4\}$. This relation allows us to resolve the last two levels in the $D_r$ Hasse diagram, cf.~Figure~\ref{fig:Hasse2} and Figure~\ref{fig:hasse4} for the cases $D_3$ and $D_4$, respectively. A more detailed explanation of the elements of the Hasse diagram can be found in Appendix~\ref{app:hasse}. Let us note that the $D_3$ Hasse diagram of Figure~\ref{fig:Hasse2} is (up to a gauge transformation setting $Q_\emptyset$ to $1$) the same as the $A_3$ one, this is not surprising since the two algebras are isomorphic. The $D_4$ Hasse diagram, on the other hand, is new and gives a clear idea of the higher rank picture.    
Here we used the directions of the arrows in the Hasse diagram to distinguish from the QQ-relations \eqref{eq:QQrel} used for the last nodes as depicted in Figure~\ref{fig:Hasse1}.

\newcommand{\dis}{0.15}
\newcommand{\diss}{0.1}
% \begin{landscape}
% \begin{sidewaysfigure}
\begin{figure}[htb]
\begin{center}
\begin{tikzpicture}[thick,scale=0.6, every node/.style={transform shape}]

\tikzstyle{yellowRectangle} = [
    circle,
    node distance=1 cm,
    text width=1.1 em,
    text centered,
    rounded corners,
    minimum height=0.2 cm,
    minimum width=0.3 cm,
    fill=gray!20,
    thick
]
\tikzstyle{syellowRectangle} = [
    rectangle,
    node distance=0.5 cm,
    text width=3 em,
    text centered,
    rounded corners,
    minimum height=0.2 cm,
    minimum width=0.4 cm,
    fill=gray!30,
    thick
]
    \node [empty](origin){};

    \node [yellowRectangle, above=1cm of origin,xshift=0.3cm] (q12) {${12}$};
    \node [yellowRectangle, right=\dis cm of q12] (q13) {${13}$};
    \node [yellowRectangle, right=\dis cm of q13] (q14) {${14}$};
    \node [yellowRectangle, right=\dis cm of q14] (q23) {${23}$};
    \node [yellowRectangle, right=\dis cm of q23] (q24) {${24}$};
    \node [yellowRectangle, right=\dis cm of q24] (q34) {${34}$};
    
    \node [yellowRectangle, right=0.5cm of q34] (q15) {${14'}$};
    \node [yellowRectangle, right=\dis cm of q15] (q16) {${13'}$};
    \node [yellowRectangle, right=\dis cm of q16] (q17) {${12'}$};
    \node [yellowRectangle, right=\dis cm of q17] (q25) {${24'}$};
    \node [yellowRectangle, right=\dis cm of q25] (q26) {${23'}$};
    \node [yellowRectangle, right=\dis cm of q26] (q28) {${21'}$};
    \node [yellowRectangle, right=\dis cm of q28] (q35) {${34'}$};
    \node [yellowRectangle, right=\dis cm of q35] (q37) {${32'}$};
    \node [yellowRectangle, right=\dis cm of q37] (q38) {${31'}$};
    \node [yellowRectangle, right=\dis cm of q38] (q46) {${43'}$};
    \node [yellowRectangle, right=\dis cm of q46] (q47) {${42'}$};
    \node [yellowRectangle, right=\dis cm of q47] (q48) {${41'}$};
    
    \node [yellowRectangle, right=0.5cm of q48] (q56) {${4'3'}$};
    \node [yellowRectangle, right=\dis cm of q56] (q57) {${4'2'}$};
    \node [yellowRectangle, right=\dis cm of q57] (q58) {${4'1'}$};
    \node [yellowRectangle, right=\dis cm of q58] (q67) {${3'2'}$};
    \node [yellowRectangle, right=\dis cm of q67] (q68) {${3'1'}$};
    \node [yellowRectangle, right=\dis cm of q68] (q78) {${2'1'}$};

    \node [yellowRectangle, below=2.2cm of origin,xshift=6cm] (q1) {${1}$};
    \node [yellowRectangle, right=1cm of q1] (q2) {${2}$};
    \node [yellowRectangle, right=1cm of q2] (q3) {${3}$};
    \node [yellowRectangle, right=1cm of q3] (q4) {${4}$};
    \node [yellowRectangle, right=1cm of q4] (q5) {${4'}$};
    \node [yellowRectangle, right=1cm of q5] (q6) {${3'}$};
    \node [yellowRectangle, right=1cm of q6] (q7) {${2'}$};
    \node [yellowRectangle, right=1cm of q7] (q8) {${1'}$};
    
    \node [syellowRectangle, above=8cm of origin,xshift=0cm] (s1) {\footnotesize${ ++++}$};
    \node [syellowRectangle, right=\diss cm of s1] (s2) {\footnotesize${ ++--}$};
    \node [syellowRectangle, right=\diss cm of s2] (s3) {\footnotesize${ +-+-}$};
    \node [syellowRectangle, right=\diss cm of s3] (s4) {\footnotesize${ +--+}$};
    \node [syellowRectangle, right=\diss cm of s4] (s5) {\footnotesize${ -++-}$};
    \node [syellowRectangle, right=\diss cm of s5] (s6) {\footnotesize${ -+-+}$};
    \node [syellowRectangle, right=\diss cm of s6] (s7) {\footnotesize${ --++}$};
    \node [syellowRectangle, right=\diss cm of s7] (s8) {\footnotesize${----}$};
    
    \node [syellowRectangle, above=8cm of origin,xshift=14.7cm] (ss1) {\footnotesize${ +++-}$};
    \node [syellowRectangle, right=\diss cm of ss1] (ss2) {\footnotesize${ ++-+}$};
    \node [syellowRectangle, right=\diss cm of ss2] (ss3) {\footnotesize${ +-++}$};
    \node [syellowRectangle, right=\diss cm of ss3] (ss4) {\footnotesize${ +---}$};
    \node [syellowRectangle, right=\diss cm of ss4] (ss5) {\footnotesize${ -+++}$};
    \node [syellowRectangle, right=\diss cm of ss5] (ss6) {\footnotesize${ -+--}$};
    \node [syellowRectangle, right=\diss cm of ss6] (ss7) {\footnotesize${ --+-}$};
    \node [syellowRectangle, right=\diss cm of ss7] (ss8) {\footnotesize${ ---+}$};
    
    \node [yellowRectangle, above=11cm  of origin,xshift=5.5cm] (one) {$\emptyset_+$};
    \node [yellowRectangle, above=11cm  of origin,xshift=19.9cm] (two) {$\emptyset_-$};

    \node [yellowRectangle, below=5cm  of origin,xshift=12.5cm] (zet) {$\emptyset$};

    \path [line] (q1) --  (q12);
    \path [line] (q1) --  (q13);
    \path [line] (q1) --  (q14);
    \path [line] (q1) --  (q15);
    \path [line] (q1) --  (q16);
    \path [line] (q1) --  (q17);
    \path [line] (q2) --  (q12);
    \path [line] (q2) --  (q23);
    \path [line] (q2) --  (q24);
    \path [line] (q2) --  (q25);
    \path [line] (q2) --  (q26);
    \path [line] (q2) --  (q28);
    \path [line] (q3) --  (q13);
    \path [line] (q3) --  (q23);
    \path [line] (q3) --  (q34);
    \path [line] (q3) --  (q35);
    \path [line] (q3) --  (q37);
    \path [line] (q3) --  (q38);
    \path [line] (q4) --  (q14);
    \path [line] (q4) --  (q24);
    \path [line] (q4) --  (q34);
    \path [line] (q4) --  (q46);
    \path [line] (q4) --  (q47);
    \path [line] (q4) --  (q48);
    \path [line] (q5) --  (q15);
    \path [line] (q5) --  (q25);
    \path [line] (q5) --  (q35);
    \path [line] (q5) --  (q56);
    \path [line] (q5) --  (q57);
    \path [line] (q5) --  (q58);
    \path [line] (q6) --  (q16);
    \path [line] (q6) --  (q26);
    \path [line] (q6) --  (q46);
    \path [line] (q6) --  (q56);
    \path [line] (q6) --  (q67);
    \path [line] (q6) --  (q68);
    \path [line] (q7) --  (q17);
    \path [line] (q7) --  (q37);
    \path [line] (q7) --  (q47);
    \path [line] (q7) --  (q57);
    \path [line] (q7) --  (q67);
    \path [line] (q7) --  (q78);
    \path [line] (q8) --  (q28);
    \path [line] (q8) --  (q38);
    \path [line] (q8) --  (q48);
    \path [line] (q8) --  (q58);
    \path [line] (q8) --  (q68);
    \path [line] (q8) --  (q78);

    \path [line] (zet) --  (q1);
    \path [line] (zet) --  (q2);
    \path [line] (zet) --  (q3);
    \path [line] (zet) --  (q4);
    \path [line] (zet) --  (q5);
    \path [line] (zet) --  (q6);
    \path [line] (zet) --  (q7);
    \path [line] (zet) --  (q8);

\tikzstyle{line} = [
    draw,
%     -latex'->,
->,
    thick, Blue!60,
]

    \path [line] (s1) --  (q12);
    \path [line] (s1) --  (q13);
    \path [line] (s1) --  (q14);
    \path [line] (s1) --  (q23);
    \path [line] (s1) --  (q24);
    \path [line] (s1) --  (q34);

    \path [line] (s2) --  (q12);
    \path [line] (s2) --  (q15);
    \path [line] (s2) --  (q16);
    \path [line] (s2) --  (q25);
    \path [line] (s2) --  (q26);
    \path [line] (s2) --  (q56);
    
    \path [line] (s3) --  (q13);
    \path [line] (s3) --  (q15);
    \path [line] (s3) --  (q17);
    \path [line] (s3) --  (q35);
    \path [line] (s3) --  (q37);
    \path [line] (s3) --  (q57);

    \path [line] (s4) --  (q14);
    \path [line] (s4) --  (q16);
    \path [line] (s4) --  (q17);
    \path [line] (s4) --  (q46);
    \path [line] (s4) --  (q47);
    \path [line] (s4) --  (q67);

    \path [line] (s5) --  (q23);
    \path [line] (s5) --  (q25);
    \path [line] (s5) --  (q28);
    \path [line] (s5) --  (q35);
    \path [line] (s5) --  (q38);
    \path [line] (s5) --  (q58);

    \path [line] (s6) --  (q24);
    \path [line] (s6) --  (q26);
    \path [line] (s6) --  (q28);
    \path [line] (s6) --  (q46);
    \path [line] (s6) --  (q48);
    \path [line] (s6) --  (q68);

    \path [line] (s7) --  (q34);
    \path [line] (s7) --  (q37);
    \path [line] (s7) --  (q38);
    \path [line] (s7) --  (q47);
    \path [line] (s7) --  (q48);
    \path [line] (s7) --  (q78);

    \path [line] (s8) --  (q56);
    \path [line] (s8) --  (q57);
    \path [line] (s8) --  (q58);
    \path [line] (s8) --  (q67);
    \path [line] (s8) --  (q68);
    \path [line] (s8) --  (q78);
    
    \path [line] (one) --  (s1);
    \path [line] (one) --  (s2);
    \path [line] (one) --  (s3);
    \path [line] (one) --  (s4);
    \path [line] (one) --  (s5);
    \path [line] (one) --  (s6);
    \path [line] (one) --  (s7);
    \path [line] (one) --  (s8);

\tikzstyle{line} = [
    draw,
%     -latex'->,
->,
    thick, Red!60,
]
    \path [line] (two) --  (ss1);
    \path [line] (two) --  (ss2);
    \path [line] (two) --  (ss3);
    \path [line] (two) --  (ss4);
    \path [line] (two) --  (ss5);
    \path [line] (two) --  (ss6);
    \path [line] (two) --  (ss7);
    \path [line] (two) --  (ss8);

    \path [line] (ss1) --  (q12);
    \path [line] (ss1) --  (q13);
    \path [line] (ss1) --  (q23);
    \path [line] (ss1) --  (q15);
    \path [line] (ss1) --  (q25);
    \path [line] (ss1) --  (q35);
    
    \path [line] (ss2) --  (q12);
    \path [line] (ss2) --  (q14);
    \path [line] (ss2) --  (q24);
    \path [line] (ss2) --  (q16);
    \path [line] (ss2) --  (q26);
    \path [line] (ss2) --  (q46);
    
    \path [line] (ss3) --  (q13);
    \path [line] (ss3) --  (q14);
    \path [line] (ss3) --  (q34);
    \path [line] (ss3) --  (q17);
    \path [line] (ss3) --  (q37);
    \path [line] (ss3) --  (q47);
    
    \path [line] (ss4) --  (q15);
    \path [line] (ss4) --  (q16);
    \path [line] (ss4) --  (q17);
    \path [line] (ss4) --  (q56);
    \path [line] (ss4) --  (q57);
    \path [line] (ss4) --  (q67);

    \path [line] (ss5) --  (q23);
    \path [line] (ss5) --  (q24);
    \path [line] (ss5) --  (q34);
    \path [line] (ss5) --  (q28);
    \path [line] (ss5) --  (q38);
    \path [line] (ss5) --  (q48);

    \path [line] (ss6) --  (q56);
    \path [line] (ss6) --  (q58);
    \path [line] (ss6) --  (q68);
    \path [line] (ss6) --  (q26);
    \path [line] (ss6) --  (q25);
    \path [line] (ss6) --  (q28);

    \path [line] (ss7) --  (q57);
    \path [line] (ss7) --  (q58);
    \path [line] (ss7) --  (q78);
    \path [line] (ss7) --  (q35);
    \path [line] (ss7) --  (q37);
    \path [line] (ss7) --  (q38);
    
    \path [line] (ss8) --  (q67);
    \path [line] (ss8) --  (q68);
    \path [line] (ss8) --  (q78);
    \path [line] (ss8) --  (q46);
    \path [line] (ss8) --  (q47);
    \path [line] (ss8) --  (q48);
    
\end{tikzpicture}
% \end{sidewaysfigure}
\caption{Hasse diagram of mixed orientation for $D_4$. Here, the level 1 and level 2  Q-operators $Q_I$ are abbreviated by their index set $I$. The third level contains the spinorial Q-operators $S_{\vec\alpha}$ which are abbreviated by $\vec \alpha$. Finally, we have $Q_\emptyset$ (denoted by $\emptyset$) at the  lowest level and $S_{\pm,\emptyset}$ (denoted by $\emptyset_\pm$)  at the highest level. These are proportional to the identity and can be fixed via \eqref{eq:polfix}.}\label{fig:hasse4}
\end{center}
\end{figure}
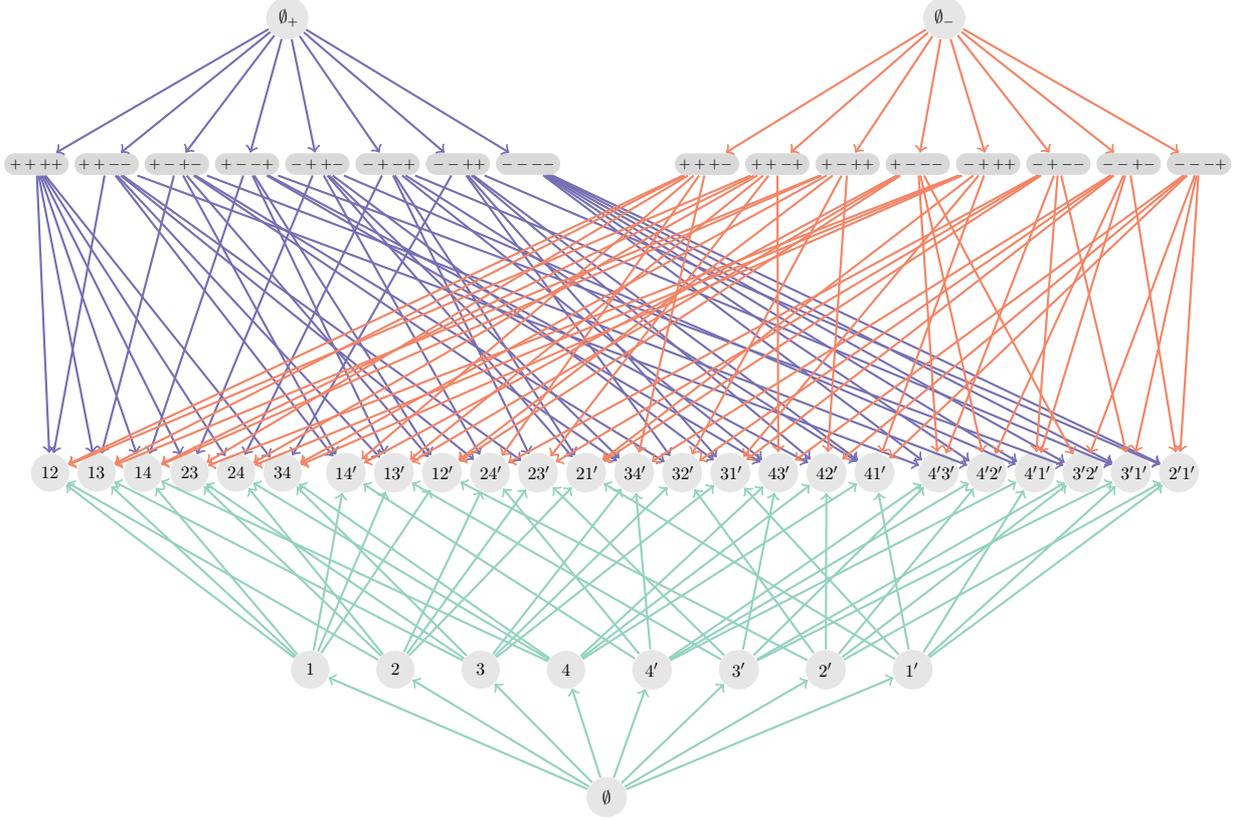

Using the QQ-relations in \eqref{eq:QQrel} we can express all Q-functions $Q_I$ in terms of Casoratian determinants of single-index Q-functions. We find
\begin{equation}\label{eq:detform}
Q_{\{i_1,\ldots, i_k\}}=\frac{(\sqrt{\tau_{i_1}\cdots\tau_{i_k}})^{\,k-1}}{\prod_{1\leqslant a<b \leqslant k}\left(\tau_{i_a}-\tau_{i_b}\right)}
\frac{\left|Q_{\{i_a\}}^{[k+1-2b]}\right|_k}{ \prod_{l=1}^{k-1}Q_\emptyset^{[k-2l]}}
\end{equation} 
with $i_a\neq i_b$, $i_a\neq i_b'$ and $\tau_{i}=\tau_{i'}^{-1}$ for $i>r$.
Similar formulas exist with spinorial Q-functions: if $I$ is an acceptable set of order $k\leqslant r-2$ and $i_{k+1},\dots,i_{r}$ are such that $I_r = I\cup\{i_{k+1},\dots,i_{r}\}$ is acceptable of order $r$ then one has
\begin{equation}\label{Qdetspinor}
Q_{I}=\frac{(\sqrt{\tau_{i_{k+1}}\cdots\tau_{i_r}})^{\,r-k-1}}{\prod_{k+1\leqslant a<b \leqslant r}\left(\tau_{i_b}-\tau_{i_a}\right)}
\frac{\left|\begin{array}{cccc}
S_{I\cup\{i'_{k+1},i_{k+2},\dots,i_r\}}^{[r-k-1]} & S_{I\cup\{i'_{k+1},i_{k+2},\dots,i_r\}}^{[r-k-3]} & \cdots & S_{I\cup\{i'_{k+1},i_{k+2},\dots,i_r\}}^{[1+k-r]}\\
S_{I\cup\{i_{k+1},i'_{k+2},\dots,i_r\}}^{[r-k-1]} & S_{I\cup\{i_{k+1},i'_{k+2},\dots,i_r\}}^{[r-k-3]} & \cdots & S_{I\cup\{i_{k+1},i'_{k+2},\dots,i_r\}}^{[1+k-r]}\\
\vdots & \vdots & \ddots  & \vdots\\
S_{I\cup\{i_{k+1},\dots,i_{r-1},i'_r\}}^{[r-k-1]} & S_{I\cup\{i_{k+1},\dots,i_{r-1},i'_r\}}^{[r-k-3]} & \cdots & S_{I\cup\{i_{k+1},\dots,i_{r-1},i'_r\}}^{[1+k-r]}
\end{array}\right|}{ \prod_{l=1}^{r-k-2}S_{I_r}^{[r-k-1-2l]}}\, .
\end{equation}

\paragraph{Gauge transformation}

The QQ-system as written above corresponds to a particular choice of gauge. In order to describe this gauge freedom, we draw inspiration from the $r=3$ case, see Appendix~\ref{app:r=3}. One needs to introduce two new Q-functions $S_{\pm,\emptyset}$, \eqref{eq:QQrel} and \eqref{eq:fac1} remain unchanged while \eqref{eq:fac2} and \eqref{SS relation} become
\begin{equation}\label{QSSS}
Q_{I} = S^{[+1]}_{I} S_{I}^{[-1]} S_{-\epsilon(I),\emptyset}
\end{equation} 
and 
\begin{equation}\label{eq:qqspin}
 S^{[+1]}_{I} S^{[-1]}_{J}- S^{[-1]}_{I} S^{[+1]}_{J}=\frac{\tau_{i_{r-1}}\tau_{i_r}-1}{\sqrt{\tau_{i_{r-1}}\tau_{i_r}}}Q_{I\cap J}\, S_{\epsilon(I),\emptyset}
\end{equation}
where $I = \{i_1,\ldots,i_r\}$ and $J = \{i_1,\ldots,i_{r-2},i'_{r-1},i'_r\}$ are acceptable sets of order $r$ and we define $\epsilon(I) = \prod_{i=1}^r \alpha_i = \epsilon(\vec\alpha)$ with $\vec\alpha$ associated to $I$ according to \eqref{Itoalpha}.
These QQ-relations  remain unchanged if one applies the gauge transformation, depending on three arbitrary functions $g$, $g_{+}$ and $g_{-}$, given by
\begin{equation}
    S_{+,\emptyset} \mapsto \frac{g_{+}^{[+3]} g_{-}^{[-1]}}{g_{+}^{[+1]} g_{-}^{[-3]}} S_{+,\emptyset}\, ,\quad S_{-,\emptyset} \mapsto \frac{g_{-}^{[+3]} g_{+}^{[-1]}}{g_{-}^{[+1]} g_{+}^{[-3]}} S_{-,\emptyset}\, ,
\end{equation}
\begin{equation}
    S_{\vec\alpha}\mapsto \frac{g_{+}^{[+2]} g_{-}}{g_{+} g_{-}^{[-2]}} g S_{\vec\alpha}\, ,\quad\text{if}\quad \epsilon(\vec\alpha) = +\, ,
\end{equation}
\begin{equation}
    S_{\vec\alpha}\mapsto \frac{g_{-}^{[+2]} g_{+}}{g_{-} g_{+}^{[-2]}} g S_{\vec\alpha}\, ,\quad\text{if}\quad \epsilon(\vec\alpha) = -\, ,
\end{equation}
\begin{equation}
    Q_I\mapsto \frac{g_{+}^{[|I|+3-r]} g_{-}^{[|I|+3-r]}}{g_{+}^{[r-3-|I|]} g_{-}^{[r-3-|I|]}} g^{[r-1-|I|]} g^{[|I|+1-r]} Q_I
\end{equation}
for $I$ acceptable. In this paper we work in the "spin chain" gauge
\begin{equation}\label{eq:polfix}
    Q_\emptyset(x) = x^N\, , \quad S_{\pm,\emptyset}(x) = 1
\end{equation}
and the Q-functions are polynomials in the spectral parameter up to twist-dependent exponential prefactors.

\section{Transfer matrix in terms of fundamental Q's}

In Section~\ref{sec:diag} we gave the transfer matrix in terms of one single Q-function for each nesting level. We can use the Casoratian formula \eqref{eq:detform} to express the transfer matrix only in terms of $Q_{\emptyset}$ and a half the number of fundamental Q-functions $Q_{\{i\}}$. We will show in this section that the transfer matrix is then given by
\begin{equation}\label{Tdet}
 T = Q_{\emptyset}^{[r-1]} Q_{\emptyset}^{[3-r]} \frac{|Q_{\{i_a\}}^{[r+2-2b-2\delta_{b,r}]}|_r}{|Q_{\{i_a\}}^{[r+2-2b]}|_r} +Q_{\emptyset}^{[1-r]} Q_{\emptyset}^{[r-3]} \frac{|Q_{\{i_a\}}^{[2b-r-2+2\delta_{b,r}]}|_r}{|Q_{\{i_a\}}^{[2b-r-2]}|_r}
\end{equation} 
with $i_a\neq i_b$ and $i_a\neq i_b'$ for all $a\neq b$.

This formula fulfills, at least for the fundamental T-function, one of the main purposes of our paper  -- to derive the Weyl-type expressions for the transfer matrices of spin chains based on $D_r$ algebra, ``quantizing" in this way the classical Weyl character determinant formula. The latter can be restored in the classical limit $x\rightarrow\infty$. In that limit $Q_{\{j\}}(x)\,\underset{x\to\infty}{\sim}\, \tau_j^{x} x^{J_j + N}$ while the fundamental $T$  behaves as  $ x^{2N}\sum_{j=1}^r(\tau_j+\frac{1}{\tau_j})$.

\subsection{Induction}\label{sec:induction}

We can prove the formula \eqref{Tdet} by expressing the transfer matrix in terms of the first $r$ fundamental Q-functions, as in   \eqref{eq:detform}, and  inserting it into \eqref{eq:abaT}. We obtain
\begin{equation}\label{eq:TfromABA}
 T_\pm = Q_{\emptyset}^{[\pm(r-1)]} Q_{\emptyset}^{[\pm(3-r)]} \sum_{k=1}^r \frac{\left|Q_{\{i\}}^{[\pm(2k-r-2j+2)]}\right|_{k-1}}{\left|Q_{\{i\}}^{[\pm(2k-r-2j)]}\right|_{k-1}} \frac{\left|Q_{\{i\}}^{[\pm(2k-r-2j)]}\right|_k}{\left|Q_{\{i\}}^{[\pm(2k-r-2j+2)]}\right|_k}\\
\end{equation} 
The desired expression \eqref{Tdet} for the transfer matrix (in the case $i_a = a$) follows from \eqref{eq:TfromABA} using the identity
\begin{equation}
 \sum_{k=1}^r\frac{\left|Q_{\{i\}}^{[\pm(2k-r-2j+2)]}\right|_{k-1}}{\left|Q_{\{i\}}^{[\pm(2k-r-2j)]}\right|_{k-1}}\frac{\left|Q_{\{i\}}^{[\pm(2k-r-2j)]}\right|_{k}}{\left|Q_{\{i\}}^{[\pm(2k-r-2j+2)]}\right|_{k}}=\frac{ \left|Q_{\{i\}}^{[\pm(r+2-2j-2\delta_{j,r})]}\right|_{r}}{\left|Q_{\{i\}}^{[\pm(r+2-2j)]}\right|_r}
\end{equation} 
which can be shown by induction on $r$. It obviously holds true for $r=1$. It remains to show that 
\begin{equation}
\frac{\left|Q_{\{i\}}^{[\pm(r+3-2j-2\delta_{j,r})]}\right|_{r+1}}{\left|Q_{\{i\}}^{[\pm(r+3-2j)]}\right|_{r+1}} = \frac{ \left|Q_{\{i\}}^{[\pm(r+1-2j-2\delta_{j,r})]}\right|_{r}}{\left|Q_{\{i\}}^{[\pm(r+1-2j)]}\right|_r} + \frac{\left|Q_{\{i\}}^{[\pm(r-2j+3)]}\right|_{r}}{\left|Q_{\{i\}}^{[\pm(r-2j+1)]}\right|_{r}} \frac{\left|Q_{\{i\}}^{[\pm(r-2j+1)]}\right|_{r+1}}{\left|Q_{\{i\}}^{[\pm(r-2j+3)]}\right|_{r+1}}\,,
\end{equation} 
or equivalently (assuming the determinants are non-vanishing)
\begin{equation}\label{identity determinants}
\begin{split}
  \left|Q_{\{i\}}^{[\mp(2j-1+2\delta_{j,r+1})]}\right|_{r+1} \left|Q_{\{i\}}^{[\mp(2j+1)]}\right|_{r}
 = \left|Q_{\{i\}}^{[\mp(2j-1)]}\right|_{r+1} &\left|Q_{\{i\}}^{[\mp(2j+1+2\delta_{j,r})]}\right|_{r}\\& +\left|Q_{\{i\}}^{[\mp(2j-1)]}\right|_{r}  \left|Q_{\{i\}}^{[\mp(2j+1)]}\right|_{r+1}\,.
 \end{split}
\end{equation} 
The latter identity can be proven as follows: one first expands each of the $(r+1)\times (r+1)$ determinants with respect to the row involving $Q_{\lbrace r+1\rbrace}$. Both sides become linear combination of $Q_{\{r+1\}}^{[\mp(2j-1)]}$ for $1\leqslant j\leqslant r+2$ and one just has to check that the coefficients on each side are the same. For $j\in\{1,r+1,r+2\}$ this is completely trivial whereas for $j\in\{2,\dots,r\}$ this becomes~\footnote{We use here another notation for determinants: if $M$ is a $p\times p$ matrix with columns $M_1,\dots,M_p$, we write $\det M = |M_1,\dots ,M_p|$.}
\begin{multline}
    |C_1,\dots,C_{j-1},C_{j+1},\dots,C_{r},C_{r+2}| |C_2,\dots,C_{r+1}|\\
    = |C_1,\dots,C_{j-1},C_{j+1},\dots,C_{r},C_{r+1}| |C_2,\dots,C_{r},C_{r+2}|\\
    - |C_1,\dots,C_{r}| |C_2,\dots,C_{j-1},C_{j+1},\dots,C_{r+2}|
\end{multline}
where $C_j$ is the transpose of the row vector $\left(Q_{\{1\}}^{[\mp(2j-1)]},\dots,Q_{\{r\}}^{[\mp(2j-1)]}\right)$. This last equality is a particular case of a Plücker identity (or Sylvester's lemma): if $M$ and $N$ are two matrices of the same size with columns $M_1,\dots,M_r$ and $N_1,\dots,N_r$ respectively then the following identity holds for any $k\in \{1,\dots,r\}$,
\begin{equation}\label{Plucker1}
    \det M \det N = \sum_{l=1}^r |M_1,\dots,M_{k-1},N_{l},M_{k+1},\dots,M_r||N_1,\dots,N_{l-1},M_{k},N_{l+1},\dots,N_r|\, .
\end{equation}
In our case $M = (C_1,\dots,C_{j-1},C_{j+1},\dots,C_{r},C_{r+2})$ and $N = (C_2,\dots,C_{r+1})$ have many columns in common so that if we decide to exchange $M_r = C_{r+2}$ only two terms survive in the sum (when $l=j-1$ or $l=r$) and they give exactly what we want.

\subsection{Reshuffling Q-functions in the transfer matrix }\label{sec:shuffle}
Here we show that the expression for the transfer matrix \eqref{Tdet} in terms of  $r$ fundamental Q-functions is invariant under the replacement $Q_{i_a}\mapsto Q_{i_a'}$ for any $a$. By obvious  symmetry with respect to permutations of the functions $Q_i,\,i\in \{1,2,\dots,r\}$  it suffices to show that the transfer matrix \label{eq:Tdet} is invariant under $Q_{i_r}\mapsto Q_{i'_r}$. This is the case if 
\begin{equation}\label{eq:cond}
\frac{ Q_{\emptyset}^{[r-1]} Q_{\emptyset}^{[3-r]}}{Q_{\emptyset}^{[r-3]} Q_{\emptyset}^{[1-r]}} = - \frac{\check T_-^{\{ i_1,\ldots,i_r\}} - \check T_-^{\{i_1,\ldots,i_r'\}}}{\check T_+^{\{i_1,\ldots,i_r\}} - \check T_+^{\{i_1,\ldots,i_r'\}}}
\end{equation} 
where we defined 
\begin{equation}
 \check T_\pm^{\{a_1,\ldots,a_r\rbrace}=\frac{\left|Q_{\{a_i\}}^{[\mp(2j-r-2+2\delta_{j,r})]}\right|_r}{\left|Q_{\{a_i\}}^{[\mp(2j-r-2)]}\right|_r}\, .
\end{equation} 
Using the Jacobi identity on determinants, one can rewrite the numerator and the denominator in the previous condition as
\begin{equation}
    \check T_-^{\{i_1,\ldots,i_r\}}-\check T_-^{\{i_1,\ldots,i_r'\}} = (-1)^{1+\lfloor\frac{r}{2}\rfloor} \frac{\W^{[-2]}_{i_1,\ldots,i_{r-1}} \W_{i_r',i_1,\ldots,i_{r-1},i_r}}{\W^{[-1]}_{i_1,\ldots,i_{r-1},i_r} \W^{[-1]}_{i_1,\ldots,i_{r-1},i_r'}}
\end{equation}
and
\begin{equation}
    \check T_+^{\{i_1,\ldots,i_r\}}-\check T_+^{\{i_1,\ldots,i_r'\}} = (-1)^{1+\lfloor\frac{r-1}{2}\rfloor+r} \frac{\W^{[-2]}_{i_1,\ldots,i_{r-1}} \W_{i_r',i_1,\ldots,i_{r-1},i_r}}{\W^{[+1]}_{i_1,\ldots,i_{r-1},i_r} \W^{[+1]}_{i_1,\ldots,i_{r-1},i_r'}}
\end{equation}
with 
\begin{equation}
 W_{i_1,\ldots,i_k}:= \left|Q_{\{i_a\}}^{[k+1-2b]}\right|_k\, .
\end{equation} 
The condition \eqref{eq:cond} then reads
\begin{equation}\label{eq:WWWW}
 \frac{
Q_{\emptyset}^{[r-1]}Q_{\emptyset}^{[3-r]}}{
Q_{\emptyset}^{[r-3]}Q_{\emptyset}^{[1-r]}}=
 \frac{\W^{[-2]}_{i_1,\ldots,i_{r-1}}\W^{[+1]}_{i_1,\ldots,i_{r-1},i_r}\W^{[+1]}_{i_1,\ldots,i_{r-1},i_r'}}{\W^{[+2]}_{i_1,\ldots,i_{r-1}}\W^{[-1]}_{i_1,\ldots,i_{r-1},i_r}\W^{[-1]}_{i_1,\ldots,i_{r-1},i_r'}}
\end{equation}
which is indeed satisfied due to the trivial relation
\begin{equation}\label{eq:QQQQ}
 \frac{Q_{\{i_1,\ldots,i_{r-1},i_r\}}Q_{\{i_1,\ldots,i_{r-1},i_r'\}}}{Q^{[+1]}_{\{i_1,\ldots,i_{r-1}\}}Q^{[-1]}_{\{i_1,\ldots,i_{r-1}\}}}=1\,,
\end{equation} 
following immediately from the factorisation properties of the Q-functions \eqref{eq:fac1} and \eqref{eq:fac2}.

\section{Bethe ansatz equations of Wronskian type}\label{sec:wronski}

We propose here a Wronskian relation on $r+1$ Q-functions which could serve for finding the Bethe roots and, eventually, the energy of the state. We call it the Wronskian BAE, in analogy to the very useful Wronskian BAE for the  $A_r$ Heisenberg XXX spin chain which has the form
\begin{equation}\label{WronskianBAEAr}
    |Q_j(x+r-2k+2)|_{r+1}=x^{N}\,\,\prod_{1\leqslant i<j\leqslant r+1}(\tau_i-\tau_{j})
\end{equation}
where Q-functions have the form \eqref{QofA}. Solving the Wronskian relation above is often more efficient than solving the Bethe equations.  This alternative method of finding Bethe roots was proposed in \cite{Pronko:1999gh} and further extended in \cite{Kazakov:2015efa,Kazakov:2018ugh,Marboe:2016yyn}. 

A similar relation for $D$-type spin chain is not as simple. In the following, we propose to use for this purpose the equation \eqref{Qdetspinor} when $I=\emptyset$:
\begin{equation}\label{WronskianBAE}
\left|\begin{array}{cccc}
S_{\{i'_{1},i_{2},\dots,i_r\}}^{[r-1]} & S_{\{i'_{1},i_{2},\dots,i_r\}}^{[r-3]} & \cdots & S_{\{i'_{1},i_{2},\dots,i_r\}}^{[1-r]}\\
S_{\{i_{1},i'_{2},\dots,i_r\}}^{[r-1]} & S_{\{i_{1},i'_{2},\dots,i_r\}}^{[r-3]} & \cdots & S_{\{i_{1},i'_{2},\dots,i_r\}}^{[1-r]}\\
\vdots & \vdots & \ddots  & \vdots\\
S_{\{i_{1},\dots,i_{r-1},i'_r\}}^{[r-1]} & S_{\{i_{1},\dots,i_{r-1},i'_r\}}^{[r-3]} & \cdots & S_{\{i_{1},\dots,i_{r-1},i'_r\}}^{[1-r]}
\end{array}\right| = \frac{\prod_{1\leqslant a<b \leqslant r}\left(\tau_{i_b}-\tau_{i_a}\right)}{(\sqrt{\tau_{i_{1}}\cdots\tau_{i_r}})^{\,r-1}}
Q_\emptyset \prod_{l=1}^{r-2}S_{I_r}^{[r-1-2l]}
\end{equation}
where we recall that $I_r = \{i_1,\dots,i_r\}\subset\{1,\dots,2r\}$ is such that $\{i_a,i'_a\}\cap\{i_b,i'_b\} = \emptyset$ for all $a\neq b$. The spinorial Q-functions are polynomials up to a twist-dependent exponential prefactor, their leading asymptotic behaviour is completely determined by the global charges $J_i = -J_{i'} =\sum_{k=1}^NJ_{ii}^{(k)}$, it is given by 
\begin{equation}
    S_{\{i_1,\dots,i_r\}}(x)\,\underset{x\to\infty}{\sim}\, \left(\prod_{a=1}^r\tau_{i_a}\right)^{\frac{x}{2}} x^{\frac{1}{2}\left(\sum_{k=1}^r J_{i_a} + N\right)}\, .
\end{equation}
The hope would be that, once the global charges are fixed, it suffices to solve equation \eqref{WronskianBAE} for the unknowns that are the coefficients of the polynomial parts of the spinorial Q-functions. In the $A_r$ case, we thus get exactly all the eigenstates with such a weight. However, this does not seem to be the case here. First of all, one should notice that there are $2^r$ equations of the type \eqref{WronskianBAE} (as many as there are spinorial Q-functions $S_{I_r}$). For a given choice of $I_r$, the number of unknown coefficients can be easily computed to be $N + \frac{r-1}{2}\left(\sum_{a=1}^r J_{i_a} + N\right)$, it thus seems natural to chose $I_r$ such that $\sum_{a=1}^r J_{i_a}$ is minimal. Nonetheless, since the degree of the polynomials on each side of the equation is $N + \frac{r-2}{2}\left(\sum_{a=1}^r J_{i_a} + N\right)$, as soon as $\sum_{a=1}^r J_{i_a} > -N$, there does not seem to be enough equations to fix all the coefficients. This is understandable if one looks at the case $r=3$: the proposed equation does not coincide with \eqref{WronskianBAEAr}, it is instead the expression of Q-functions with three indices in terms of single-index Q-functions. A possible way to resolve this issue would be to solve \eqref{WronskianBAE} for different choices of $I_r$ and to look for common sets of solutions.

Once the Wronskian BAE has been solved, we have enough spinorial Q-functions to recover $r$ single-index Q-functions using \eqref{Qdetspinor}. Any of them can be used to compute the energy of the state through \eqref{Energy}.

A Wronskian BAE for single-index Q-functions would be \eqref{eq:QQQQ}, which also reads
\begin{equation}\label{WronskianBAEQ}
    Q_{\emptyset}^{[r-2]}Q_{\emptyset}^{[2-r]}\left|Q_{\{i_a\}}^{[r+1-2b]}\right|_{r-1}\left|Q_{\{i_a\}}^{[r-1-2b]}\right|_{r-1} = \prod_{a=1}^{r-1}\frac{\tau_{i_a}}{\left(\tau_{i_a}-\tau_{i_r}\right)\left(\tau_{i_a}-\tau_{i_r'}\right)} \left|Q_{\{i_a\}}^{[r+1-2b]}\right|_{r} \left|Q_{\{j_a\}}^{[r+1-2b]}\right|_{r}
\end{equation}
where $j_a = i_a$ for $1\leqslant a\leqslant r-1$ and $j_r = i'_r$ and the asymptotic behaviour of the relevant functions is given by
\begin{equation}
    Q_{\{i\}}(x)\,\underset{x\to\infty}{\sim}\, \tau_i^x x^{N+J_i}\, .
\end{equation}
Once again there are $2^r$ equations of this type but they are of higher order than \eqref{WronskianBAE}.

\section{QQ$'$-type formulas for T-functions}
\label{sec:QQ'}

In this section we first present QQ$'$-formulas for the symmetric and spinorial T-operators. The reasoning behind our rather heuristic derivation is in analogy to \cite{Bazhanov:2010jq} where the BGG resolution \cite{bernstein1975differential} was used. Here we give arguments  on the level of characters, see also \cite{Kazakov:2010iu}, which we take as hints to obtain the actual BGG-type relation for  the fundamental and spinorial transfer matrices. The final formulas have been checked in several examples for small finite lengths.  Further we provide a consistency check. Namely, we recover the Weyl-type expression for the fundamental transfer matrix \eqref{Tdet} by reducing there the number of used single-index Q-functions from $2r$ to $r$. In the final subsection we introduce the Hirota equation \cite{Kuniba:1993cn} and solve it using the QQ$'$-formulas for symmetric T-operators. This yields QQ$'$-type formulas for any rectangular transfer matrix $T_{a,s}$.

\subsection{Symmetric transfer matrices} \label{sec:qqf}

In \cite{Frassek:2020nki} it was argued that the product of Lax matrices can be brought to the form
\begin{equation}\label{eq:qlax}
 L_i^{(1)}(x+x_i)L_{i'}^{(2)}(x-x_i)=S_i\mathfrak{L}_i^{+,(1)}(x)G_i^{(2)}S_i^{-1}
\end{equation} 
where  the Lax operator $L_i$ are defined via $L_i(x)=\tilde B_{1,i}\,L(x)\,\tilde B_{1,i}$ for $i=1,\ldots,r$ and $L_i(x)=\IDb L_{i'}(x)\IDb$ for $i=r+1,\ldots,2r$, see Section~\ref{sec:firstfun}. The superscripts $(1,2)$ indicate two different families of oscillators. The letter $S_i$ denotes a similarity transformation in the oscillators space and $G_i$ a dummy matrix that does not depend on the spectral parameter and commutes with the Lax matrix $\mathfrak{L}_i^+(x)$. Their precise form is given in \cite{Frassek:2020nki}.
We identify the Lax matrix $\mathfrak{L}^+_i(x)$ as a realisation of  \eqref{eq:fundlax}. The parameter $x_i$ then plays the role of the representation label. We stress that the term linear in the spectral parameter is given by the generators $J_{ij}$, cf.~\eqref{eq:fundlax}. In the case \eqref{eq:qlax}, the representation of $\mathfrak{so}(2r)$ is infinite-dimensional in the oscillators space and becomes reducible for certain values of the parameter $x_i$.
The infinite-dimensional representation of $\mathfrak{so}(2r)$ is characterised by its character.  
For example for $i=1$ the Cartan elements are of the form
\begin{align}
 J_{1,1}&=1-r+2x_1-\sum_{k=2}^{2r-1}\oad_k\oa_k\,,\\
 J_{i,i}&=\oad_i\oa_i-\oad_{i'}\oa_{i'}\,,\qquad 2\leqslant i\leqslant  r\,.
%  J_{2r,2r}&=\sum_{k=2}^{2r-1}N_k-1+r-2x_1
\end{align} 
The character can then be computed via
\begin{equation}
\begin{split}
\chi^+_1(x_1)\equiv \tr \prod_{i=1}^r\tau_{i}^{J_{ii}}&
 =\tau_1^{2x_1}\prod_{k=2}^{r}\frac{\tau_{1}}{(\tau_1-\tau_k)(\tau_1-\tau_{k'})}\,.
 \end{split}
\end{equation} 
We find similar formulas for the product of Lax matrices $L_i(x+x_i)L_{i'}(x-x_i)$ by exchanging $\tau_1\leftrightarrow\tau_i$ and $x_1\rightarrow x_i$ for $1\leqslant i\leqslant r$ and $\tau_j\rightarrow\tau_j^{-1}$, $x_i\rightarrow x_{i'}$ for $i>r$, cf Section~\ref{sec:firstfun}. 
The twist dependent prefactor is invariant under $\tau_i\to \tau_i^{-1}$. We find
\begin{equation}
\chi_i^+(x_i)=
\begin{cases}\tau_i^{+2x_i}\prod_{k\neq i}\frac{\tau_{i}}{(\tau_i-\tau_k)(\tau_i-\tau_{k'})},\qquad& 1\leqslant i\leqslant r\\
 \tau_{i'}^{-2x_{i}}\prod_{k\neq i'}^{r}\frac{\tau_{i'}}{(\tau_{i'}-\tau_k)(\tau_{i'}-\tau_{k'})},\qquad & r<i\leqslant 2r\\
\end{cases}\,.
\end{equation}  
% \begin{equation}
%  \chi_i^+=\tau_1^{2x_1+r-1}\prod_{k=2}^{r}\frac{1}{(\tau_1-\tau_k)(\tau_1-\tau_{k'})}\,.
% \end{equation} 
The finite dimensional characters are related to the one above by the sum formula
\begin{equation}\label{Chista}
 \chi_s=\sum_{i=1}^{2r}\chi_i^+\left(\frac{s+r-1}{2}\right)=
\sum_{i=1}^r\left[
 \prod _{j\neq i}\frac{\tau_i}{ \left(\tau_i-\tau _{j'}\right) \left(\tau_i-\tau_j\right)}\right]\left(\tau^{s+r-1}_i+\tau_{i'}^{s+r-1}\right)\,.
\end{equation} 
From our results for finite length and the discussion above we find that the formula can be lifted to transfer matrices and Q-operators. It reads
\begin{equation}\label{TsQQ}
T_{1,s}(x)=
\sum_{i=1}^r\left[
 \prod _{j\neq i}\frac{\tau _i}{ \left(\tau _i-\tau _{j'}\right) \left(\tau _i-\tau _j\right)}\right]\left(Q_{\{i\}}^{[s+r-1]}Q_{\{i'\}}^{[1-r-s]}+Q_{\{i\}}^{[1-r-s]}Q_{\{i'\}}^{[s+r-1]}\right)\,.
\end{equation} 
Notice that in the limit $x\to\infty$ \eqref{TsQQ} becomes \eqref{Chista}, as it should be.

\subsection{Spinorial transfer matries}\label{sec:qqs}
A similar factorisation formula as \eqref{eq:qlax} exists for the spinorial Lax matrices \eqref{eq:laxlin}. It reads
\begin{equation}\label{eq:slax}
\check  L_{\vec \alpha}^{(1)}(x+x_{\vec \alpha})\check L_{-\vec \alpha}^{(2)}(x-x_{\vec \alpha}-\kappa)=\check S_{\vec \alpha}\check{\mathfrak{L}}^{+,(1)}_{\vec \alpha}(x)\check G^{(2)}_{\vec \alpha}\check S^{-1}_{\vec \alpha}\,.
\end{equation} 
Here we defined $\check L_{\vec\alpha}(x)=B(\vec\alpha)\check L(x)B(\vec\alpha)$ and use a notation similar as above in \eqref{eq:qlax}. The similarity transformation $\check S_{\vec \alpha}$ only depends on the oscillators and $\check G_{\vec \alpha}$ is a matrix that is independent of the spectral parameter and commutes with the Lax matrix $\check{\mathfrak{L}}^+_{\vec \alpha}$. The latter denotes an infinite-dimensional realisation of the spinorial Lax matrix 
\begin{equation}\label{eq:spilax}
 \check{\mathfrak{L}}(x)=z\ID+\F_{ij}\otimes E_{ji}\,,
\end{equation} 
where $J_{ij}$ denote the generators of a spinorial representation. Again the parameter $x_{\vec \alpha}$ in \eqref{eq:slax} has the role of the representation label. As before we compute the character of the oscillator representation. In the case ${\vec \alpha}=(+,\ldots,+)$ we find
\begin{equation}
\chi_{(+,\ldots,+)}\equiv \tr\prod_{i=1}^r \tau_i^{J_{ii}}=\prod_{i=1}^r\tau_i^{x_{(+,\ldots,+)}}\prod_{1\leqslant j<k\leqslant r}\frac{\tau_j\tau_k}{\tau_j\tau_k-1}\,,
\end{equation} 
where 
\begin{equation}
 J_{ii}=x_{(+,\ldots,+)}-\sum_{j=i+1}^r\oad_{i,j'}\oa_{j',i}-\sum_{j=1}^{i-1}\oad_{j,i'}\oa_{i',j}\,,\qquad 1\leqslant i\leqslant r\,.
\end{equation} 
The general formula can be obtained using the relations among the spinorial Lax matrices as presented in Section~\ref{sec:spinq}. We get 
\begin{equation}
\chi_{\vec\alpha}^+\left(x_{\vec\alpha}\right)=\prod_{i=1}^r\tau_i^{\alpha_i x_{\vec\alpha}}\prod_{1\leqslant j<k\leqslant r}\frac{\tau_j^{\alpha_j}\tau_k^{\alpha_k}}{\tau_j^{\alpha_j}\tau_k^{\alpha_k}-1}\,,
\end{equation} 
The characters of the finite-dimensional spinor representations $\pm$ with 
$ f=(s/2,\ldots,s/2,\pm s/2)$ can then be written as
\begin{equation}
 \chi_{\pm,s}= \sum_{{\{\alpha_i\}}_\pm}\chi_{\vec \alpha}^+\left(\frac{s}{2}\right)=
 \sum_{{\{\alpha_i\}}_\pm}\prod_{i=1}^r\tau_i^{\frac{s}{2}\alpha_i}\prod_{1\leqslant j<k\leqslant r}\frac{\tau_j^{\alpha_j}\tau_k^{\alpha_k}}{\tau_j^{\alpha_j}\tau_k^{\alpha_k}-1}\, .
\end{equation} 
Here the sum is taken over all configurations $\{\vec \alpha\}_\pm$ such that $\prod_i\alpha=\pm1$.

On the level of monodromies, we propose the formula
\begin{equation}\label{TpmS}
\begin{split}
T_{\pm,s}&=
 \sum_{\{\alpha_i\}_\pm}   \prod_{1\leqslant j<k\leqslant r}\frac{\tau_j^{\alpha_j}\tau_k^{\alpha_k}}{\tau_j^{\alpha_j}\tau_k^{\alpha_k}-1}\prod_{i=1}^r\tau_i^{-\frac{\kappa}{2}\alpha_i} S_{{\vec\alpha}}^{[r+s-1]}S_{-\vec\alpha}^{[1-s-r]}\,.
 \end{split}
\end{equation} 
This formula  has been verified for small finite lengths by comparing to the transfer matrices directly constructed within the quantum inverse scattering method using the Lax matrices in \eqref{eq:spilax} for finite-dimensional spinor representations.

\subsection{Derivation of Weyl-type formula for $T_{1,1}$ from QQ$'$-relations}\label{sec:deriveWeyl}

Let us write \eqref{TsQQ} as
\begin{equation}\label{T1s 2r}
    T_{1,s} = \sum_{i=1}^{2r} h_i \,Q_{\{i\}}^{[s+r-1]}Q_{\{i'\}}^{[1-r-s]}
\end{equation}
where $h_i = \prod_{j(\neq i, i')}^r (u_i - u_j)^{-1}$ and $u_j=\tau_j+1/\tau_j$. We further assume that when $s\in \lbrace 1-r,\dots,0\rbrace $ the identity is still verified if one sets
\begin{equation}\label{Tsnegative}
    T_{1,0} = Q_\emptyset^{[r-2]} Q_{\emptyset}^{[2-r]}\quad \text{and}\quad  T_{1,s} = 0\quad\text{for} \quad 1-r\leqslant s\leqslant -1\, .
\end{equation}
We show here that the conditions \eqref{T1s 2r} and \eqref{Tsnegative} are enough to recover the expression \eqref{Tdet} giving $T_{1,1}$ in terms of only $r$ of the single-index Q-functions, and so are consistent with it. We also show in Appendix~\ref{app:wronski} how to retrieve the Wronskian equation \eqref{WronskianBAEQ} from these conditions.

One simply has to notice that $\eqref{T1s 2r}$ implies that there exist some Q-dependent coefficients $C_{j,k',k}$ (defined for $0\leqslant k'\leqslant k\leqslant r$ and $0\leqslant j \leqslant k-k'$) such that
\begin{equation}\label{CTrelation}
\sum_{k'=0}^{k}\sum_{j=0}^{k-k'} C_{j,k',k} T_{1,k'+1-r}^{[2j + k' - k]} = \sum_{i=1}^r h_i \left|\begin{array}{llll}
Q_{1}^{[-k]} & Q_{1}^{[-k+2]} & \cdots & Q_{1}^{[k]}\\
\vdots & \vdots &   & \vdots\\
Q_{k}^{[-k]} & Q_{k}^{[-k+2]} & \cdots & Q_{k}^{[k]}\\
Q_{i}^{[-k]} & Q_{i}^{[-k+2]} & \cdots & Q_{i}^{[k]}
\end{array}\right| \left|\begin{array}{llll}
Q_{1}^{[-k]} & Q_{1}^{[-k+2]} & \cdots & Q_{1}^{[k]}\\
\vdots & \vdots &   & \vdots\\
Q_{k}^{[-k]} & Q_{k}^{[-k+2]} & \cdots & Q_{k}^{[k]}\\
Q_{i'}^{[-k]} & Q_{i'}^{[-k+2]} & \cdots & Q_{i'}^{[k]}
\end{array}\right|\,
\end{equation}
It suffices indeed to expand the determinants with respect to their last row and perform the sum over $i$. One has for instance
\begin{equation}
    C_{0,k,k} = (-1)^k\left|\begin{array}{ccc}
Q_{1}^{[-k]} & \cdots & Q_{1}^{[k-2]}\\
\vdots &   & \vdots\\
Q_{k}^{[-k]} & \cdots & Q_{k}^{[k-2]}\\
\end{array}\right| \left|\begin{array}{ccc}
Q_{1}^{[-k+2]} & \cdots & Q_{1}^{[k]}\\
\vdots &   & \vdots\\
Q_{k}^{[-k+2]} & \cdots & Q_{k}^{[k]}\\
\end{array}\right|\, .
\end{equation}
In particular, plugging the constraints $\eqref{Tsnegative}$ in the previous relation when $k=r$ gives us
\begin{equation}
    C_{0,r-1,r}Q_\emptyset^{[r-3]} Q_{\emptyset}^{[1-r]} + C_{1,r-1,r}Q_\emptyset^{[r-1]} Q_{\emptyset}^{[3-r]} + C_{0,r,r} T_{1,1} = 0\, .
\end{equation}
Since
\begin{multline}
    C_{0,r-1,r} = (-1)^{r+1}\left|Q_{i}^{[-r+2j]}\right|_r\times \left|Q_{i}^{[-r+2j-2+2\delta_{j,r} ]}\right|_r\\
    \quad\text{and}\quad C_{1,r-1,r} = (-1)^{r+1}\left|Q_{i}^{[-r+2j-2]}\right|_r\times \left|Q_{i}^{[-r+2j-2\delta_{j,1} ]}\right|_r
\end{multline}
we recover \eqref{Tdet} in the case $i_a = a$. Notice that with this derivation, the symmetry under $Q_{\{i\}}\leftrightarrow Q_{\{i'\}}$ is immediate because the equations we started from were already symmetric.

\subsection{Transfer matrices for general rectangular representations}

In this section we propose relatively simple formulas for T-functions in rectangular representations in terms of bi-linear expressions involving Wronskians of both types of single-index Q-functions,  $Q_{i}$ and $Q_{i'}$, where $i=1,2,\dots,r$. These formulas follow from \eqref{TsQQ} when solving the Hirota equations \cite{Kuniba:1993cn} satisfied by the T-functions. These equations read as follows ($s\in\mathbb{N}^*$):
\begin{equation}\label{Hirota tail}
    T_{a,s}^{[+1]}\,T_{a,s}^{[-1]} = T_{a,s+1}\,T_{a,s-1} + T_{a-1,s}\,T_{a+1,s}
\end{equation}
for $1\leqslant a\leqslant r-3$,
\begin{equation}
    T_{r-2,s}^{[+1]}\, T_{r-2,s}^{[-1]} = T_{r-2,s+1}\,T_{r-2,s-1} + T_{r-3,s}\,T_{+,s}\,T_{-,s}\, ,
\end{equation}
which can be written in the same form as the previous equation if one sets $T_{r-1,s} = T_{+,s}T_{-,s}$, and
\begin{equation}
    T_{\pm,s}^{[+1]} \,T_{\pm,s}^{[-1]} = T_{\pm,s+1}\,T_{\pm,s-1} + T_{r-2,s}\, .
\end{equation}
%\begin{equation}
%    T_{-,s}^{[+1]} T_{-,s}^{[-1]} = T_{-,s+1}T_{-,s-1} + T_{r-2,s}
%\end{equation}
The boundary conditions are ($0\leqslant a\leqslant r-2$, $s\in\mathbb{N}$)
\begin{equation}\label{boundaryT}
    T_{a,0} = Q_\emptyset^{[r-a-1]}Q_\emptyset^{[a+1-r]}\, ,\qquad T_{0,s} = Q_\emptyset^{[r+s-1]}Q_\emptyset^{[1-r-s]}\, ,
\end{equation}
 and
\begin{equation}
    T_{\pm,0}(x) = Q_\emptyset(x)\, .
\end{equation}

 We shall  determine here the QQ$'$-type relations for $T_{a,s}$ for $1\leqslant a \leqslant r-1$, but not for $T_{\pm,s}$. For these spinorial transfer matrices, the spinorial Q-functions seem more suitable, see equation \eqref{TpmS}. We start from
\begin{equation}
    T_{1,s}^{[+1]}T_{1,s}^{[-1]} - T_{1,s-1}T_{1,s+1} = \sum_{1\leqslant i_1<i_2\leqslant2r}h_{i_1} h_{i_2} \left|\begin{array}{cc}
Q_{\{i_1\}}^{[s+r]} & Q_{\{i_1\}}^{[s+r-2]}\\
Q_{\{i_2\}}^{[s+r]} & Q_{\{i_2\}}^{[s+r-2]}\\
\end{array}\right|
\left|\begin{array}{cc}
Q_{\{i'_1\}}^{[2-s-r]} & Q_{\{i'_1\}}^{[-s-r]}\\
Q_{\{i'_2\}}^{[2-s-r]} & Q_{\{i'_2\}}^{[-s-r]}\\
\end{array}\right|
\end{equation}
which can be also written, if the transfer matrices satisfy the Hirota equation \eqref{Hirota tail} with boundary conditions \eqref{boundaryT}, as follows
\begin{equation}
    T_{1,s}^{[+1]}\,T_{1,s}^{[-1]} - T_{1,s-1}\,T_{1,s+1} = T_{0,s}\,T_{2,s} = Q_\emptyset^{[r+s-1]}Q_\emptyset^{[1-r-s]} T_{2,s}.
\end{equation}
Putting the two expressions together yields the following expression for the second row of transfer matrices:
\begin{equation}
    T_{2,s} = \frac{1}{Q_\emptyset^{[r+s-1]}Q_\emptyset^{[1-r-s]}} \sum_{1\leqslant i_1<i_2\leqslant2r}h_{i_1} h_{i_2} \left|\begin{array}{cc}
Q_{\{i_1\}}^{[s+r]} & Q_{\{i_1\}}^{[s+r-2]}\\
Q_{\{i_2\}}^{[s+r]} & Q_{\{i_2\}}^{[s+r-2]}\\
\end{array}\right|
\left|\begin{array}{cc}
Q_{\{i'_1\}}^{[2-s-r]} & Q_{\{i'_1\}}^{[-s-r]}\\
Q_{\{i'_2\}}^{[2-s-r]} & Q_{\{i'_2\}}^{[-s-r]}\\
\end{array}\right|\, .
\end{equation}

This procedure can be continued for $1\leqslant a \leqslant r-1$, it yields
\begin{equation}\label{TasQQ'}
    T_{a,s} = \frac{1}{\prod_{k=1}^{a-1}Q_\emptyset^{[r+s+2k-a-1]}Q_\emptyset^{[1+a-r-s-2k]}} \sum_{1\leqslant i_1<\ldots<i_a\leqslant2r}h_{i_1} \cdots h_{i_a} W_{i_1,\dots,i_a}^{[s+r-1]} W_{i'_1,\dots,i'_a}^{[1-s-r]}\, ,
\end{equation}
where we recall that $W_{i_1,\ldots,i_k} = \left|Q_{\{i_a\}}^{[k+1-2b]}\right|_k$. The proof of this formula, which we present in Appendix \ref{app:proofTasQQ'}, boils down to verifying the relation
\begin{multline}\label{WW'Hirota}
    \frac{1}{2}\sum_{\substack{1\leqslant i_1<\dots<i_a\leqslant2r \\ 1\leqslant j_1<\dots<j_a\leqslant2r}} \Bigg|\begin{array}{cc}
W_{i_1,\dots,i_a}^{[s+r]} & W_{i_1,\dots,i_a}^{[s+r-2]}\\
W_{j_1,\dots,j_a}^{[s+r]} & W_{j_1,\dots,j_a}^{[s+r-2]}\\
\end{array}\Bigg|\,\,
\Bigg|\begin{array}{cc}
W_{i'_1,\dots,i'_a}^{[2-s-r]} & W_{i'_1,\dots,i'_a}^{[-s-r]}\\
W_{j'_1,\dots,j'_a}^{[2-s-r]} & W_{j'_1,\dots,j'_a}^{[-s-r]}\\
\end{array}\Bigg|\\
=  \left(\sum_{1\leqslant i_1<\dots<i_{a-1}\leqslant2r} W_{i_1,\dots,i_{a-1}}^{[s+r-1]} W_{i'_1,\dots,i'_{a-1}}^{[1-s-r]}\right)\left(\sum_{1\leqslant i_1<\dots<i_{a+1}\leqslant2r} W_{i_1,\dots,i_{a+1}}^{[s+r-1]} W_{i'_1,\dots,i'_{a+1}}^{[1-s-r]}\right)\, .
\end{multline}

\section{Weyl-type formulas for T-functions from tableaux representations }\label{sec:Weyl-type}

The tableaux sum formulas of \cite{Kuniba:1994na} give expressions for the transfer matrices of any rectangular representation $T_{a,s}$ through the single terms in the sum of the transfer matrix \eqref{eq:t} as given in \eqref{eq:abaT}. In total there are $2r$ different terms (boxes), $r$ for $T_{+}$ and $r$ for $T_-$. Instead of using the summands in the form \eqref{eq:abaT} involving $Q$'s of different levels, we shall express them either in terms of $Q_\emptyset$ and $r$ single-index Q-functions as in \eqref{eq:TfromABA} or in terms of $r+1$ spinorial Q-functions. This will yield new expressions for totally symmetric $T_{1,s}$ and totally antisymmetric $T_{a,1}$ T-functions.

We start from the expressions \eqref{eq:TfromABA} for $T_\pm$ such that
\begin{equation}
    T_{1,1} = T_+ + T_- = \sum_{k=1}^{2r} b_{k,r}
\end{equation}
where $b_{k,r}$ denotes a box as given in \cite{Kuniba:1994na} for $D_r$ with index $k$. The expression above, in the character limit $x\to\infty$, allows to identify $b_{k,r}$ in terms of the single-index Q-functions. We get
\begin{equation}\label{eq:box1}
b_{k,r} = Q_{\emptyset}^{[r-1]} Q_{\emptyset}^{[3-r]} \frac{\left|Q_{\{i\}}^{[2k-r-2j+2]}\right|_{k-1}}{ \left|Q_{\{i\}}^{[2k-r-2j]}\right|_{k-1}} \frac{\left|Q_{\{i\}}^{[2k-r-2j]}\right|_{k}}{\left|Q_{\{i\}}^{[2k-r-2j+2]}\right|_{k}}
%= Q_{\emptyset}^{[r-1]} Q_{\emptyset}^{[3-r]} \frac{W_{1,\dots,k-1}^{[k+2-r]}}{W_{1,\dots,k-1}^{[k-r]}} \frac{W_{1,\dots,k}^{[k-1-r]}}{W_{1,\dots,k}^{[k+1-r]}}
\end{equation} 
for $1\leqslant k\leqslant r$ and
\begin{equation}\label{eq:box2}
b_{k,r} = Q_{\emptyset}^{[1-r]} Q_{\emptyset}^{[r-3]} \frac{\left|Q_{\{i\}}^{[r-2j-2]}\right|_{k'-1}}{\left|Q_{\{i\}}^{[r-2j]}\right|_{k'-1}} \frac{\left|Q_{\{i\}}^{[r+2-2j]}\right|_{k'}}{\left|Q_{\{i\}}^{[r-2j]}\right|_{k'}}
%= Q_{\emptyset}^{[1-r]} Q_{\emptyset}^{[r-3]} \frac{W_{1,\dots,k'-1}^{[r-k'-2]}}{W_{1,\dots,k'-1}^{[r-k']}} \frac{W_{1,\dots,k'}^{[r+1-k']}}{W_{1,\dots,k'}^{[r-1-k']}}
\end{equation} 
for $r+1\leqslant k\leqslant 2r$, and we recall that $k' = 2r-k+1$.

The simplest examples of the tableaux sum formulas beyond $T_{1,1}$ are for $T_{1,2}$ and $T_{2,1}$. They arise when writing
\begin{equation}
    T_{1,1}^-T_{1,1}^+ = \left[\left(\sum_{1\leqslant i\leqslant j\leqslant 2r} b_{i,r}^{[-1]} b_{j,r}^{[+1]}\right) - b_{r,r}^{[-1]} b_{r',r}^{[+1]}\right] + \left[\left(\sum_{1\leqslant i< j\leqslant 2r}b_{i,r}^{[+1]} b_{j,r}^{[-1]} \right)+ b_{r',r}^{[+1]} b_{r,r}^{[-1]}\right]
\end{equation}
and identifying the terms in the brackets with $T_{1,0}T_{1,2}$ and $T_{0,1}T_{2,1}$ from the Hirota equation \eqref{Hirota tail}, so that
\begin{equation}
T_{1,2}\,Q_\emptyset^{[r-2]}Q_\emptyset^{[2-r]} = \sum_{1\leqslant i\leqslant j\leqslant 2r} b_{i,r}^{[-1]} b_{j,r}^{[+1]} - b_{r,r}^{[-1]} b_{r',r}^{[+1]}\,,
\end{equation} 
and
\begin{equation}
 T_{2,1}\,Q_{\emptyset}^{[+r]}Q_{\emptyset}^{[-r]}= \sum_{1\leqslant i< j\leqslant 2r}b_{i,r}^{[+1]} b_{j,r}^{[-1]}+ b_{r',r}^{[+1]} b_{r,r}^{[-1]}\, .
\end{equation} 
As we see, it is independent of the actual representation of the box terms $b_{k,r}$. As we will see in the following, substituting \eqref{eq:box1} and \eqref{eq:box2} will yield new expressions for the transfer matrices that only depend on $r$ single-index Q-functions and $Q_\emptyset$.

\subsection{Symmetric representations}

The transfer matrices for generic symmetric representations are given by~\cite{Kuniba:1994na}
\begin{equation}\label{symmetric boxes}
    T_{1,s} = \frac{1}{\prod_{k=1}^{s-1}Q_\emptyset^{[r-s-2+2k]}Q_\emptyset^{[-(r-s-2+2k)]}}\underset{1\leqslant i_1\leqslant \cdots\leqslant i_s\leqslant 2r}{\sum\,'} b_{i_1,r}^{[1-s]} \cdots b_{i_s,r}^{[s-1]}
\end{equation}
where the symbol $\sum'$ stands for a sum in which we do not allow for $r$ and $r+1$ to appear at the same time. The denominator appears as a consequence of our boundary conditions for the Hirota equation.

\subsubsection{General symmetric sum}

Let us define
\begin{equation}
    \tilde{b}_k = \frac{\left|Q_{\{i\}}^{[2k-2j+2]}\right|_{k-1}}{\left|Q_{\{i\}}^{[2k-2j]}\right|_{k-1}}\frac{\left|Q_{\{i\}}^{[2k-2j]}\right|_{k}}{\left|Q_{\{i\}}^{[2k-2j+2]}\right|_{k}}
\end{equation}
for $1\leqslant k\leqslant r$ and
\begin{equation}
    \tilde{b}_k = \frac{\left|Q_{\{i\}}^{[-(2k'-2j+2)]}\right|_{k'-1}}{\left| Q_{\{i\}}^{[-(2k'-2j)]}\right|_{k'-1}}\frac{\left|Q_{\{i\}}^{[-(2k'-2j)]}\right|_{k'}}{\left|Q_{\{i\}}^{[-(2k'-2j+2)]}\right|_{k'}}
\end{equation}
for $r+1\leqslant k\leqslant 2r$ such that
\begin{equation}
    b_{k,r} = Q_{\emptyset}^{[r-1]}Q_{\emptyset}^{[3-r]} \tilde{b}_k^{[-r]} \quad \text{if} \quad k\leqslant r\quad \text{and}\quad b_{k,r} = Q_{\emptyset}^{[1-r]}Q_{\emptyset}^{[r-3]} \tilde{b}_k^{[r]} \quad \text{if} \quad r+1\leqslant k\, .
\end{equation}

For $l\geqslant 1$, one has
\begin{equation}\label{sum boxes}
    \sum_{1\leqslant i_1\leqslant\ldots\leqslant i_l\leqslant r} \tilde{b}_{i_1}^{[-2l+1]} \cdots \tilde{b}_{i_l}^{[-1]} = \frac{\left|Q_{\{i\}}^{[2r+1-2j-2l\delta_{j,r}]}\right|_{r}}{\left|Q_{\{i\}}^{[2r+1-2j]}\right|_{r}}
\end{equation}
and
\begin{equation}\label{sum boxes 2}
    \sum_{r+1\leqslant i_1\leqslant\ldots\leqslant i_l\leqslant 2r} \tilde{b}_{i_1}^{[1]} \cdots \tilde{b}_{i_l}^{[2l-1]} = \frac{\left|Q_{\{i\}}^{[-(2r+1-2j-2l\delta_{j,r})]}\right|_{r}}{\left|Q_{\{i\}}^{[-(2r+1-2j)]}\right|_{r}}\, .
\end{equation}

The  two identities are equivalent, so it is enough to prove the first one. We do it by induction in $r$. It is trivial when $r=1$. If it is true for some $r_0\geqslant 1$ then let us show by induction on $l$ that it is also true for $r_0+1$: the case $l=1$ has been proven earlier in Section \ref{sec:induction} so we assume that the identity holds for some $l_0\geqslant 1$. We then write
\begin{multline}
    \sum_{1\leqslant i_1\leqslant\ldots\leqslant i_{l_0+1}\leqslant r_0+1} \tilde{b}_{i_1}^{[-2l_0-1]}\cdots\tilde{b}_{i_{l_0+1}}^{[-1]} = \sum_{1\leqslant i_1\leqslant\ldots\leqslant i_{l_0+1}\leqslant r_0} \tilde{b}_{i_1}^{[-2l_0-1]}\cdots\tilde{b}_{i_{l_0+1}}^{[-1]}\\
    + \tilde{b}_{r_0+1}^{[-1]}\sum_{1\leqslant i_1\leqslant\ldots\leqslant i_{l_0}\leqslant r_0+1} \tilde{b}_{i_1}^{[-2l_0-1]}\cdots\tilde{b}_{i_{l_0}}^{[-3]}\, .
\end{multline}
Since we have assumed that the identity holds for $r_0$ and any $l$, for $(r_0+1,l_0)$ we can write
\begin{multline}
    \sum_{1\leqslant i_1\leqslant\ldots\leqslant i_{l_0+1}\leqslant r_0+1} \tilde{b}_{i_1}^{[-2l_0-1]} \cdots \tilde{b}_{i_{l+1}}^{[-1]} = \frac{\left|Q_{\{i\}}^{[2r_0+1-2j-2(l_0+1)\delta_{j,r_0}]}\right|_{r_0}}{\left| Q_{\{i\}}^{[2r_0+1-2j]}\right|_{r_0}}\\
    + \frac{\left|Q_{\{i\}}^{[2r_0-2j+3]}\right|_{r_0}}{\left|Q_{\{i\}}^{[2r_0-2j+1]}\right|_{r_0}} \frac{\left|Q_{\{i\}}^{[2r_0+1-2j-2l_0\delta_{j,r_0+1}]}\right|_{r_0+1}}{\left| Q_{\{i\}}^{[2r_0-2j+3]}\right|_{r_0+1}}\, .
\end{multline}
Consequently, for \eqref{sum boxes} to hold for $(r_0+1,l_0+1)$, one only has to show that
\begin{multline}
    \left|Q_{\{i\}}^{[-2j]}\right|_{r_0} \left|Q_{\{i\}}^{[2-2j-2(l_0+1)\delta_{j,r_0+1}]}\right|_{r_0+1} = \left|Q_{\{i\}}^{[2-2j]}\right|_{r_0+1} \left|Q_{\{i\}}^{[-2j-2(l_0+1)\delta_{j,r_0}]}\right|_{r_0}\\
    + \left|Q_{\{i\}}^{[2-2j]}\right|_{r_0} \left|Q_{\{i\}}^{[-2j-2l_0\delta_{j,r_0+1}]}\right|_{r_0+1}\, .
\end{multline}
This last relation can be proven in much the same way as \eqref{identity determinants} which itself corresponds to the case $l_0=0$.

\subsubsection{Application to the computation of transfer matrices}

In order to apply the summation formulas \eqref{sum boxes} and \eqref{sum boxes 2} we first rewrite equation \eqref{symmetric boxes} as
\begin{multline}
    T_{1,s} = \sum_{l=0}^s Q_\emptyset^{[2j+r-s-2]} Q_\emptyset^{[2+2j-r-s]} \sum_{\substack{1\leqslant i_1\leqslant\ldots\leqslant i_l\leqslant r \\ r+1\leqslant i_{l+1}\leqslant\ldots\leqslant i_s\leqslant 2r}} \left(\tilde{b}_{i_1}^{[1-s-r]}\cdots\tilde{b}_{i_l}^{[2l-s-r-1]}\right)\\
    \times\left(\tilde{b}_{i_{l+1}}^{[2l-s+r+1]}\cdots\tilde{b}_{i_s}^{[s+r-1]}\right) - \sum_{l=1}^{s-1} Q_\emptyset^{[2l+r-s-2]} Q_\emptyset^{[2+2l-r-s]} \tilde{b}_{r}^{[2l-s-r-1]}\tilde{b}_{r+1}^{[2l-s+r+1]}\\
    \times \sum_{\substack{1\leqslant i_1\leqslant\ldots\leqslant i_{l-1}\leqslant r \\ r+1\leqslant i_{l+2}\leqslant\ldots\leqslant i_s\leqslant 2r}} \left(\tilde{b}_{i_1}^{[1-s-r]}\cdots\tilde{b}_{i_{l-1}}^{[2l-s-r-3]}\right)\left(\tilde{b}_{i_{l+2}}^{[2l-s+r+3]}\cdots\tilde{b}_{i_s}^{[s+r-1]}\right)\, .
\end{multline}
In virtue of \eqref{sum boxes} and \eqref{sum boxes 2}, this gives
\begin{multline}
    T_{1,s} = \sum_{l=0}^s Q_\emptyset^{[2l+r-s-2]} Q_\emptyset^{[2+2l-r-s]} \frac{\left| Q_{\{i\}}^{[2l+r-s+1-2j-2l\delta_{j,r}]}\right|_{r}}{\left|Q_{\{i\}}^{[2l+r-s+1-2j]}\right|_{r}} \frac{ \left|Q_{\{i\}}^{[-(r+s+1-2l-2j-2(s-l)\delta_{j,r})]}\right|_{r}}{\left|Q_{\{i\}}^{[-(r+s+1-2l-2j)]}\right|_{r}}\\
    - \sum_{l=1}^{s-1} Q_\emptyset^{[2l+r-s-2]} Q_\emptyset^{[2+2l-r-s]} \frac{\left|Q_{\{i\}}^{[2l+r-s-1-2j-2(l-1)\delta_{j,r}]}\right|_{r}}{\left|Q_{\{i\}}^{[2l+r-s+1-2j]}\right|_{r}} \frac{\left|Q_{\{i\}}^{[-(r+s-1-2l-2j-2(s-1-l)\delta_{j,r})]}\right|_{r}}{\left|Q_{\{i\}}^{[-(r+s+1-2l-2j)]}\right|_{r}}\, .
\end{multline}
The terms for $1\leqslant l\leqslant s-1$ of each sum can be combined thanks to a Plücker identity, to give an explicit and concise Weyl-type representation of symmetric T-functions for $D_r$ algebra
\begin{equation}\label{Tsdet}
    T_{1,s} = \sum_{l=0}^s Q_\emptyset^{[2l+r-s-2]} Q_\emptyset^{[2+2l-r-s]} \frac{\left| Q_{\{i\}}^{[2l+r-s+1-2j+2(s-l)\delta_{j,1}-2l\delta_{j,r}]}\right|_{r}}{\left|Q_{\{i\}}^{[2l+r-s+1-2j]}\right|_{r}}\, .
\end{equation}
Once again there are $2^r$ formulas of this type depending on which set of $r$ single-index Q-functions we use in the right-hand side. Finally, let us mention that we also checked that one can get the same formula starting from the conditions \eqref{T1s 2r} and \eqref{Tsnegative} using a method similar to that of Section~\ref{sec:deriveWeyl} (or more directly for low ranks, see Appendix~\ref{app:r=2} for the case $r=2$).

%\begin{multline}
%    T_{1,s} = \sum_{l=0}^s Q_\emptyset^{[2l+r-s-2]} Q_\emptyset^{[2+2l-r-s]} \frac{\left| Q_{\{i\}}^{[2l-s+r+1-2j-2l\delta_{j,r}]}\right|_{r}}{\left|Q_{\{i\}}^{[2l-s+r+1-2j]}\right|_{r}} \frac{ \left|Q_{\{i\}}^{[-(s+r+1-2l-2j-2(s-l)\delta_{j,r})]}\right|_{r}}{\left|Q_{\{i\}}^{[-(s+r+1-2l-2j)]}\right|_{r}}\\
%    - \sum_{l=1}^{s-1} Q_\emptyset^{[2l+r-s-2]} Q_\emptyset^{[2+2l-r-s]} \frac{\left|Q_{\{i\}}^{[2l+r-s+1-2j]}\right|_{r-1}}{\left|Q_{\{i\}}^{[2l+r-s-1-2j]}\right|_{r-1}} \frac{\left|Q_{\{i\}}^{[-(r+s+1-2l-2j)]}\right|_{r-1}}{\left| Q_{\{i\}}^{[-(r+s-1-2l-2j)]}\right|_{r-1}} \\
%    \times \frac{\left|Q_{\{i\}}^{[2l+r-s-1-2j-2(l-1)\delta_{j,r}]}\right|_{r}}{\left|Q_{\{i\}}^{[2l+r-s+1-2j]}\right|_{r}} \frac{\left|Q_{\{i\}}^{[-(r+s-1-2l-2j-2(s-1-l)\delta_{j,r})]}\right|_{r}}{\left|Q_{\{i\}}^{[-(r+s+1-2l-2j)]}\right|_{r}}\, .
%\end{multline}

\subsection{Antisymmetric representations}

The transfer matrices for generic antisymmetric representations are given by~\cite{Kuniba:1994na}
\begin{multline}
    T_{a,1} = \frac{1}{\prod_{k=1}^{a-1} Q_{\emptyset}^{[r-a+2k]} Q_{\emptyset}^{[-(r-a+2k)]}}\sum_{\substack{1\leqslant i_1<\dots<i_k\leqslant r \\ r+1\leqslant j_1<\dots<j_l\leqslant 2r \\ a-k-l\in 2\mathbb{N}}} b_{i_1,r}^{[a-1]}\cdots b_{i_k,r}^{[a+1-2k]}\\ \times b_{r+1,r}^{[a-1-2k]} b_{r,r}^{[a-3-2k]} \cdots b_{r+1,r}^{[2l+3-a]} b_{r,r}^{[2l+1-a]} b_{j_1,r}^{[2l-a-1]}\cdots b_{j_l,r}^{[1-a]}\, .
\end{multline}

As it happens, in order to obtain nice expressions for these transfer matrices involving a reduced number of Q-functions, it is more convenient to turn to spinorial Q-functions. If, in order to shorten the notations, we write
\begin{equation}
    S_{I_r} = S_{\{1,\dots,r\}}\, ,\quad S_i = S_{\{1,\dots,r-i,r+i,r-i+2,\dots,r\}}\quad \text{for}\quad i\in\{1,\dots,r\}
\end{equation}
then, according to \eqref{Qdetspinor}, the boxes are also given by
\begin{equation}\label{eq:sbox1}
b_{k,r} = Q_{\emptyset}^{[r-1]} Q_{\emptyset}^{[1-r]}\frac{S_{I_r}^{[-2]}}{S_{I_r}} \frac{\left|S_{i}^{[-2j]}\right|_{r-k}}{ \left|S_{i}^{[2-2j]}\right|_{r-k}} \frac{\left|S_{i}^{[4-2j]}\right|_{r+1-k}}{\left|S_{i}^{[2-2j]}\right|_{r+1-k}}
\end{equation} 
for $1\leqslant k\leqslant r$ and
\begin{equation}\label{eq:sbox2}
b_{k,r} = Q_{\emptyset}^{[r-1]} Q_{\emptyset}^{[1-r]}\frac{S_{I_r}^{[+2]}}{S_{I_r}} \frac{\left|S_{i}^{[2j]}\right|_{k-r-1}}{ \left|S_{i}^{[2j-2]}\right|_{k-r-1}} \frac{\left|S_{i}^{[2j-4]}\right|_{k-r}}{\left|S_{i}^{[2j-2]}\right|_{k-r}}
\end{equation} 
for $r+1\leqslant k\leqslant 2r$. The relevant summation formulas read
\begin{equation}
    \sum_{1\leqslant i_1<\ldots< i_l\leqslant r} b_{i_1,r}^{[-1]} \cdots b_{i_l,r}^{[1-2l]} = \frac{S_{I_r}^{[-1-2l]}}{S_{I_r}^{[-1]}} \left(\prod_{a=1}^l Q_{\emptyset}^{[r-2l-2+2a]} Q_{\emptyset}^{[-r-2l+2a]}\right) \frac{\left|S_i^{[1-2j+2\theta(l-j))]}\right|_{r}}{\left|S_i^{[1-2j]}\right|_{r}}
\end{equation}
and
\begin{equation}
    \sum_{r+1\leqslant i_1<\ldots< i_l\leqslant 2r} b_{i_1,r}^{[2l-1]} \cdots b_{i_l,r}^{[1]} = \frac{S_{I_r}^{[2l+1]}}{S_{I_r}^{[+1]}} \left(\prod_{a=1}^l Q_{\emptyset}^{[r-2+2a]} Q_{\emptyset}^{[-r+2a]}\right) \frac{\left|S_i^{[2r-1-2j+2\theta(r-l-j)]}\right|_{r}}{\left|S_i^{[2r+1-2j]}\right|_{r}}
\end{equation}
where we used the Heaviside function $\theta$ that is $0$ for negative arguments and $1$ for non-negative ones. These formulas can be proven in much the same way as \eqref{sum boxes} and \eqref{sum boxes 2}.

This permits us to write
\begin{multline}\label{Tadet}
    T_{a,1} = \frac{Q_{\emptyset}^{[r-a]} Q_{\emptyset}^{[a-r]}}{S_{I_r}^{[a-1]}S_{I_r}^{[1-a]}}\sum_{\substack{ 0\leqslant k,l \leqslant a \\ a-k-l\in 2\mathbb{N}}} S_{I_r}^{[a+1-2k]} S_{I_r}^{[2l-1-a]} \frac{\left|S_i^{[a+1-2j+2\theta(k-j)]}\right|_{r}}{\left|S_i^{[a+1-2j]}\right|_{r}}\\
    \times \frac{\left|S_i^{[2r-a-1-2j+2\theta(r-l-j)]}\right|_{r}}{\left|S_i^{[2r-a+1-2j]}\right|_{r}}\, .
\end{multline}
This equation should be compared with the much more complicated expression given in Appendix~\ref{app:uglyWeyl-type} for the same quantity but in terms of single-index Q-functions. Similarly, the expression for $T_{1,s}$ in terms of spinorial Q-functions is not as simple as \eqref{Tsdet}.

In the particular case $a=1$ the previous expression reads
\begin{equation}
    T_{1,1} = \frac{Q_{\emptyset}^{[r-1]} Q_{\emptyset}^{[1-r]}}{S_{I_r}}\left(S_{I_r}^{[-2]} \frac{\left|S_i^{[2-2j+2\delta_{1,j}]}\right|_{r}}{\left|S_i^{[2-2j]}\right|_{r}} + S_{I_r}^{[+2]} \frac{\left|S_i^{[2r-2j-2\delta_{j,r}]}\right|_{r}}{\left|S_i^{[2r-2j]}\right|_{r}}\right)\, .
\end{equation}
It should be compared with \eqref{Tdet}. When $a=r-1$, as  expected from the fact that $T_{r-1,1} = T_{+,1} T_{-,1}$, there is a factorization:
\begin{multline}\label{eq:factorT}
    T_{r-1,1} = \frac{1}{S_{I_r}^{[r-2]}S_{I_r}^{[2-r]}} \frac{Q_{\emptyset}^{[+1]} Q_{\emptyset}^{[-1]}}{\left|S_i^{[r+2-2j]}\right|_{r} \left|S_i^{[r-2j]}\right|_{r}} \left(\sum_{\substack{k=0 \\ k\, \text{even}}}^{r} S_{I_r}^{[r-2k]} \left|S_i^{[r-2j+2\theta(k-j)]}\right|_{r}\right) \\
    \times \left(\sum_{\substack{k=0 \\ k\, \text{odd}}}^{r} S_{I_r}^{[r-2k]} \left|S_i^{[r-2j+2\theta(k-j)]}\right|_{r}\right)\, .
\end{multline}

%For instance, when $a=2$ it reads
%\begin{multline}
%    T_{2,1} = Q_\emptyset^{[r-4]} Q_\emptyset^{[4-r]} \frac{\left|Q_{\{i\}}^{[r+3-2j-2\delta_{j,r}]}\right|_r \left|Q_{\{i\}}^{[r-1-2j+2\delta_{j,1}]}\right|_r}{\left|Q_{\{i\}}^{[r+3-2j]}\right|_r \left|Q_{\{i\}}^{[r-1-2j]}\right|_r}\\
%    + \frac{Q_\emptyset^{[r-2]} Q_\emptyset^{[2-r]}}{Q_\emptyset^{[r]} Q_\emptyset^{[-r]}} \frac{1}{\left|Q_{\{i\}}^{[r+1-2j]}\right|_r} \Bigg(Q_\emptyset^{[r-2]} Q_\emptyset^{[2-r]} \frac{\left|Q_{\{i\}}^{[r+3-2j]}\right|_r \left|Q_{\{i\}}^{[r-1-2j]}\right|_r}{\left|Q_{\{i\}}^{[r+1-2j]}\right|_r}\\
%    + Q_\emptyset^{[r]} Q_\emptyset^{[4-r]} \left|Q_{\{i\}}^{[r+1-2j-2\theta(j+1-r)]}\right|_r + Q_\emptyset^{[r-4]} Q_\emptyset^{[-r]} \left|Q_{\{i\}}^{[r+1-2j+2\theta(2-j)]}\right|_r\Bigg)\, .
%\end{multline}

%The formulas \eqref{Tsdet} for symmetric transfer matrices look much  simpler than for the antisymmetric ones. It would be good to have the CBR type representation involving them for $D_r$ algebra, but it does not seem to exist.

% =======
% \begin{multline}
%     T_{r-1,1} = \frac{1}{S_{I_r}^{[r-2]}S_{I_r}^{[2-r]}} \frac{Q_{\emptyset}^{+} Q_{\emptyset}^{-}}{\left|S_i^{[r+2-2j]}\right|_{r} \left|S_i^{[r-2j]}\right|_{r}} \left(\sum_{\substack{k=0 \\ k\, \text{even}}}^{r} S_{I_r}^{[r-2k]} \left|S_i^{[r-2j+2\theta(k-j)]}\right|_{r}\right) \\
%     \times \left(\sum_{\substack{k=0 \\ k\, \text{odd}}}^{r} S_{I_r}^{[r-2k]} \left|S_i^{[r-2j+2\theta(k-j)]}\right|_{r}\right)\, .
% >>>>>>> da4f3c5dfcee5e88319ab42ee0f4d780d44adf00

\subsection{Spinorial representations}
Following \cite{Kuniba:2010ir} we express the spinorial T-functions $T_{\pm,1}$ in terms of the Q-functions along a nesting path, cf.~Section~\ref{sec:diag}. One finds
\begin{equation}
T_{\pm,1}
 =\sum_{|\alpha|=\pm1}Q_\emptyset^{[-\alpha_1]} \left(\frac{S^{[\rho_+(\vec\alpha)+1]}_{(+,\ldots,+)}}{S^{[\rho_+(\vec\alpha)-1]}_{(+,\ldots,+)}}\right)^{\frac{\alpha_{r-1}+\alpha_{r}}{2}}\left(\frac{S^{[\rho_{-}(\vec\alpha)+1]}_{(+,\ldots,+,-)}}{S^{[\rho_{-}(\vec\alpha)-1]}_{(+,\ldots,+,-)}}\right)^{\frac{\alpha_{r-1}-\alpha_{r}}{2}}\prod_{k=1}^{r-2} \left(\frac{Q^{[\rho_k(\vec\alpha)+1]}_{\{1,\ldots,k\}}}{Q^{[\rho_k(\vec\alpha)-1]}_{\{1,\ldots,k\}}}\right)^{\frac{\alpha_{k}-\alpha_{k+1}}{2}}
\end{equation} 
where $Q_\emptyset=x^N$ and the shifts are determined via
\begin{equation}
\begin{split}
 \rho_k(\vec\alpha)&=\alpha_1+\ldots+\alpha_{k-1}+\frac{\alpha_k-\alpha_{k+1}}{2}\qquad \text{for}\qquad 1\leq k\leq r-2\, ,\\
 \rho_\pm(\vec\alpha)&=\alpha_1+\ldots+\alpha_{r-2}+\frac{\alpha_{r-1}\pm\alpha_{r}}{2}\,.
 \end{split}
\end{equation} 
Expressing all Q-functions in terms of spinorial ones using \eqref{Qdetspinor} we obtain determinant formulas for $T_{\pm,1}$. We find  
\begin{equation}\label{eq:tp1}
 T_{+,1}=\frac{(\sqrt{\tau_{i_{1}}\cdots\tau_{i_r}})^{\,r-1}}{\prod_{1\leqslant a<b \leqslant r}\left(\tau_{i_b}-\tau_{i_a}\right)}
\frac{1}{ \prod_{l=1}^{r-1}S_{I_r}^{[r-2l]}}\, \sum_{\substack{k=0 \\ k\, \text{even}}}^{r} S_{I_r}^{[r-2k]} \left|S_i^{[2j-r-2\theta(r-k-j)]}\right|_{r}
\end{equation} 
and
\begin{equation}\label{eq:tm1}
 T_{-,1}=\frac{(\sqrt{\tau_{i_{1}}\cdots\tau_{i_r}})^{\,r-1}}{\prod_{1\leqslant a<b \leqslant r}\left(\tau_{i_b}-\tau_{i_a}\right)}\frac{1}{ \prod_{l=1}^{r-1}S_{I_r}^{[r-2l]}}\,  \sum_{\substack{k=0 \\ k\, \text{odd}}}^{r}  S_{I_r}^{[r-2k]} \left|S_i^{[2j-r-2\theta(r-k-j))]}\right|_{r}
\end{equation} 
These expressions have been verified for $r=3,4,5$ for a particular choice of $I_r$ and we are missing a generic proof. However, the formulas are consistent with the factorisation $T_{r-1,1}=T_{+,1}T_{-,1}$ in \eqref{eq:factorT} expected from the Hirota relations. In principle, one can now generate, from \eqref{Tadet}, \eqref{eq:tp1} and \eqref{eq:tm1} above, all transfer matrices of rectangular representations using Cherednik-Bazhanov-Reshetikhin type formulas written for $D_r$ symmetry in \cite{Kuniba:2010ir}.

%We further remark that the spinorial transfer matrices are expressed in terms of $r+1$ Q-functions. In this sense \eqref{eq:tp1} and \eqref{eq:tm1} are the analogs of the Weyl-type formula \eqref{Tdet} for the fundamental transfer matrix.

\section{Discussion}

In this work, we proposed the full system of Baxter Q-functions -- the QQ-system -- for the spin chains with $SO(2r)$~symmetry. This QQ-system is described by a novel type of Hasse diagram presented for various ranks on Figures~\ref{fig:Hasse2}, \ref{fig:hasse4} and \ref{fig:HasseD2}. We also found Weyl-type formulas for transfer matrices (T-functions) of symmetric and antisymmetric representations in terms of sums of ratios of determinants of $r$ basic Q-functions. We also proposed QQ$'$-type formulas expressing the T-functions through $2r$ basic single-index Q-functions. These could be a powerful tool for the study of spin chains and sigma models with $D_{r}$ symmetry. We also reformulated the Bethe ansatz equations in the form of a single Wronskian relation on $r+1$ basic Q-functions. It is the analogue of a similar Wronskian relation for spin chains with $A_r$ symmetry. However, apart from the Bethe roots our equation contain extra solutions whose role has yet to be clarified.

Our main assumptions in this article are the Pl\"ucker QQ-relations \eqref{eq:QQrel} and \eqref{SS relation}, as well as the QQ$'$-relations \eqref{TsQQ} and \eqref{TpmS}. The QQ-relations  are motivated by the asymptotics of the Q-operators and the QQ$'$-type relations by the factorisation formulas for the Lax matrices for Q-operators and the corresponding  character formulas, as discussed in Section~\ref{sec:qqf} and~\ref{sec:qqs}. Both relations remain to be proven but we have tested them explicitly for several examples of small finite length T and Q-operators. The QQ-relations \eqref{eq:QQrel} allow to express the fundamental transfer matrix, for which an expression in terms of one Q-function of each level is known from the algebraic Bethe ansatz, in terms of $r$ single- index Q-functions and $Q_\emptyset$, cf.~\eqref{Tdet}. This Weyl-type expression has been independently obtained from the QQ$'$-type relations \eqref{TsQQ}, see Section~\ref{sec:deriveWeyl}. We take this as a consistency check. The new formulas for  $T_{a,s}$ are obtained from the Hirota equation and the tableaux formulas of \cite{Kuniba:1994na}. They can be seen as consequences of the Weyl-type formula for the fundamental transfer matrix in terms of $r+1$ Q-functions and the QQ$'$-type relations.

Unlike the well understood QQ-system of $A$-type, in the $D$-type QQ-system there are still  many questions left and issues  to be clarified. The questions exists already on the operator level: the Yangian for $SO(2r)$ spin chains is constructed only for ``rectangular" representations and  the R-matrix -- the main building block for Q and T-functions is known only for the symmetric and spinorial  representations~\cite{Shankar:1978rb,Reshetikhin:1986vd,Frassek:2020nki}. A full classification of Lax matrices including the ones for the Q-operators was recently given in \cite{Frassek:2018try} for $A$-type. This may shed some light upon the transfer matrices for general rectangular representations and beyond.  It would be interesting, using the tableaux formulas (which look quite involved~\cite{nakai2007paths}) to find the Weyl-type determinant formulas for the arbitrary rectangular representations, generalizing our formulas for symmetric and antisymmetric representations. These could also be interesting for the study of  the Q-system and its relation to cluster algebras, see e.g.  \cite{Kedem:2007zz,DiFrancesco:2008mc}. Moreover, a solid proof of our QQ-system and our Wronskian formulas for $T$'s in symmetric and antisymmetric representations is yet to be found on both the analytic and operator level. It may be possible using the BGG resolution or the analogue of coderivative method proposed in \cite{Kazakov:2007na} and used for this purpose in \cite{Kazakov:2010iu}, see also  \cite{Leurent:2012xc} for a review. 
Unfortunately, we do not know yet a suitable analogue of Baxter TQ equations (quantum spectral determinant) which appeared to be so useful for the spin chains with $A_r$ symmetry~\cite{Baxter:1982zz,Bazhanov:1996dr}, see \cite{Kazakov:2015efa} for the modern description in terms of forms as well as \cite{Chervov:2006xk} in terms of the quantum determinant. It is possible that the QQ$'$-type formulas for transfer matrices proposed in this work can replace the Baxter equations for $D_{r}$ algebra.

It would be interesting to generalize our approach to the study of spin chains with open boundary conditions and to the non-compact, highest weight and principal series representations of $D$ algebras. For a much better studied case of these aspects in $A$-type integrable system we refer the reader to \cite{Frassek:2015mra,Baseilhac:2017hoz,Vlaar:2020jww,Frassek:2017bfz,Derkachov:2006fw}. One encounters non-compact representations in sigma models~\cite{Balog:2005yz} and spin chains~\cite{Chicherin:2012yn} with principal series representations of the  $d$-dimensional conformal group $SO(2,d)$~\cite{Chicherin:2012yn}. They recently appeared in the study of $d$-dimensional fishnet CFT~\cite{Kazakov:2018ugh,Gurdogan:2015csr,Basso:2019xay} and the associated planar graphs (of the shape of regular 2-dimensional lattice)~\cite{Zamolodchikov:1980mb}. Whether as for the lowest, 4-dimensional conformal  $SO(2,4)$ symmetry one can use its isomorphism to the $A$ type $SU(2,2)$ group to construct the suitable QQ-system and Baxter TQ~system for the efficient study of Fishnet CFT~\cite{Gromov:2017blm,Gromov:2019bsj}, for $d>4$  we have to find an alternative approach which can be based on the $D$-type QQ-system  constructed in this work. The structure of the QQ-system does not depend on the choice of real section of the orthogonal group but  the Wronskian, Weyl-type formulas for $T$ do depend. So one could try to construct the quantum spectral curve (QSC  formalism for $d>4$ fishnet CFT in analogy to the $d=4$ case~\cite{Gromov:2016rrp}.    

Finally, we hope that our methods can be generalized to  $B$, $C$ and exceptional  types of algebras and their deformations, as well as to superalgebras such as $\mathfrak{osp}(m|2n)$ where the QQ-system and T-functions are yet to be constructed. This includes the case relevant for  $AdS_4/CFT_3$ for which the QSC has recently been studied in  \cite{Cavaglia:2014exa,Bombardelli:2017vhk,Bombardelli:2018bqz}.  A first step could be the evaluation of the oscillators type Lax matrices for Q-operators using the results of \cite{Braverman:2016pwk,Nakajima:2019olw} as done in \cite{Frassek:2020lky} for type~$A$.

\paragraph{Acknowledgments}
We thank Andrea Cavagli\`{a}, Nikolay Gromov, S\'ebastien Leurent, Vidas Regelskis, Nikolay Reshetikhin,  Zengo Tsuboi and  Alexander Tsymbaliuk for useful discussions. 
 RF is supported by the German research foundation (Deutsche Forschungsgemeinschaft DFG) Research Fellowships Programme 416527151 ``Baxter Q-operators and supersymmetric gauge theories''.

\appendix
\addtocontents{toc}{\protect\setcounter{tocdepth}{1}}

\section{$D_r$ Kirillov-Reshetikhin modules and characters}\label{app:Characters}

The Kirillov-Reshetikhin modules form a family $\lbrace {\rm{W}}_{a,s}(x) | a\in\{1,\dots,r-2,+,-\}, s\in\mathbb{N}^*, x\in\mathbb{C}\rbrace$ of modules of the Yangian $Y(D_r)$ that were first introduced in \cite{kirillov1990representations}. When restricted to $D_r\subset Y(D_r)$ they decompose into irreducible representations of $D_r$ according to
\begin{equation}
    {\rm{W}}_{a,s}(x) \simeq \bigoplus_{\substack{n_i\in\mathbb{N} \\ n_1 + n_3 + \cdots + n_a = s}} {\rm{V}}(n_1\omega_1 + n_3\omega_3 + \cdots + n_a\omega_a)
\end{equation}
for odd $a\leqslant r-2$,
\begin{equation}
    {\rm{W}}_{a,s}(x) \simeq \bigoplus_{\substack{n_i\in\mathbb{N} \\ n_0 + n_2 + \cdots + n_a = s}} {\rm{V}}(n_0\omega_0 + n_2\omega_2 + \cdots + n_a\omega_a)
\end{equation}
for even $a\leqslant r-2$,
\begin{equation}
    {\rm{W}}_{+,s}(x) \simeq {\rm{V}}(s\,\omega_{r-1})\quad\text{and}\quad {\rm{W}}_{-,s}(x) \simeq {\rm{V}}(s\,\omega_r)\, .
\end{equation}
Here $\omega_0 = 0$, while $\omega_1,\dots,\omega_r$ are the fundamental weights of $D_r$, ${\rm{V}}(f)$ denotes the irreducible $D_r$-module with highest weight $f$. Notice that the previous decompositions are independent of the spectral parameter $x$. The characters are expressed in terms of weights $\m_1,\dots, \m_r$ that are related to the non-negative integers $n_1,\dots, n_r$ (Dynkin labels) via
\begin{equation}
    \m_a = n_a + \cdots + n_{r-2} + \frac{1}{2}\left(n_{r-1}+n_r\right)
\end{equation}
for $1\leqslant a\leqslant r-2$ and
\begin{equation}
    \m_{r-1} = \frac{1}{2}\left(n_{r-1} + n_r\right)\, ,\qquad \m_{r} = \frac{1}{2}\left(n_{r-1} - n_r\right)\, .
\end{equation}

The finite-dimensional irreducible representations of $SO(2r)$ are in one-to-one correspondence with $(\m_1,\dots,\m_r)$ such that 
\begin{equation}\label{lambdaSO}
\m_1 \geqslant\cdots\geqslant |\m_r|\geqslant0 \quad \text{and}\quad \left\{
\begin{array}{ll}
\forall i\in\lbrace1,\dots,r\rbrace,\,\m_i\in\mathbb{Z}\\
\text{or}\quad \forall i\in\lbrace1,\dots,r\rbrace,\,\m_i\in\frac{1}{2}+\mathbb{Z}
\end{array}
\right.\,.
\end{equation}
The characters of these irreducible representations are given by ($\h_j = \m_j + r - j$)
\begin{equation}
\chi^{SO(2r)}_{\m}(\tau) = \frac{|\tau_i^{\h_j}+\tau_i^{-\h_j}|_r + |\tau_i^{\h_j}-\tau_i^{-\h_j}|_r}{|\tau_i^{r-j}+\tau_i^{-r+j}|_r} = \frac{|\tau_i^{\h_j}+\tau_i^{-\h_j}|_r + |\tau_i^{\h_j}-\tau_i^{-\h_j}|_r}{2\prod_{1\leqslant i < j \leqslant r} (\tau_i +\tau_i^{-1} - \tau_i - \tau_j^{-1})}\, .
\end{equation}
One should notice that, when $\m_r = 0$, the second determinant in the numerator is $0$ because its last column vanishes.

Since the Kirillov-Reshetikhin modules for the symmetric representations $(\m_1,\dots,\m_r) = (s,0,\dots,0)$ coincide with the usual irreducible $D_r$ modules, so do the characters. They are given by
\begin{equation}
    \chi_s(\tau) = h_s(\tau_1,\dots,\tau_r,\tau_1^{-1},\dots,\tau_r^{-1}) - h_{s-2}(\tau_1,\dots,\tau_r,\tau_1^{-1},\dots,\tau_r^{-1})
\end{equation}
where $h_{-2} = h_{-1} = 0$ and $h_s$ for $s\geqslant0$ is the homogeneous symmetric polynomial defined by
\begin{equation}
    h_s(x_1,\dots,x_p) = \sum_{1\leqslant i_1\leqslant\dots\leqslant i_s\leqslant p} x_{i_1}\cdots x_{i_s}\, .
\end{equation}
We also have the following generating series:
\begin{equation}
    \sum_{s=0}^{+\infty} t^s h_s(x_1,\dots,x_p) = \frac{1}{\prod_{k=1}^p (1-t x_k)}\, ,\quad \sum_{s=0}^{+\infty} t^s \chi_s(\tau) = \frac{1-t^2}{\prod_{k=1}^r (1-t \tau_k) (1-t \tau_k^{-1})}\, .
\end{equation}

\section{Q-function example: one site}\label{app:explicitQ}

For $N=1$ the Q-operators are diagonal and we can read off the Q-functions. For $Q_1(x)$ we find
\begin{equation}
\begin{split}
 \left(Q_1(x)\right)_{11}&=\tau_1^x\Bigg[x^2-x\sum_{k=2}^{r}\left(1+\frac{\tau_k}{\tau_1-\tau_k}+\frac{\tau_k^{-1}}{\tau_1-\tau_k^{-1}}\right)\\&\qquad\qquad+\sum_{k=2}^r\left[\frac{1}{(\tau_1-\tau_k)(\tau_1-\tau_k^{-1})}+\frac{\tau_k}{2(\tau_1-\tau_k)}+\frac{\tau_k^{-1}}{2(\tau_1-\tau_k^{-1})}\right]+\frac{2r-3}{4}\Bigg]\,,
 \end{split}
\end{equation} 
\begin{equation}
 \left(Q_1(x)\right)_{ii}=\tau_1^x\Bigg[x-\frac{1}{2}+ \frac{\tau_1^{-1}}{\tau_1^{-1}- \tau_i}\Bigg]\,,\qquad 1<i\leq r\,,
\end{equation} 
\begin{equation}
 \left(Q_1(x)\right)_{ii}=\tau_1^x\Bigg[ x+\frac{1}{2}-\frac{\tau_{i'}}{\tau_1-\tau_{i'}}\Bigg]\,,\qquad r<i\leq 2r-1\,,
\end{equation} 
\begin{equation}
 \left(Q_{1}(x)\right)_{2r2r}=\tau_1^x\,.
\end{equation}

\section{Crossing relations}
\subsection{Crossing symmetry of transfer matrix}\label{app:sym}

The transfer matrix \eqref{eq:transferm} satisfies the crossing relations
\begin{equation}
 T_{1,s}(x)=\left.T_{1,s}^t(-x)\right|_{\tau_i\to \tau_i^{-1}}
\end{equation} 
We further note that when defining reflection matrix 
\begin{equation}\label{eq:invar}
\IDb =\left(\begin{array}{ccc}
                         0&0&1\\
                         0&\iddots&0\\
                         1&0&0\\
                        \end{array}
\right)\,.
\end{equation} 
 the twist parameters of the transfer matrix  exchange: $\IDb T_{1,s}(x)\IDb=\left.T_{1,s}(x)\right|_{\tau_i\to \tau_i^{-1}}$.
 It thus follows that
\begin{equation}
 T_{1,s}(x)=\IDb T_{1,s}^t(-x)\IDb=T_{1,s}'(-x)\,.
\end{equation}

\subsection{Crossing symmetry of single-index Q-operators}

In this appendix we discuss the derivation of the crossing relation for the single-index Q-operators. The corresponding Lax matrices satisfy
\begin{equation}
L^t(-z-1)|_{\text{p.h.}}=L(z)G\,.
\end{equation}
here $t$ denotes the transpose in the matrix space and ``p.h.'' denotes the particle hole transformation
\begin{equation}
 (\oa_i,\oad_i)|_{\text{p.h.}}=(-\oad_i,\oa_i)
\end{equation} 
and $G$ is the diagonal matrix
\begin{equation}
 G=
 \left(\begin{BMAT}[5pt]{c|c|c}{c|c|c}
 1&
0&0\\0&-\ID&0\\
0&0&1\\
      \end{BMAT}
\right)\,.
\end{equation} 
Using the symmetries of the twist in the Q-operator
\begin{equation}
\left. D\right|_{\tau_i\to\tau_i^{-1}}=\left(\tau_1^{-2}\right)^r D|_{\text{p.h.}}
\end{equation} 
we find that the normalised trace is independent of particle hole transformation. The extra factor above drops. We obtain
\begin{equation}
\left.Q_1^t(-x)\right|_{\tau_i\to\tau_i^{-1}}= Q_1(x)\left(G\otimes\ldots\otimes G\right)
\end{equation} 
Such equation holds for any $Q_i$. On the level of eigenvalues the transformation $G\otimes\ldots\otimes G$ only yields a possible sign, depending on the magnon number.

It further follows that 
\begin{equation}
\begin{split}
Q_1'(-x)&=\left(\IDb\otimes\ldots\otimes\IDb\right)
 Q_1^t(-x)\left(\IDb\otimes\ldots\otimes\IDb\right)\\
 &= \left(\IDb\otimes\ldots\otimes\IDb\right)\left. Q_1(x)\right|_{\tau_j\to\tau_j^{-1}}\left(\IDb\otimes\ldots\otimes\IDb\right)\left(G\otimes\ldots\otimes G\right)\\
 &=  Q_{1'}(x)\left(G\otimes\ldots\otimes G\right)\,.
 \end{split}
\end{equation}

\section{QQ-system of $A_3\simeq D_3$}\label{app:r=3}

We show in this appendix that, as is expected from the isomorphism $A_3\simeq D_3$, the known QQ-system for $A_3$ can be interpreted as the QQ-system for $D_3$, albeit in a particular gauge. We start with a reminder of the QQ-system for $A_3$: in order to avoid confusion, we shall denote $\QQ_I$ for $I\subset\{1,2,3,4\}$ the Q-functions for $A_3$ and the $SL(4)$ twists will be $z_1$, $z_2$, $z_3$ and $z_4$ such that $z_1z_2z_3z_4 = 1$. The following relations hold (neither $i$ nor $j$ belongs to $I$):
\begin{equation}
\QQ_{I\cup \{ i\}}^{[+1]} \QQ_{I\cup\{j\}}^{[-1]} -  \QQ_{I\cup\{i\}}^{[-1]} \QQ_{I\cup\{j\}}^{[+1]}= \frac{z_i-z_j}{\sqrt{z_i z_j}} \QQ_{I} \QQ_{I\cup\{i,j\}}\, .
\end{equation}
From these relations, one can also deduce that
\begin{equation}\label{QQQA3full}
\left|\begin{array}{lll}
\QQ_{\lbrace i,j\rbrace}^{[-2]} & \QQ_{\lbrace i,j\rbrace} & \QQ_{\lbrace i,j\rbrace}^{[+2]} \\
\QQ_{\lbrace i,k\rbrace}^{[-2]} & \QQ_{\lbrace i,k\rbrace} & \QQ_{\lbrace i,k\rbrace}^{[+2]} \\
\QQ_{\lbrace i,l\rbrace}^{[-2]} & \QQ_{\lbrace i,l\rbrace} & \QQ_{\lbrace i,l\rbrace}^{[+2]}
\end{array}
\right| = \frac{(z_j-z_k)(z_j-z_l)(z_k-z_l)}{z_j z_k z_l} \QQ_{\lbrace i\rbrace}^{[-1]} \QQ_{\lbrace i\rbrace}^{[+1]} \QQ_{\lbrace 1,2,3,4\rbrace}
\end{equation}
and
\begin{equation}\label{QQQA3empty}
\left|\begin{array}{lll}
\QQ_{\lbrace i,j\rbrace}^{[-2]} & \QQ_{\lbrace i,j\rbrace} & \QQ_{\lbrace i,j\rbrace}^{[+2]} \\
\QQ_{\lbrace j,k\rbrace}^{[-2]} & \QQ_{\lbrace j,k\rbrace} & \QQ_{\lbrace j,k\rbrace}^{[+2]} \\
\QQ_{\lbrace i,k\rbrace}^{[-2]} & \QQ_{\lbrace i,k\rbrace} & \QQ_{\lbrace i,k\rbrace}^{[+2]}
\end{array}
\right| = \frac{(z_j-z_i)(z_j-z_k)(z_i-z_k)}{z_i z_j z_k} \QQ_{\{ i,j,k\}}^{[-1]} \QQ_{\{ i,j,k\}}^{[+1]} \QQ_{\emptyset}\, .
\end{equation}
Both of these equations are identified with equation \eqref{QSSS}. More generally, both QQ-systems are the same if one makes the following identification between the two sets of Q-functions:
\begin{equation}
    Q_{\{1\}} = \QQ_{\{1,2\}}\, ,\quad Q_{\{2\}} = \QQ_{\{1,3\}}\, ,\quad Q_{\{3\}} = \QQ_{\{1,4\}}\, ,
\end{equation}
\begin{equation}
    Q_{\{1'\}} = \QQ_{\overline{\{1,2\}}} = \QQ_{\{3,4\}}\, ,\quad Q_{\{2'\}} = \QQ_{\{2,4\}}\, ,\quad Q_{\{3'\}} = \QQ_{\{2,3\}}\, ,
\end{equation}
\begin{equation}
    S_{(+,+,+)} = \QQ_{\{1\}}\, ,\quad S_{(+,-,-)} = \QQ_{\{2\}}\, ,\quad S_{(-,+,-)} = \QQ_{\{3\}}\, ,\quad S_{(-,-,+)} = \QQ_{\{4\}}\, ,
\end{equation}
\begin{equation}
    S_{(-,-,-)} = \QQ_{\{2,3,4\}}\, ,\quad S_{(+,+,-)} = \QQ_{\{1,2,3\}}\, ,\quad S_{(+,-,+)} = \QQ_{\{1,2,4\}}\, ,\quad S_{(-,+,+)} = \QQ_{\{1,3,4\}}\, .
\end{equation}
The twists are related via
\begin{equation}
    \tau_1 = z_1z_2 = \frac{1}{z_3z_4}\, , \quad  \tau_2 = z_1z_3 = \frac{1}{z_2z_4}\, , \quad  \tau_3 = z_1z_4 = \frac{1}{z_2z_3}
\end{equation}
while the remaining Q-functions are
\begin{equation}
    Q_\emptyset = 1\, ,\quad S_{+,\emptyset} = \QQ_\emptyset\, \quad \text{and} \quad S_{-,\emptyset} = \QQ_{\{1,2,3,4\}}\, .
\end{equation}
The previous equation shows that in identifying the two QQ-systems we had to partly fix the gauge for $D_3$. This explains why in the $A_3$ QQ-system there are only two gauge degrees of freedom \cite{Kazakov:2015efa} while there are three of them for the $D_3$ one.

\section{Elements of the Hasse diagram}\label{app:hasse}
In this appendix we give a more detailed explanation of the Hasse diagrams in Figure~\ref{fig:Hasse2} and Figure~\ref{fig:hasse4}.
 
\paragraph{QQ-relations along the tail} 

\begin{figure}[h]
 \begin{center}
% \vspace{-1cm}
% \resizebox{0.6\textwidth}{!}{
\begin{tikzpicture}
    \node [empty](origin){};
    \node [yellowRectangle, left=1cm of origin] (qsc1) {$Q_{I\cup\{i\}}$};
    \node [yellowRectangle, above=1cm of origin](q2){$Q_{I\cup\{i\}\cup\{j\}}$};
    \node [yellowRectangle, right=1cm of origin](qsc3){$Q_{I\cup \{j\}}$};
    \node [yellowRectangle, below=1cm of origin](q0){$Q_{I}$};
    \path [line] (qsc1) --  (q2);
    \path [line] (qsc3) -- (q2);
    \path [line] (q0) --  (qsc1);
    \path [line] (q0) -- (qsc3);
\end{tikzpicture}
% }
\end{center}
\caption{Plaquette for QQ-relations along the tail depicted in green.}
\label{fig:QQ}
\end{figure}
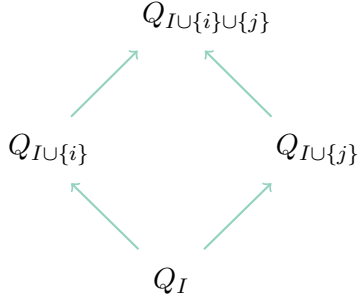

Along the tail of the Dynkin diagram the QQ-relations are given in \eqref{eq:QQrel}. They are depicted by the plaquette in Figure~\ref{fig:QQ}.
% $$
%  Q_{J\cup \{i\}}^{[+1]}Q_{J\cup\{j\}}^{[-1]}- Q_{J\cup\{i\}}^{[-1]}Q_{J\cup\{j\}}^{[+1]}= \frac{\tau_i-\tau_j}{\sqrt{\tau_i\tau_j}}Q_{J} Q_{J\cup\{i,j\}}
% $$
% with 
% $Q_\emptyset(x)=x^N$.
% 
% but exclude Q-functions with $i,i'\in I$

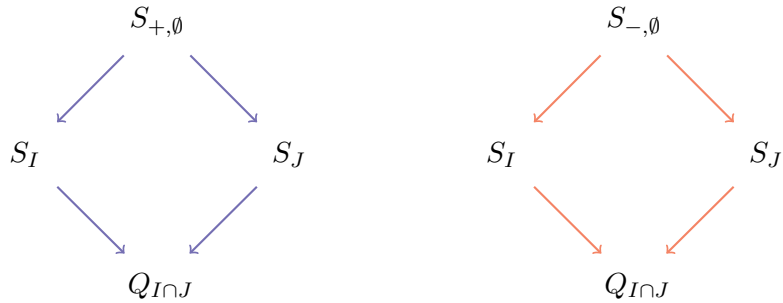
\begin{figure}[h]
\begin{center}
% \resizebox{\textwidth}{!}{
\begin{tikzpicture}
% [thick, scale = 0.6, transform canvas={scale=0.6}]

    \node [empty](origin){};
    \node [yellowRectangle, left=1cm of origin] (qsc1) {$S_I$};
    \node [yellowRectangle, above=1cm of origin](q2){\mbox{${S_{+,\emptyset}}$}};
    \node [yellowRectangle, right=1cm of origin](qsc3){$S_J$};
    \node [yellowRectangle, below=1cm of origin](q0){$Q_{I\cap J}$};
    
    \node [empty, right=6cm of origin](origin2){};
        \node [yellowRectangle, left=1cm of origin2] (qsc13) {$S_I$};
    \node [yellowRectangle, above=1cm of origin2](q23){\mbox{${S_{-,\emptyset}}$}};
    \node [yellowRectangle, right=1cm of origin2](qsc33){$S_J$};
    \node [yellowRectangle, below=1cm of origin2](q03){$Q_{I\cap J}$};

\tikzstyle{line} = [
    draw,
%     -latex'->,
->,
    thick, Blue!60,
]

    {\color{red}
    \path [line] (q2) --  (qsc1);
    \path [line] (q2) -- (qsc3);
    \path [line] (qsc1) --  (q0);
    \path [line] (qsc3) -- (q0);
    }

\tikzstyle{line} = [
    draw,
%     -latex'->,
->,
    thick, Red!60,
]

    {\color{red}
    \path [line] (q23) --  (qsc13);
    \path [line] (q23) -- (qsc33);
    \path [line] (qsc13) --  (q03);
    \path [line] (qsc33) -- (q03);
    }
\end{tikzpicture}
% }
\end{center}
 \caption{Spinorial QQ-relations for spinor nodes $\pm$ depicted in blue and red respectively.}
 \label{fig:sppla}
\end{figure}

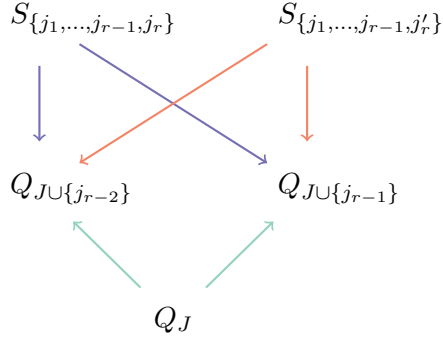
\begin{figure}[h]
  \begin{center}
  \begin{tikzpicture}

    \node [empty](origin){};

    \node [yellowRectangle, left=1cm of origin] (q1) {$Q_{J\cup \{j_{r-2}\}}$};
    \node [yellowRectangle, right=1cm of origin] (q2) {$Q_{J\cup \{j_{r-1}\}}$};

    \node [yellowRectangle, above=1cm of q1] (ppp) {$S_{\{j_1,\ldots,j_{r-1},j_r\}}$};
    \node [yellowRectangle, above=1cm of q2] (pmm) {$S_{\{j_1,\ldots,j_{r-1},j_r'\}}$};
    
%     \node [yellowRectangle, below=2cm  of origin,xshift=8.25cm] (zet) {$Q_\emptyset$};
    \node [yellowRectangle, below=1cm  of origin] (zet) {$Q_J$};

\tikzstyle{line} = [
    draw,
%     -latex'->,
->,
    thick, Blue!60,
]

    \path [line] (ppp) --  (q1);
    \path [line] (ppp) --  (q2);
    \tikzstyle{line} = [
    draw,
%     -latex'->,
->,
    thick, Red!60,
]
    \path [line] (pmm) --  (q2);
    \path [line] (pmm) --  (q1);

\tikzstyle{line} = [
    draw,
%     -latex'->,
->,
    thick, Red!60,
]

\tikzstyle{line} = [
    draw,
%     -latex'->,
->,
    thick, SeaGreen!60,
]

 \path [line] (zet) --  (q1);
    \path [line] (zet) --  (q2);
\end{tikzpicture}
 \end{center}
 \caption{QQ-relations at the fork. To avoid confusion with the plaquette we introduce a different direction for the arrows pointing away from spinorial nodes.}
 \label{fig:fork}
\end{figure}

\paragraph{QQ-relations at the spinor nodes}
The QQ-relations for the spinorial nodes were introduced in \eqref{SS relation} and \eqref{eq:qqspin}. We depict them as the QQ-relations along the tail by a plaquette, see Figure~\ref{fig:sppla}. Here we chose the opposite orientation of the arrows to avoid confusion with plaquettes at the fork, cf.~Figure~\ref{fig:fork}. Further depending on the spinor node, we choose a blue or red color for the arrows.

\paragraph{QQ-relations at the fork}

At the fork we have the QQ-relations \eqref{eq:QQrel} with $|J|=r-3$. In this case the Q-function at level $r-1$ factorises into two spinorial Q-functions, see \eqref{eq:fac1}. For  $J=\{j_1,\ldots,j_{r-3}\}$ we can write the QQ-relations as
$$
{\small
 Q_{J\cup \{j_{r-2}\}}^{[+1]}Q_{J\cup\{j_{r-1}\}}^{[-1]}- Q_{J\cup\{j_{r-2}\}}^{[-1]}Q_{J\cup\{j_{r-1}\}}^{[+1]}= \frac{\tau_{j_{r-2}}-\tau_{j_{r-1}}}{\sqrt{\tau_{j_{r-2}}\tau_{j_{r-1}}}}Q_{J} S_{\{j_1,\ldots,j_{r-1},j_r\}}S_{\{j_1,\ldots,j_{r-1},j_r'\}}\,,
}
$$
where $S_{\{j_1,\ldots,j_{r-1},j_r\}}$ and $S_{\{j_1,\ldots,j_{r-1},j_r'\}}$ belong to different spinor nodes. We denote these relations by the ``cat'' shaped diagram in Figure~\ref{fig:fork}. To avoid confusion with the plaquettes in Figure~\ref{fig:QQ} and Figure~\ref{fig:sppla} we chose all arrows to be ingoing at level $r-2$.

\section{Details for the computations of Section \ref{sec:QQ'}}

\subsection{Wronskian condition from QQ$'$-type constraints}\label{app:wronski}

Plugging the constraints $\eqref{Tsnegative}$ into equation \eqref{CTrelation} for $k=r-1$ we get
\begin{equation}
    C_{0,r-1,r-1} Q_\emptyset^{[r-2]} Q_{\emptyset}^{[2-r]} = h_r\left|\begin{array}{llll}
Q_{1}^{[-r+1]} & Q_{1}^{[-r+3]} & \cdots & Q_{1}^{[r-1]}\\
\vdots & \vdots &   & \vdots\\
Q_{r-1}^{[-r+1]} & Q_{r-1}^{[-r+3]} & \cdots & Q_{r-1}^{[r-1]}\\
Q_{r}^{[-r+1]} & Q_{r}^{[-r+3]} & \cdots & Q_{r}^{[r-1]}
\end{array}\right| \left|\begin{array}{llll}
Q_{1}^{[-r+1]} & Q_{1}^{[-r+3]} & \cdots & Q_{1}^{[r-1]}\\
\vdots & \vdots &   & \vdots\\
Q_{r-1}^{[-r+1]} & Q_{r-1}^{[-r+3]} & \cdots & Q_{r-1}^{[r-1]}\\
Q_{r'}^{[-r+1]} & Q_{r'}^{[-r+3]} & \cdots & Q_{r'}^{[r-1]}
\end{array}\right|\, .
\end{equation}
Using the explicit expression of $C_{0,r-1,r-1}$ gives
\begin{equation}
    W_{1,\dots,r-1}^{[-]}W_{1,\dots,r-1}^{[+]} Q_\emptyset^{[r-2]} Q_{\emptyset}^{[2-r]} = \frac{1}{\prod_{j=1}^{r-1}(u_j -u_r)} W_{1,\dots,r-1,r}W_{1,\dots,r-1,r'}
\end{equation}
where we used the notation $W_{i_1,\ldots,i_k} = \left|Q_{\{i_a\}}^{[k+1-2b]}\right|_k$. The derivation makes it clear that the previous identity still holds if one exchanges some $Q_{\{i\}}$ with $Q_{\{i'\}}$ so that one may actually write
\begin{equation}
    W_{i_1,\dots,i_{r-1}}^{[-]}W_{i_1,\dots,i_{r-1}}^{[+]} Q_\emptyset^{[r-2]} Q_{\emptyset}^{[2-r]} = \frac{1}{\prod_{j\neq i_r,i'_r}^{r}(u_j -u_r)} W_{i_1,\dots,i_{r-1},i_r}W_{i_1,\dots,i_{r-1},i'_r}\, .
\end{equation}
where we only assume that for all $1\leqslant a\neq b \leqslant r$ one has $\{i_a,i'_a\}\cap\{i_b,i'_b\} = \emptyset$. This is exactly equation \eqref{WronskianBAEQ}.

\subsection{Proof of equation \eqref{TasQQ'}}\label{app:proofTasQQ'}

We prove here the following claim: if $T_{a,s}$ satisfy the  Hirota equations and $T_{1,s}$ is given by equation \eqref{T1s 2r} then $T_{a,s}$ for $a\leqslant r-1$ is given by equation \eqref{TasQQ'}.

The proof is made by induction: the claim is true for $a=1$ by assumption and we have also shown, in the main text, that it is true for $a=2$. For higher $a$, the claim is clearly equivalent to equation \eqref{WW'Hirota} which is itself a particular case of the following identity:
\begin{multline}\label{WWtildeHirota}
    \frac{1}{2}\sum_{\substack{1\leqslant i_1<\dots<i_a\leqslant2r \\ 1\leqslant j_1<\dots<j_a\leqslant2r}} \Bigg|\begin{array}{cc}
W_{i_1,\dots,i_a}^{[+1]} & W_{i_1,\dots,i_a}^{[-1]}\\
W_{j_1,\dots,j_a}^{[+1]} & W_{j_1,\dots,j_a}^{[-1]}\\
\end{array}\Bigg|\,\,
\Bigg|\begin{array}{cc}
\WW_{i_1,\dots,i_a}^{[+1]} & \WW_{i_1,\dots,i_a}^{[-1]}\\
\WW_{j_1,\dots,j_a}^{[+1]} & \WW_{j_1,\dots,j_a}^{[-1]}\\
\end{array}\Bigg|\\
=  \left(\sum_{1\leqslant i_1<\dots<i_{a-1}\leqslant2r} W_{i_1,\dots,i_{a-1}} \WW_{i_1,\dots,i_{a-1}}\right)\left(\sum_{1\leqslant i_1<\dots<i_{a+1}\leqslant2r} W_{i_1,\dots,i_{a+1}} \WW_{i_1,\dots,i_{a+1}}\right)\, .
\end{multline}
where $W_{i_1,\dots,i_a} = \left|Q_{i_j}^{[a+1-2k]}\right|_a$ and $\WW_{i_1,\dots,i_a} = \left|P_{i_j}^{[a+1-2k]}\right|_a$ for $\{Q_i\}_{1\leqslant i \leqslant 2r}$ and $\{P_i\}_{1\leqslant i \leqslant 2r}$ two sets of arbitrary functions. In this appendix, most of the summation indices run from $1$ to $2r$ so we will not write these bounds under each summation symbols in the following. The only indices for which it will be different will be called $m$, $n$, $\tilde{m}$ or $\tilde{n}$, the values they may take will be indicated each time.

We shall now prove \eqref{WWtildeHirota}. Let us start from the left-hand side, we expand each of the determinants $W$ and $\WW$ with respect to the columns with shifts $\pm a$, for instance: $W_{i_1,\dots,i_a}^{[+1]} = \sum_{m=1}^a (-1)^{m+1}Q_{i_m}^{[a]} W_{i_1,\dots,\widehat{i_m},\dots i_a}$ and $W_{j_1,\dots,j_a}^{[-1]} = \sum_{n=1}^a (-1)^{n+a}Q_{j_n}^{[-a]} W_{j_1,\dots,\widehat{j_n},\dots j_a}$ where the hat over an index means that we omit it. We thus obtain
\begin{multline}
   \sum_{\substack{ i_1<\ldots<i_a \\ j_1<\ldots<j_a}} \Bigg|\begin{array}{cc}
W_{i_1,\dots,i_a}^{[+1]} & W_{i_1,\dots,i_a}^{[-1]}\\
W_{j_1,\dots,j_a}^{[+1]} & W_{j_1,\dots,j_a}^{[-1]}\\
\end{array}\Bigg|\,\,
\Bigg|\begin{array}{cc}
\WW_{i_1,\dots,i_a}^{[+1]} & \WW_{i_1,\dots,i_a}^{[-1]}\\
\WW_{j_1,\dots,j_a}^{[+1]} & \WW_{j_1,\dots,j_a}^{[-1]}\\
\end{array}\Bigg|\\
= \frac{1}{(a!)^2}\sum_{\substack{i_1,\dots,i_a \\  j_1,\dots,j_a}} \Bigg|\begin{array}{cc}
W_{i_1,\dots,i_a}^{[+1]} & W_{i_1,\dots,i_a}^{[-1]}\\
W_{j_1,\dots,j_a}^{[+1]} & W_{j_1,\dots,j_a}^{[-1]}\\
\end{array}\Bigg|\,\,
\Bigg|\begin{array}{cc}
\WW_{i_1,\dots,i_a}^{[+1]} & \WW_{i_1,\dots,i_a}^{[-1]}\\
\WW_{j_1,\dots,j_a}^{[+1]} & \WW_{j_1,\dots,j_a}^{[-1]}\\
\end{array}\Bigg|\\
= \frac{1}{(a!)^2}\sum_{\substack{ i_1,\dots,i_a \\  j_1,\dots,j_a \\ 1\leqslant m,n,\tilde{m},\tilde{n} \leqslant a}} (-1)^{m+n+\tilde{m}+\tilde{n}} \Bigg|\begin{array}{cc}
Q_{i_m}^{[a]} & Q_{i_m}^{[-a]}\\
Q_{j_n}^{[a]} & Q_{j_n}^{[-a]}\\
\end{array}\Bigg|\,\,
\Bigg|\begin{array}{cc}
P_{i_{\tilde{m}}}^{[a]} & P_{i_{\tilde{m}}}^{[-a]}\\
P_{j_{\tilde{n}}}^{[a]} & P_{j_{\tilde{n}}}^{[-a]}\\
\end{array}\Bigg|\\
\times W_{i_1,\dots,\widehat{i_m},\dots i_a} W_{j_1,\dots,\widehat{j_n},\dots j_a}
\WW_{i_1,\dots,\widehat{i_{\tilde{m}}},\dots i_a} \WW_{j_1,\dots,\widehat{j_{\tilde{n}}},\dots j_a} = L_1 + L_2 + L_3
\end{multline}
where we have split the sums over $m$, $n$, $\tilde{m}$ and $\tilde{n}$ into three contributions $L_1$, $L_2$ and $L_3$. $L_1$ contains all the terms with $m=\tilde{m}$ and $n=\tilde{n}$, $L_2$ all the terms with $m=\tilde{m}$ and $n\neq \tilde{n}$ or $m\neq \tilde{m}$ and $n = \tilde{n}$ while $L_3$ contains all the terms with $m\neq \tilde{m}$ and $n\neq \tilde{n}$. In each of the three cases the remaining sums (over $i$'s and $j$'s) do not depend on the actual values of $m$, $n$, $\tilde{m}$ and $\tilde{n}$ anymore so that we can perform the sums over these latter indices. We thus get
\begin{equation}\label{L1}
    L_1 = \left(\frac{1}{(a-1)!}\sum_{i_1,\dots,i_{a-1}} W_{i_1,\dots,i_{a-1}} \WW_{i_1,\dots,i_{a-1}}\right)^2  \sum_{ i,j}\Bigg|\begin{array}{cc}
Q_{i}^{[a]} & Q_{i}^{[-a]}\\
Q_{j}^{[a]} & Q_{j}^{[-a]}\\
\end{array}\Bigg|\,\,
\Bigg|\begin{array}{cc}
P_{i}^{[a]} & P_{i}^{[-a]}\\
P_{j}^{[a]} & P_{j}^{[-a]}\\
\end{array}\Bigg|\, ,
\end{equation}
\begin{multline}\label{L2}
    L_2 = \frac{2}{(a-2)!}\left(\frac{1}{(a-1)!}\sum_{i_1,\dots,i_{a-1}} W_{i_1,\dots,i_{a-1}} \WW_{i_1,\dots,i_{a-1}}\right)\\
\times\left(-\sum_{ i,j,k,i_1,\dots,i_{a-2}} \Bigg|\begin{array}{cc}
Q_i^{[a]} & Q_i^{[-a]}\\
Q_j^{[a]} & Q_j^{[-a]}\\
\end{array}\Bigg|\,\,
\Bigg|\begin{array}{cc}
P_i^{[a]} & P_i^{[-a]}\\
P_k^{[a]} & P_{k}^{[-a]}\\
\end{array}\Bigg| W_{k,i_1,\dots,i_{a-2}} \WW_{j,i_1,\dots,i_{a-2}}\right)\, ,
\end{multline}
%The contribution coming from $m=\tilde{m}$ and $n\neq \tilde{n}$ is the same as that from $m\neq \tilde{m}$ and $n = \tilde{n}$ so that their sum is
%\begin{multline}
%    \frac{2}{(a!)^2}\sum_{\substack{1\leqslant i_1,\dots,i_a\leqslant2r \\ 1\leqslant j_1,\dots,j_a\leqslant2r \\ 1\leqslant m,n\neq\tilde{n} \leqslant a}} (-1)^{n+\tilde{n}} \Bigg|\begin{array}{cc}
%Q_{i_m}^{[a]} & Q_{i_m}^{[-a]}\\
%Q_{j_n}^{[a]} & W_{j_n}^{[-a]}\\
%\end{array}\Bigg|\,\,
%\Bigg|\begin{array}{cc}
%P_{i_m}^{[a]} & P_{i_m}^{[-a]}\\
%P_{j_{\tilde{n}}}^{[a]} & P_{j_{\tilde{n}}}^{[-a]}\\
%\end{array}\Bigg|\\
%\times W_{i_1,\dots,\widehat{i_m},\dots i_a} W_{j_1,\dots,\widehat{j_n},\dots j_a}
%\WW_{i_1,\dots,\widehat{i_m},\dots i_a} \WW_{j_1,\dots,\widehat{j_{\tilde{n}}},\dots j_a}\\
%= \frac{2}{(a-2)!}\left(\frac{1}{(a-1)!}\sum_{1\leqslant i_1,\dots,i_{a-1}\leqslant2r} W_{i_1,\dots,i_{a-1}} \WW_{i_1,\dots,i_{a-1}}\right)\\
%\times\left(-\sum_{1\leqslant i,j,k,i_1,\dots,i_{a-1}\leqslant 2r} \Bigg|\begin{array}{cc}
%Q_i^{[a]} & Q_i^{[-a]}\\
%Q_j^{[a]} & Q_j^{[-a]}\\
%\end{array}\Bigg|\,\,
%\Bigg|\begin{array}{cc}
%P_i^{[a]} & P_i^{[-a]}\\
%P_k^{[a]} & P_{k}^{[-a]}\\
%\end{array}\Bigg| W_{k,i_1,\dots,i_{a-1}} \WW_{j,i_1,\dots,i_{a-1}}\right)
%\end{multline}
%The remaining terms are those for which $m\neq \tilde{m}$ and $n\neq \tilde{n}$. Once again the sums over indices $i_k$'s and $j_k$'s do not depend on the actual value of $m$, $n$, $\tilde{m}$ and $\tilde{n}$ so that we get the same term $a^2(a-1)^2$ times, the contribution is thus written
\begin{multline}
    L_3 = \frac{1}{((a-2)!)^2}\sum_{\substack{i_1,\dots,i_{a-2}\\ j_1,\dots,j_{a-2}\\ i,j,k,l }} \Bigg|\begin{array}{cc}
Q_i^{[a]} & Q_i^{[-a]}\\
Q_j^{[a]} & Q_j^{[-a]}\\
\end{array}\Bigg|\,\,
\Bigg|\begin{array}{cc}
P_k^{[a]} & P_k^{[-a]}\\
P_l^{[a]} & P_l^{[-a]}\\
\end{array}\Bigg|\\
\times W_{k,i_1,\dots,i_{a-2}} W_{l,j_1,\dots,j_{a-2}} \WW_{i,i_1,\dots,i_{a-2}} \WW_{j,j_1,\dots,j_{a-2}}\, .
\end{multline}
One can rewrite $L_3$ using the Plücker identity \eqref{Plucker1}. We first use it to write
\begin{equation}
    W_{k,i_1,\dots,i_{a-2}}W_{l,j_1,\dots,j_{a-1}} = W_{l,i_1,\dots,i_{a-2}}W_{k,j_1,\dots,j_{a-2}} + \sum_{p=1}^{a-2} (-1)^{p-1} W_{k,l,i_1,\dots,\widehat{i_p},\dots,i_{a-2}} W_{i_p,j_1,\dots,j_{a-2}}
\end{equation}
which we then plug in the expression for $L_3$, after some renaming of the indices this yields
\begin{multline}
    L_3 = -L_3 + \frac{1}{(a-3)!(a-2)!}\sum_{\substack{i_1,\dots,i_{a-3}\\ j_1,\dots,j_{a-1}\\ i,j,k,l }} \Bigg|\begin{array}{cc}
Q_i^{[a]} & Q_i^{[-a]}\\
Q_j^{[a]} & Q_j^{[-a]}\\
\end{array}\Bigg|\,\,
\Bigg|\begin{array}{cc}
P_k^{[a]} & P_k^{[-a]}\\
P_l^{[a]} & P_l^{[-a]}\\
\end{array}\Bigg|\\
\times W_{k,l,i_1,\dots,i_{a-3}} W_{j_1,\dots,j_{a-1}} \WW_{i,j_1,i_1,\dots,i_{a-3}} \WW_{j,j_2,\dots,j_{a-1}}\, .
\end{multline}
This means that
\begin{multline}
    L_3 = \frac{1}{2(a-3)!(a-2)!}\sum_{\substack{i_1,\dots,i_{a-3}\\ j_1,\dots,j_{a-1}\\ i,j,k,l }} \Bigg|\begin{array}{cc}
Q_i^{[a]} & Q_i^{[-a]}\\
Q_j^{[a]} & Q_j^{[-a]}\\
\end{array}\Bigg|\,\,
\Bigg|\begin{array}{cc}
P_k^{[a]} & P_k^{[-a]}\\
P_l^{[a]} & P_l^{[-a]}\\
\end{array}\Bigg|\\
\times W_{k,l,i_1,\dots,i_{a-3}} W_{j_1,\dots,j_{a-1}} \WW_{i,j_1,i_1,\dots,i_{a-3}} \WW_{j,j_2,\dots,j_{a-1}}\, .
\end{multline}
We now apply again the Plücker identity:
\begin{equation}
    \WW_{i,j_1,i_1,\dots,i_{a-3}} \WW_{j,j_2,\dots,j_{a-1}} = \WW_{i,j,i_1,\dots,i_{a-3}} \WW_{j_1,j_2,\dots,j_{a-1}} + \sum_{p=2}^{a-1} (-1)^p \WW_{i,j_p,i_1,\dots,i_{a-3}} \WW_{j,j_1,j_2,\dots,\widehat{j_p},\dots,j_{a-1}}
\end{equation}
so that
\begin{multline}
    L_3 = \frac{1}{2(a-3)!(a-2)!}\sum_{\substack{i_1,\dots,i_{a-3}\\ j_1,\dots,j_{a-1}\\ i,j,k,l }} \Bigg|\begin{array}{cc}
Q_i^{[a]} & Q_i^{[-a]}\\
Q_j^{[a]} & Q_j^{[-a]}\\
\end{array}\Bigg|\,\,
\Bigg|\begin{array}{cc}
P_k^{[a]} & P_k^{[-a]}\\
P_l^{[a]} & P_l^{[-a]}\\
\end{array}\Bigg|\\
\times W_{k,l,i_1,\dots,i_{a-3}} W_{j_1,\dots,j_{a-1}} \WW_{i,j,i_1,\dots,i_{a-3}} \WW_{j_1,\dots,j_{a-1}} - (a-2)L_3\, .
\end{multline}
Finally, we arrive at the following expression:
\begin{multline}\label{L3}
    L_3 = \frac{1}{2(a-3)!}\left(\frac{1}{(a-1)!}\sum_{i_1,\dots,i_{a-1}} W_{i_1,\dots,i_{a-1}} \WW_{i_1,\dots,i_{a-1}}\right)\\
    \times\left(\sum_{i,j,k,l,i_1,\dots,i_{a-3}} \Bigg|\begin{array}{cc}
Q_i^{[a]} & Q_i^{[-a]}\\
Q_j^{[a]} & Q_j^{[-a]}\\
\end{array}\Bigg|\,\,
\Bigg|\begin{array}{cc}
P_k^{[a]} & P_k^{[-a]}\\
P_l^{[a]} & P_l^{[-a]}\\
\end{array}\Bigg| W_{k,l,i_1,\dots,i_{a-3}} \WW_{i,j,i_1,\dots,i_{a-3}}\right)\, .
\end{multline}

In order to prove \eqref{WWtildeHirota} we need to show that
\begin{multline}
    \frac{L_1+L_2+L_3}{2} = \left(\frac{1}{(a-1)!}\sum_{i_1,\dots,i_{a-1}} W_{i_1,\dots,i_{a-1}} \WW_{i_1,\dots,i_{a-1}}\right)\\
    \times\left(\frac{1}{(a+1)!}\sum_{i_1,\dots,i_{a+1}} W_{i_1,\dots,i_{a+1}} \WW_{i_1,\dots,i_{a+1}}\right)\, .
\end{multline}
From expressions \eqref{L1}, \eqref{L2} and \eqref{L3} this is equivalent to showing that
\begin{multline}\label{auxiliaryWWtilde}
    \frac{(a+1)a}{2} \sum_{i,j,i_1,\dots,i_{a-1}} \Bigg|\begin{array}{cc}
Q_i^{[a]} & Q_i^{[-a]}\\
Q_j^{[a]} & Q_j^{[-a]}\\
\end{array}\Bigg|\,\,
\Bigg|\begin{array}{cc}
P_i^{[a]} & P_i^{[-a]}\\
P_j^{[a]} & P_j^{[-a]}\\
\end{array}\Bigg| W_{i_1,\dots,i_{a-1}} \WW_{i_1,\dots,i_{a-1}}\\
    -(a+1)a(a-1) \sum_{i,j,k,i_1,\dots,i_{a-2}} \Bigg|\begin{array}{cc}
Q_i^{[a]} & Q_i^{[-a]}\\
Q_j^{[a]} & Q_j^{[-a]}\\
\end{array}\Bigg|\,\,
\Bigg|\begin{array}{cc}
P_i^{[a]} & P_i^{[-a]}\\
P_k^{[a]} & P_{k}^{[-a]}\\
\end{array}\Bigg| W_{k,i_1,\dots,i_{a-2}} \WW_{j,i_1,\dots,i_{a-2}}\\
+\frac{(a+1)a(a-1)(a-2)}{4} \sum_{i,j,k,l,i_1,\dots,i_{a-3}} \Bigg|\begin{array}{cc}
Q_i^{[a]} & Q_i^{[-a]}\\
Q_j^{[a]} & Q_j^{[-a]}\\
\end{array}\Bigg|\,\,
\Bigg|\begin{array}{cc}
P_k^{[a]} & P_k^{[-a]}\\
P_l^{[a]} & P_l^{[-a]}\\
\end{array}\Bigg| W_{k,l,i_1,\dots,i_{a-3}} \WW_{i,j,i_1,\dots,i_{a-3}}\\
= \sum_{i_1,\dots,i_{a+1}} W_{i_1,\dots,i_{a+1}} \WW_{i_1,\dots,i_{a+1}}\, .
\end{multline}

This last identity is proven by expanding the determinants in the right-hand side with respect to their first and last columns:
\begin{equation}
    W_{i_1,\dots,i_{a+1}} = \sum_{1\leqslant m<n\leqslant a+1} (-1)^{m+n+a}\Bigg|\begin{array}{cc}
Q_{i_m}^{[a]} & Q_{i_m}^{[-a]}\\
Q_{i_n}^{[a]} & Q_{i_n}^{[-a]}\\
\end{array}\Bigg| W_{i_1,\dots,\widehat{i_m},\dots,\widehat{i_n},\dots,i_{a+1}}
\end{equation}
and
\begin{equation}
    \WW_{i_1,\dots,i_{a+1}} = \sum_{1\leqslant \tilde{m}<\tilde{n}\leqslant a+1} (-1)^{\tilde{m}+\tilde{n}+a} \Bigg|\begin{array}{cc}
P_{i_{\tilde{m}}}^{[a]} & P_{i_{\tilde{m}}}^{[-a]}\\
P_{i_{\tilde{n}}}^{[a]} & P_{i_{\tilde{n}}}^{[-a]}\\
\end{array}\Bigg| \WW_{i_1,\dots,\widehat{i_{\tilde{m}}},\dots,\widehat{i_{\tilde{n}}},\dots,i_{a+1}}\, .
\end{equation}
We then once again group the terms depending on the values of $m$, $n$, $\tilde{m}$ and $\tilde{n}$ and recover exactly the identity \eqref{auxiliaryWWtilde}. There are indeed  $\frac{(a+1)a}{2}$ terms with $(m,n) = (\tilde{m},\tilde{n})$, $(a+1)a(a-1)$ terms with $m=\tilde{m}$ and $n\neq \tilde{n}$ or $m\neq \tilde{m}$ and $n = \tilde{n}$, and $\frac{(a+1)a(a-1)(a-2)}{4}$ terms with $m\neq \tilde{m}$ and $n\neq \tilde{n}$.

\section{More on Weyl-type formulas}

\subsection{From QQ$'$-relations to Weyl-type formulas for $D_2\simeq A_1\oplus A_1$}\label{app:r=2}

Here we demonstrate for the examples of $D_2$ spin chains how to use the QQ$'$-relations to recover the Weyl-type formulas for T-functions. The Hasse diagram is depicted in Figure~\ref{fig:HasseD2}.%This case can be compared with the known formulas for $A_2$ algebras using the isomorphisms $SO(4)\sim SU(2)\otimes SU(2)$.

\tikzstyle{yellowRectangle2} = [
    rectangle,
    node distance=1 cm,
    text width=5 em,
    text centered,
    rounded corners,
    minimum height=0.2 cm,
    minimum width=0.3 cm,
    fill=gray!30,
    thick
]

\tikzstyle{yellowRectangle1} = [
    circle,
    node distance=1 cm,
    text width=1.1 em,
    text centered,
    rounded corners,
    minimum height=0.2 cm,
    minimum width=0.3 cm,
    fill=gray!30,
    thick
]
\begin{figure}
 \begin{center}
  \begin{tikzpicture}[scale=0.7, transform shape]

    \node [empty](origin){};
      
    \node [yellowRectangle2, below=0cm of origin] (zet) {$x^N=Q_\emptyset$};
    \node [yellowRectangle2, right=7cm  of zet] (one) {$\ID=S_{\emptyset,-}$};
    \node [yellowRectangle2, left=7cm  of zet] (two) {$\ID=S_{\emptyset,+}$}; 
    
    \node [yellowRectangle2, right=2.5cm of zet, yshift=3cm] (pp) {$Q_1=S_{+,+}$};
    \node [yellowRectangle2, right=2.5cm of zet, yshift=-3cm] (mm)  {$Q_2=S_{-_,-}$};
    \node [yellowRectangle2, left=2.5cm of zet, yshift=-3cm] (pm) {$Q_{2'}=S_{+,-}$};
    \node [yellowRectangle2, left=2.5cm of zet, yshift=3cm] (mp) {$Q_{1'}=S_{-,+}$};
        
    \path [line] (one) --  (pp);
    \path [line] (one) --  (mm);
    \path [line] (two) --  (pm);
    \path [line] (two) --  (mp);

 \path [line] (pp) --  (zet);
 \path [line] (mm) --  (zet);
 \path [line] (pm) --  (zet);
 \path [line] (mp) --  (zet);

\end{tikzpicture}
 \end{center}
\caption{Hasse diagram for $D_2\simeq A_1\oplus A_1$}
 \label{fig:HasseD2}
\end{figure}
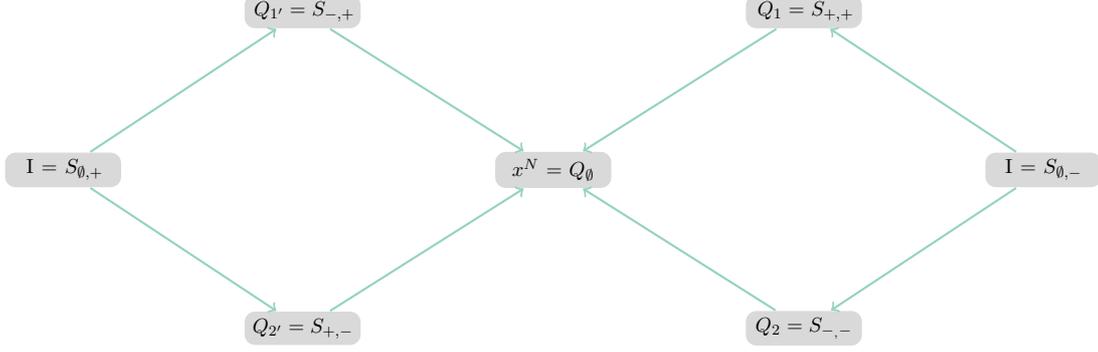

From the two constraints
\begin{equation}
Q_{1}Q_{1'}+Q_{2}Q_{2'}=0  \,,
\end{equation} 
and 
\begin{equation}
Q_{1}^{[1]}Q_{1'}^{[-1]}+Q_{1}^{[-1]}Q_{1'}^{[1]}+Q_{2}^{[1]}Q_{2'}^{[-1]}+Q_{2}^{[-1]}Q_{2'}^{[1]}=Q_{\emptyset}^2\,,
\end{equation}
cf.~\eqref{Tsnegative}, we obtain
\begin{equation}\label{F1F2}
\left(\frac{Q_{1'}^{[1]}}{Q_{2}^{[1]}}-\frac{Q_{1'}^{[-1]}}{Q_{2}^{[-1]}}\right)=\frac{Q_{\emptyset}^2}{W_1}
\end{equation}
where \(W_{n}=Q_{1}^{[n]}Q_{2}^{[-n]}-Q_{1}^{[-n]}Q_{2}^{[n]}\).

Further on, excluding $Q_{2'}$ from 
\begin{equation}
T_{1,s}=Q_{1}^{[s+1]}Q_{1'}^{[-s-1]}+Q_{1}^{[-s-1]}Q_{1'}^{[s+1]}+Q_{2}^{[s+1]}Q_{2'}^{[-s-1]}+Q_{2}^{[-s-1]}Q_{2'}^{[s+1]}
\end{equation} we get
\begin{equation}
T_{1,s}=\left(\frac{Q_{1'}^{[s+1]}}{Q_{2}^{[s+1]}}-\frac{Q_{1'}^{[-s-1]}}{Q_{2}^{[-s-1]}}\right)W_{s+1}\, .
\end{equation}
Excluding the  difference in the first bracket in the rhs using \eqref{F1F2} we arrive at
%\begin{equation}
%T=(Q_{\emptyset}^{[1]})^2 \frac{W_{2}}{W_{1}^{[1]}} + (Q_{\emptyset}^{[-1]})^2 \frac{W_{2}}{W_{1}^{[-1]}}\, .
%\end{equation}
%It reduces to \eqref{Tdet} for $r=2$. We can easily generalize it to higher symmetric representations:
\begin{equation}
T_{1,s} = W_{s+1} \sum_{l=0}^{s} \frac{(Q_{\emptyset}^{[2l-s]})^2}{W_{1}^{[2l-s]}}\,.
\end{equation}  This coincides with the $r=2$ case of the determinant formula~\eqref{Tsdet}. 

\subsection{Additional formulas in the general case}\label{app:uglyWeyl-type}

For the sake of completeness, we give here the Weyl-type formulas complementary to those given in Section~\ref{sec:Weyl-type}, i.e. for $T_{1,s}$ in terms of spinorial Q-functions:
\begin{multline}
    T_{1,s} = Q_\emptyset^{[r+s-2]} Q_\emptyset^{[2-r-s]}\Bigg(\sum_{l=0}^{s}\frac{S_{I_r}^{[-s-1]} S_{I_r}^{[s+1]}}{S_{I_r}^{[2l-s-1]} S_{I_r}^{[2l-s+1]}} \frac{\left|S_i^{[-(2r+s-1-2j-2l\delta_{j,r})]}\right|_{r}}{\left|S_i^{[-(2r+s-1-2j)]}\right|_{r}} \frac{\left|S_i^{[2r+s-1-2j-2(s-l)\delta_{j,r}]}\right|_{r}}{\left|S_i^{[2r+s-1-2j]}\right|_{r}}\\
    - \sum_{l=1}^{s-1}\frac{S_{I_r}^{[-s-1]} S_{I_r}^{[s+1]}}{S_{I_r}^{[2l-s-1]} S_{I_r}^{[2l-s+1]}} \frac{\left|S_i^{[-(2r+s-1-2j-2(l-1)\delta_{j,r})]}\right|_{r}}{\left|S_i^{[-(2r+s-1-2j)]}\right|_{r}} \frac{\left|S_i^{[2r+s-1-2j-2(s-1-l)\delta_{j,r}]}\right|_{r}}{\left|S_i^{[2r+s-1-2j]}\right|_{r}}\Bigg)
\end{multline}
and for $T_{a,1}$ in terms of single-index Q-functions:
\begin{multline}
    T_{a,1} = \frac{1}{\prod_{k=1}^{a-1} Q_{\emptyset}^{[r-a+2k]} Q_{\emptyset}^{[-(r-a+2k)]}} \sum_{\substack{0\leqslant k,l \leqslant a \\ a-k-l\in 2\mathbb{N}}} \left(\prod_{m=1}^k Q_{\emptyset}^{[r+a-2m]} Q_{\emptyset}^{[4+a-r-2m]}\right)\\
    \times \frac{ \left|Q_{\{i\}}^{[r+a+3-2k-2j-2\theta(j+k-r-1)]}\right|_r}{\left|Q_{\{i\}}^{[r+a+3-2k-2j]}\right|_r} \prod_{m=1}^{\frac{a-k-l}{2}} \left(Q_{\emptyset}^{[r+a-2k-4m]} Q_{\emptyset}^{[4+a-r-2k-4m]}\right)^2\\
    \times\frac{\left|Q_{\{i\}}^{[r+a+1-2k-2j]}\right|_r \left|Q_{\{i\}}^{[r+a-3-2k-2j]}\right|_r^2 \cdots  \left|Q_{\{i\}}^{[2l+r+5-a-2j]}\right|_r^2 \left|Q_{\{i\}}^{[2l+r+1-a-2j]}\right|_r}{\left|Q_{\{i\}}^{[r+a-1-2k-2j]}\right|_r^2 \cdots \left|Q_{\{i\}}^{[2l+r+3-a-2j]}\right|_r^2}\\
    \times \left(\prod_{m=1}^l Q_{\emptyset}^{[2+2l-r-a-2m]} Q_{\emptyset}^{[2l+r-a-2m-2]}\right) \frac{\left|Q_{\{i\}}^{[-(r+a+3-2l-2j-2\theta(j+l-r-1))]}\right|_r}{\left|Q_{\{i\}}^{[-(r+a+3-2l-2j)]}\right|_r}\, .
\end{multline}

Notice that the first formula, expressing  $T_{1,s}$ in terms of spinorial $Q$-functions, is much more complicated than the expression  \eqref{Tsdet}  in terms of fundamental $Q$-functions, whereas the second formula expressing  $T_{a,1}$ in terms of fundamental $Q$-functions, is more complicated than \eqref{Tadet} expressing it in terms of spinorial $Q$-functions.

%\begin{equation}
%    \sum_{1\leqslant i_1\leqslant\ldots\leqslant i_l\leqslant r} b_{i_1,r}^{[1]} \cdots b_{i_l,r}^{[2l-1]} = \frac{S_{I_r}^{[-1]}}{S_{I_r}^{[2l-1]}} \left(\prod_{a=1}^l Q_{\emptyset}^{[r-2+2a]} Q_{\emptyset}^{[-r+2a]}\right) \frac{\left|S_i^{[-(2r-1-2j-2l\delta_{j,r})]}\right|_{r}}{\left|S_i^{[-(2r-1-2j)]}\right|_{r}}\, ,
%\end{equation}
%\begin{equation}
%    \sum_{r+1\leqslant i_1\leqslant\ldots\leqslant i_l\leqslant 2r} b_{i_1,r}^{[-2l+1]} \cdots b_{i_l,r}^{[-1]} = \frac{S_{I_r}^{[+1]}}{S_{I_r}^{[-2l+1]}} \left(\prod_{a=1}^l Q_{\emptyset}^{[r-2l-2+2a]} Q_{\emptyset}^{[-r-2l+2a]}\right) \frac{\left|S_i^{[2r-1-2j-2l\delta_{j,r}]}\right|_{r}}{\left|S_i^{[2r-1-2j]}\right|_{r}}\, ,
%\end{equation}

%For $1\leqslant l\leqslant r$, one has
%\begin{equation}\label{sum boxes a}
%    \sum_{1\leqslant i_1<\ldots< i_l\leqslant r} \tilde{b}_{i_1}^{[2l-1]}\cdots\tilde{b}_{i_l}^{[1]} = \frac{ \left|Q_{\{i\}}^{[2r+3-2j-2\theta(j+l-r-1)]}\right|_r}{\left|Q_{\{i\}}^{[2r+3-2j]}\right|_r}
%\end{equation}
%and
%\begin{equation}\label{sum boxes a2}
%    \sum_{r+1\leqslant i_1<\ldots< i_l\leqslant 2r} \tilde{b}_{i_1}^{[-1]}\cdots\tilde{b}_{i_l}^{[1-2l]} = \frac{ \left|Q_{\{i\}}^{[-(2r+3-2j-2\theta(j+l-r-1))]}\right|_r}{\left|Q_{\{i\}}^{[-(2r+3-2j)]}\right|_r}
%\end{equation}

\bibliography{DrQ.bib}

\providecommand{\href}[2]{#2}\begingroup\raggedright\begin{thebibliography}{10}

\bibitem{Baxter:1982zz}
R.~Baxter, \emph{{Exactly Solved Models in Statistical Mechanics}}. 1982.

\bibitem{Bazhanov:1996dr}
V.~V. Bazhanov, S.~L. Lukyanov and A.~B. Zamolodchikov, \emph{{Integrable
  structure of conformal field theory. 2. Q operator and DDV equation}},
  \href{https://doi.org/10.1007/s002200050240}{\emph{Commun. Math. Phys.}
  {\bfseries 190} (1997) 247}
  [\href{https://arxiv.org/abs/hep-th/9604044}{{\ttfamily hep-th/9604044}}].

\bibitem{Faddeev:1994zg}
L.~Faddeev and G.~Korchemsky, \emph{{High-energy QCD as a completely integrable
  model}}, \href{https://doi.org/10.1016/0370-2693(94)01363-H}{\emph{Phys.
  Lett. B} {\bfseries 342} (1995) 311}
  [\href{https://arxiv.org/abs/hep-th/9404173}{{\ttfamily hep-th/9404173}}].

\bibitem{Derkachov:2001yn}
S.~E. Derkachov, G.~Korchemsky and A.~Manashov, \emph{{Noncompact Heisenberg
  spin magnets from high-energy QCD: 1. Baxter Q operator and separation of
  variables}}, \href{https://doi.org/10.1016/S0550-3213(01)00457-6}{\emph{Nucl.
  Phys. B} {\bfseries 617} (2001) 375}
  [\href{https://arxiv.org/abs/hep-th/0107193}{{\ttfamily hep-th/0107193}}].

\bibitem{Lipatov:1993qn}
L.~N. Lipatov, \emph{{High-energy asymptotics of multicolor QCD and
  two-dimensional conformal field theories}},
  \href{https://doi.org/10.1016/0370-2693(93)90951-D}{\emph{Phys. Lett.}
  {\bfseries B309} (1993) 394}.

\bibitem{Beisert:2010jr}
N.~Beisert et~al., \emph{{Review of AdS/CFT Integrability: An Overview}},
  \href{https://doi.org/10.1007/s11005-011-0529-2}{\emph{Lett. Math. Phys.}
  {\bfseries 99} (2012) 3} [\href{https://arxiv.org/abs/1012.3982}{{\ttfamily
  1012.3982}}].

\bibitem{Gromov:2013pga}
N.~Gromov, V.~Kazakov, S.~Leurent and D.~Volin, \emph{{Quantum Spectral Curve
  for Planar $\mathcal{N} = 4$ Super-Yang-Mills Theory}},
  \href{https://doi.org/10.1103/PhysRevLett.112.011602}{\emph{Phys. Rev. Lett.}
  {\bfseries 112} (2014) 011602}
  [\href{https://arxiv.org/abs/1305.1939}{{\ttfamily 1305.1939}}].

\bibitem{Gromov:2014caa}
N.~Gromov, V.~Kazakov, S.~Leurent and D.~Volin, \emph{{Quantum spectral curve
  for arbitrary state/operator in AdS$_{5}$/CFT$_{4}$}},
  \href{https://doi.org/10.1007/JHEP09(2015)187}{\emph{JHEP} {\bfseries 09}
  (2015) 187} [\href{https://arxiv.org/abs/1405.4857}{{\ttfamily 1405.4857}}].

\bibitem{Dorey:2007zx}
P.~Dorey, C.~Dunning and R.~Tateo, \emph{{The ODE/IM Correspondence}},
  \href{https://doi.org/10.1088/1751-8113/40/32/R01}{\emph{J. Phys. A}
  {\bfseries 40} (2007) R205}
  [\href{https://arxiv.org/abs/hep-th/0703066}{{\ttfamily hep-th/0703066}}].

\bibitem{Boos:2008rh}
H.~Boos, M.~Jimbo, T.~Miwa, F.~Smirnov and Y.~Takeyama, \emph{{Hidden Grassmann
  Structure in the XXZ Model II: Creation Operators}},
  \href{https://doi.org/10.1007/s00220-008-0617-z}{\emph{Commun. Math. Phys.}
  {\bfseries 286} (2009) 875}
  [\href{https://arxiv.org/abs/0801.1176}{{\ttfamily 0801.1176}}].

\bibitem{Kuniba:2016fpi}
A.~Kuniba, V.~Mangazeev, S.~Maruyama and M.~Okado, \emph{{Stochastic R matrix
  for $U_q(A_n^{(1)})$}},
  \href{https://doi.org/10.1016/j.nuclphysb.2016.09.016}{\emph{Nucl. Phys. B}
  {\bfseries 913} (2016) 248}
  [\href{https://arxiv.org/abs/1604.08304}{{\ttfamily 1604.08304}}].

\bibitem{Lazarescu_2014}
A.~Lazarescu and V.~Pasquier, \emph{{Bethe Ansatz and Q-operator for the open
  ASEP}}, \href{https://doi.org/10.1088/1751-8113/47/29/295202}{\emph{Journal
  of Physics A: Mathematical and Theoretical} {\bfseries 47} (2014) 295202}
  [\href{https://arxiv.org/abs/1403.6963}{{\ttfamily 1403.6963}}].

\bibitem{Frenkel:2013uda}
E.~Frenkel and D.~Hernandez, \emph{{Baxter's Relations and Spectra of Quantum
  Integrable Models}},
  \href{https://doi.org/10.1215/00127094-3146282}{\emph{Duke Math. J.}
  {\bfseries 164} (2015) 2407}
  [\href{https://arxiv.org/abs/1308.3444}{{\ttfamily 1308.3444}}].

\bibitem{Sklyanin:1984sb}
E.~K. Sklyanin, \emph{{The Quantum Toda Chain}},
  \href{https://doi.org/10.1007/3-540-15213-X_80}{\emph{Lect. Notes Phys.}
  {\bfseries 226} (1985) 196}.

\bibitem{Faddeev:1996iy}
L.~Faddeev, \emph{{How Algebraic Bethe Ansatz works for integrable model}},
  \href{https://arxiv.org/abs/hep-th/9605187}{{\ttfamily hep-th/9605187}}.

\bibitem{Bazhanov:1994ft}
V.~V. Bazhanov, S.~L. Lukyanov and A.~B. Zamolodchikov, \emph{{Integrable
  structure of conformal field theory, quantum KdV theory and thermodynamic
  Bethe ansatz}}, \href{https://doi.org/10.1007/BF02101898}{\emph{Commun. Math.
  Phys.} {\bfseries 177} (1996) 381}
  [\href{https://arxiv.org/abs/hep-th/9412229}{{\ttfamily hep-th/9412229}}].

\bibitem{Bazhanov:1998dq}
V.~V. Bazhanov, S.~L. Lukyanov and A.~B. Zamolodchikov, \emph{{Integrable
  structure of conformal field theory. 3. The Yang-Baxter relation}},
  \href{https://doi.org/10.1007/s002200050531}{\emph{Commun. Math. Phys.}
  {\bfseries 200} (1999) 297}
  [\href{https://arxiv.org/abs/hep-th/9805008}{{\ttfamily hep-th/9805008}}].

\bibitem{Antonov:1996ag}
A.~Antonov and B.~Feigin, \emph{{Quantum group representations and Baxter
  equation}}, \href{https://doi.org/10.1016/S0370-2693(96)01526-2}{\emph{Phys.
  Lett. B} {\bfseries 392} (1997) 115}
  [\href{https://arxiv.org/abs/hep-th/9603105}{{\ttfamily hep-th/9603105}}].

\bibitem{Bazhanov:2001xm}
V.~V. Bazhanov, A.~N. Hibberd and S.~M. Khoroshkin, \emph{{Integrable structure
  of W(3) conformal field theory, quantum Boussinesq theory and boundary affine
  Toda theory}},
  \href{https://doi.org/10.1016/S0550-3213(01)00595-8}{\emph{Nucl. Phys. B}
  {\bfseries 622} (2002) 475}
  [\href{https://arxiv.org/abs/hep-th/0105177}{{\ttfamily hep-th/0105177}}].

\bibitem{Rossi:2002ed}
M.~Rossi and R.~Weston, \emph{{A Generalized Q-operator for
  $U_q(\widehat{sl_2}$) Vertex Models}},
  \href{https://doi.org/10.1088/0305-4470/35/47/304}{\emph{J. Phys. A}
  {\bfseries 35} (2002) 10015}
  [\href{https://arxiv.org/abs/math-ph/0207004}{{\ttfamily math-ph/0207004}}].

\bibitem{Korff_2006}
C.~Korff, \emph{{A Q-operator for the twisted XXX model}},
  \href{https://doi.org/10.1088/0305-4470/39/13/002}{\emph{Journal of Physics
  A: Mathematical and General} {\bfseries 39} (2006) 3203–3219}
  [\href{https://arxiv.org/abs/math-ph/0511022}{{\ttfamily math-ph/0511022}}].

\bibitem{Bazhanov:2008yc}
V.~V. Bazhanov and Z.~Tsuboi, \emph{{Baxter's Q-operators for supersymmetric
  spin chains}},
  \href{https://doi.org/10.1016/j.nuclphysb.2008.06.025}{\emph{Nucl. Phys. B}
  {\bfseries 805} (2008) 451}
  [\href{https://arxiv.org/abs/0805.4274}{{\ttfamily 0805.4274}}].

\bibitem{Kojima:2008zza}
T.~Kojima, \emph{{Baxter's Q-operator for the W-algebra WN}},
  \href{https://doi.org/10.1088/1751-8113/41/35/355206}{\emph{J. Phys. A}
  {\bfseries 41} (2008) 355206}
  [\href{https://arxiv.org/abs/0803.3505}{{\ttfamily 0803.3505}}].

\bibitem{Boos:2010ss}
H.~Boos, F.~Göhmann, A.~Klümper, K.~S. Nirov and A.~V. Razumov,
  \emph{{Exercises with the universal R-matrix}},
  \href{https://doi.org/10.1088/1751-8113/43/41/415208}{\emph{J. Phys. A}
  {\bfseries 43} (2010) 415208}
  [\href{https://arxiv.org/abs/1004.5342}{{\ttfamily 1004.5342}}].

\bibitem{Boos:2013noa}
H.~Boos, F.~Göhmann, A.~Klümper, K.~S. Nirov and A.~V. Razumov,
  \emph{{Quantum groups and functional relations for higher rank}},
  \href{https://doi.org/10.1088/1751-8113/47/27/275201}{\emph{J. Phys. A}
  {\bfseries 47} (2014) 275201}
  [\href{https://arxiv.org/abs/1312.2484}{{\ttfamily 1312.2484}}].

\bibitem{Frassek:2010ga}
R.~Frassek, T.~$\L$ukowski, C.~Meneghelli and M.~Staudacher, \emph{{Oscillator
  Construction of $\mathfrak{su}(n|m)$ Q-Operators}},
  \href{https://doi.org/10.1016/j.nuclphysb.2011.04.008}{\emph{Nucl. Phys. B}
  {\bfseries 850} (2011) 175}
  [\href{https://arxiv.org/abs/1012.6021}{{\ttfamily 1012.6021}}].

\bibitem{Tsuboi:2019vvv}
Z.~Tsuboi, \emph{{A note on $q$-oscillator realizations of $U_{q}(gl(M|N))$ for
  Baxter $Q$-operators}},
  \href{https://doi.org/10.1016/j.nuclphysb.2019.114747}{\emph{Nucl. Phys. B}
  {\bfseries 947} (2019) 114747}
  [\href{https://arxiv.org/abs/1907.07868}{{\ttfamily 1907.07868}}].

\bibitem{Bazhanov:2010ts}
V.~V. Bazhanov, T.~$\L$ukowski, C.~Meneghelli and M.~Staudacher, \emph{{A
  Shortcut to the Q-Operator}},
  \href{https://doi.org/10.1088/1742-5468/2010/11/P11002}{\emph{J. Stat. Mech.}
  {\bfseries 1011} (2010) P11002}
  [\href{https://arxiv.org/abs/1005.3261}{{\ttfamily 1005.3261}}].

\bibitem{Bazhanov:2010jq}
V.~V. Bazhanov, R.~Frassek, T.~$\L$ukowski, C.~Meneghelli and M.~Staudacher,
  \emph{{Baxter Q-Operators and Representations of Yangians}},
  \href{https://doi.org/10.1016/j.nuclphysb.2011.04.006}{\emph{Nucl. Phys. B}
  {\bfseries 850} (2011) 148}
  [\href{https://arxiv.org/abs/1010.3699}{{\ttfamily 1010.3699}}].

\bibitem{Frassek:2011aa}
R.~Frassek, T.~$\L$ukowski, C.~Meneghelli and M.~Staudacher, \emph{{Baxter
  Operators and Hamiltonians for 'nearly all' Integrable Closed
  $\mathfrak{gl}(n)$ Spin Chains}},
  \href{https://doi.org/10.1016/j.nuclphysb.2013.06.006}{\emph{Nucl. Phys. B}
  {\bfseries 874} (2013) 620}
  [\href{https://arxiv.org/abs/1112.3600}{{\ttfamily 1112.3600}}].

\bibitem{Frassek:2020nki}
R.~Frassek, \emph{{Oscillator realisations associated to the D-type Yangian:
  Towards the operatorial Q-system of orthogonal spin chains}},
  \href{https://doi.org/10.1016/j.nuclphysb.2020.115063}{\emph{Nuclear Physics
  B} {\bfseries 956} (2020) 115063}
  [\href{https://arxiv.org/abs/2001.06825}{{\ttfamily 2001.06825}}].

\bibitem{Kazakov:2007na}
V.~Kazakov and P.~Vieira, \emph{{From Characters to Quantum (Super)Spin Chains
  via Fusion}},
  \href{https://doi.org/10.1088/1126-6708/2008/10/050}{\emph{JHEP} {\bfseries
  10} (2008) 050} [\href{https://arxiv.org/abs/0711.2470}{{\ttfamily
  0711.2470}}].

\bibitem{Kazakov:2010iu}
V.~Kazakov, S.~Leurent and Z.~Tsuboi, \emph{{Baxter's Q-operators and
  operatorial Backlund flow for quantum (super)-spin chains}},
  \href{https://doi.org/10.1007/s00220-012-1428-9}{\emph{Commun. Math. Phys.}
  {\bfseries 311} (2012) 787}
  [\href{https://arxiv.org/abs/1010.4022}{{\ttfamily 1010.4022}}].

\bibitem{Alexandrov:2011aa}
A.~Alexandrov, V.~Kazakov, S.~Leurent, Z.~Tsuboi and A.~Zabrodin,
  \emph{{Classical tau-function for quantum spin chains}},
  \href{https://doi.org/10.1007/JHEP09(2013)064}{\emph{JHEP} {\bfseries 09}
  (2013) 064} [\href{https://arxiv.org/abs/1112.3310}{{\ttfamily 1112.3310}}].

\bibitem{Tsuboi:2009ud}
Z.~Tsuboi, \emph{{Solutions of the T-system and Baxter equations for
  supersymmetric spin chains}},
  \href{https://doi.org/10.1016/j.nuclphysb.2009.08.009}{\emph{Nucl. Phys. B}
  {\bfseries 826} (2010) 399}
  [\href{https://arxiv.org/abs/0906.2039}{{\ttfamily 0906.2039}}].

\bibitem{Pronko:1999gh}
G.~Pronko and Y.~Stroganov, \emph{{The Complex of Solutions of the Nested Bethe
  Ansatz. The $A_2$ Spin Chain}},
  \href{https://doi.org/10.1088/0305-4470/33/46/309}{\emph{J. Phys. A}
  {\bfseries 33} (2000) 8267}
  [\href{https://arxiv.org/abs/hep-th/9902085}{{\ttfamily hep-th/9902085}}].

\bibitem{Krichever:1996qd}
I.~Krichever, O.~Lipan, P.~Wiegmann and A.~Zabrodin, \emph{{Quantum Integrable
  Systems and Elliptic Solutions of Classical Discrete Nonlinear Equations}},
  \href{https://doi.org/10.1007/s002200050165}{\emph{Commun. Math. Phys.}
  {\bfseries 188} (1997) 267}
  [\href{https://arxiv.org/abs/hep-th/9604080}{{\ttfamily hep-th/9604080}}].

\bibitem{Dorey:2000ma}
P.~Dorey, C.~Dunning and R.~Tateo, \emph{{Differential equations for general
  SU(n) Bethe ansatz systems}},
  \href{https://doi.org/10.1088/0305-4470/33/47/308}{\emph{J. Phys. A}
  {\bfseries 33} (2000) 8427}
  [\href{https://arxiv.org/abs/hep-th/0008039}{{\ttfamily hep-th/0008039}}].

\bibitem{Kazakov:2007fy}
V.~Kazakov, A.~S. Sorin and A.~Zabrodin, \emph{{Supersymmetric Bethe ansatz and
  Baxter equations from discrete Hirota dynamics}},
  \href{https://doi.org/10.1016/j.nuclphysb.2007.06.025}{\emph{Nucl. Phys. B}
  {\bfseries 790} (2008) 345}
  [\href{https://arxiv.org/abs/hep-th/0703147}{{\ttfamily hep-th/0703147}}].

\bibitem{Kazakov:2015efa}
V.~Kazakov, S.~Leurent and D.~Volin, \emph{{T-system on T-hook: Grassmannian
  Solution and Twisted Quantum Spectral Curve}},
  \href{https://doi.org/10.1007/JHEP12(2016)044}{\emph{JHEP} {\bfseries 12}
  (2016) 044} [\href{https://arxiv.org/abs/1510.02100}{{\ttfamily
  1510.02100}}].

\bibitem{Tsuboi:1997iq}
Z.~Tsuboi, \emph{{Analytic Bethe ansatz and functional equations for Lie
  superalgebra $sl(r+1|s+1)$}},
  \href{https://doi.org/10.1088/0305-4470/30/22/031}{\emph{J. Phys. A}
  {\bfseries 30} (1997) 7975}
  [\href{https://arxiv.org/abs/0911.5386}{{\ttfamily 0911.5386}}].

\bibitem{Kuniba:2010ir}
A.~Kuniba, T.~Nakanishi and J.~Suzuki, \emph{{T-systems and Y-systems in
  integrable systems}},
  \href{https://doi.org/10.1088/1751-8113/44/10/103001}{\emph{J. Phys. A}
  {\bfseries 44} (2011) 103001}
  [\href{https://arxiv.org/abs/1010.1344}{{\ttfamily 1010.1344}}].

\bibitem{cherednik1989special}
I.~V. Cherednik, \emph{Special bases of irreducible representations of a
  degenerate affine hecke algebra}, {\emph{Functional Analysis and its
  Applications} {\bfseries 20} (1986) 76}.

\bibitem{Bazhanov:1989yk}
V.~Bazhanov and N.~Reshetikhin, \emph{{Restricted Solid on Solid Models
  Connected With Simply Based Algebras and Conformal Field Theory}},
  \href{https://doi.org/10.1088/0305-4470/23/9/012}{\emph{J. Phys. A}
  {\bfseries 23} (1990) 1477}.

\bibitem{Tsuboi:1996sh}
Z.~Tsuboi and A.~Kuniba, \emph{{Solutions of a discretized Toda field equation
  for $D_r$ from Analytic Bethe Ansatz}},
  \href{https://doi.org/10.1088/0305-4470/29/23/034}{\emph{J. Phys. A}
  {\bfseries 29} (1996) 7785}
  [\href{https://arxiv.org/abs/hep-th/9608002}{{\ttfamily hep-th/9608002}}].

\bibitem{Kuniba:1994na}
A.~Kuniba and J.~Suzuki, \emph{{Analytic Bethe Ansatz for fundamental
  representations of Yangians}},
  \href{https://doi.org/10.1007/BF02101234}{\emph{Commun. Math. Phys.}
  {\bfseries 173} (1995) 225}
  [\href{https://arxiv.org/abs/hep-th/9406180}{{\ttfamily hep-th/9406180}}].

\bibitem{Reshetikhin:1983vw}
N.~Reshetikhin, \emph{{A method of functional equations in the theory of
  exactly solvable quantum systems}},
  \href{https://doi.org/10.1007/BF00400435}{\emph{Lett. Math. Phys.} {\bfseries
  7} (1983) 205}.

\bibitem{Klumper:1992vt}
A.~Klümper and P.~Pearce, \emph{{Conformal weights of RSOS lattice models and
  their fusion hierarchies}}, {\emph{Physica A} {\bfseries 183} (1992) 304}.

\bibitem{Gromov:2010km}
N.~Gromov, V.~Kazakov, S.~Leurent and Z.~Tsuboi, \emph{{Wronskian Solution for
  AdS/CFT Y-system}},
  \href{https://doi.org/10.1007/JHEP01(2011)155}{\emph{JHEP} {\bfseries 01}
  (2011) 155} [\href{https://arxiv.org/abs/1010.2720}{{\ttfamily 1010.2720}}].

\bibitem{Derkachov:2006fw}
S.~E. Derkachov and A.~N. Manashov, \emph{{$\mathcal{R}$-Matrix and Baxter
  $\mathcal{Q}$-Operators for the Noncompact SL(N,$\mathbb{C}$) Invariant Spin
  Chain}}, {\emph{SIGMA} {\bfseries 2} (2006) 084}
  [\href{https://arxiv.org/abs/nlin/0612003}{{\ttfamily nlin/0612003}}].

\bibitem{Derkachov:2010qe}
S.~E. Derkachov and A.~N. Manashov, \emph{{Noncompact sl(N) spin chains:
  BGG-resolution, $\mathcal{Q}$-operators and alternating sum representation
  for finite dimensional transfer matrices}},
  \href{https://doi.org/10.1007/s11005-011-0472-2}{\emph{Lett. Math. Phys.}
  {\bfseries 97} (2011) 185} [\href{https://arxiv.org/abs/1008.4734}{{\ttfamily
  1008.4734}}].

\bibitem{Nepomechie:2020ixi}
R.~I. Nepomechie, \emph{{The $A_m^{(1)}$ Q-system}},
  \href{https://doi.org/10.1142/s0217732320502600}{\emph{Modern Physics Letters
  A} (2020) 2050260} [\href{https://arxiv.org/abs/2003.06823}{{\ttfamily
  2003.06823}}].

\bibitem{fulton2013representation}
W.~Fulton and J.~Harris, \emph{Representation Theory: A First Course},
  vol.~129. Springer-Verlag New York, 2004.

\bibitem{okounkov1996shifted}
{Andrei Okounkov and Grigori Olshanski}, \emph{{Shifted Schur functions II.
  Binomial formula for characters of classical groups and applications}},
  {\emph{Amer. Math. Soc. Transl.} {\bfseries 181} (1998) 245}
  [\href{https://arxiv.org/abs/q-alg/9612025}{{\ttfamily q-alg/9612025}}].

\bibitem{Campoleoni:2015qrh}
A.~Campoleoni, H.~A. Gonzalez, B.~Oblak and M.~Riegler, \emph{{Rotating Higher
  Spin Partition Functions and Extended BMS Symmetries}},
  \href{https://doi.org/10.1007/JHEP04(2016)034}{\emph{JHEP} {\bfseries 04}
  (2016) 034} [\href{https://arxiv.org/abs/1512.03353}{{\ttfamily
  1512.03353}}].

\bibitem{Hatayama:1998zp}
G.~Hatayama, A.~Kuniba, M.~Okado, T.~Takagi and Y.~Yamada, \emph{{Remarks on
  Fermionic Formula}},  \href{https://arxiv.org/abs/math/9812022}{{\ttfamily
  math/9812022}}.

\bibitem{Ogievetsky:1986hu}
E.~Ogievetsky and P.~Wiegmann, \emph{{Factorized S Matrix and the Bethe Ansatz
  for Simple Lie Groups}},
  \href{https://doi.org/10.1016/0370-2693(86)91644-8}{\emph{Phys. Lett. B}
  {\bfseries 168} (1986) 360}.

\bibitem{MacKay:1990mp}
N.~J. MacKay, \emph{{New factorized S-matrices associated with SO(N)}},
  \href{https://doi.org/10.1016/0550-3213(91)90384-A}{\emph{Nucl. Phys.}
  {\bfseries B356} (1991) 729}.

\bibitem{kirillov1990representations}
A.~N. Kirillov and N.~Y. Reshetikhin, \emph{Representations of yangians and
  multiplicities of occurrence of the irreducible components of the tensor
  product of representations of simple lie algebras}, {\emph{Journal of Soviet
  Mathematics} {\bfseries 52} (1990) 3156}.

\bibitem{Ekhammar:2020enr}
S.~Ekhammar, H.~Shu and D.~Volin, \emph{{Extended systems of Baxter Q-functions
  and fused flags I: simply-laced case}},
  \href{https://arxiv.org/abs/2008.10597}{{\ttfamily 2008.10597}}.

\bibitem{Zamolodchikov:1978xm}
A.~B. Zamolodchikov and A.~B. Zamolodchikov, \emph{{Factorized S Matrices in
  Two-Dimensions as the Exact Solutions of Certain Relativistic Quantum Field
  Theory Models}},
  \href{https://doi.org/10.1016/0003-4916(79)90391-9}{\emph{Annals Phys.}
  {\bfseries 120} (1979) 253}.

\bibitem{Arnaudon2006}
D.~Arnaudon, A.~Molev and E.~Ragoucy, \emph{{On the R-matrix realization of
  Yangians and their representations}},
  \href{https://doi.org/10.1007/s00023-006-0281-9}{\emph{Annales Henri
  Poincar{\'e}} {\bfseries 7} (2006) 1269}
  [\href{https://arxiv.org/abs/math/0511481}{{\ttfamily math/0511481}}].

\bibitem{Reshetikhin:1986vd}
N.~{\relax Yu}. Reshetikhin, \emph{{Integrable models of quantum
  one-dimensional magnets with $O(n)$ and $Sp(2k)$ symmetry}},
  \href{https://doi.org/10.1007/BF01017501}{\emph{Theor. Math. Phys.}
  {\bfseries 63} (1985) 555}.

\bibitem{Isaev:2015hak}
A.~Isaev, D.~Karakhanyan and R.~Kirschner, \emph{{Orthogonal and symplectic
  Yangians and Yang--Baxter R-operators}},
  \href{https://doi.org/10.1016/j.nuclphysb.2016.01.007}{\emph{Nucl. Phys. B}
  {\bfseries 904} (2016) 124}
  [\href{https://arxiv.org/abs/1511.06152}{{\ttfamily 1511.06152}}].

\bibitem{deVega:1986xj}
H.~de~Vega and M.~Karowski, \emph{{Exact Bethe Ansatz Solution of O(2n)
  Symmetric Theories}},
  \href{https://doi.org/10.1016/0550-3213(87)90146-5}{\emph{Nucl. Phys. B}
  {\bfseries 280} (1987) 225}.

\bibitem{Martins:1997wb}
M.~Martins and P.~Ramos, \emph{{The Algebraic Bethe ansatz for rational braid -
  monoid lattice models}},
  \href{https://doi.org/10.1016/S0550-3213(97)00342-8}{\emph{Nucl. Phys. B}
  {\bfseries 500} (1997) 579}
  [\href{https://arxiv.org/abs/hep-th/9703023}{{\ttfamily hep-th/9703023}}].

\bibitem{Gerrard:2019dtc}
A.~Gerrard and V.~Regelskis, \emph{{Nested algebraic Bethe ansatz for deformed
  orthogonal and symplectic spin chains}},
  \href{https://doi.org/10.1016/j.nuclphysb.2020.115021}{\emph{Nucl. Phys. B}
  {\bfseries 956} (2020) 115021}
  [\href{https://arxiv.org/abs/1912.11497}{{\ttfamily 1912.11497}}].

\bibitem{Masoero:2015lga}
D.~Masoero, A.~Raimondo and D.~Valeri, \emph{{Bethe Ansatz and the Spectral
  Theory of Affine Lie Algebra-Valued Connections I. The simply-laced Case}},
  \href{https://doi.org/10.1007/s00220-016-2643-6}{\emph{Commun. Math. Phys.}
  {\bfseries 344} (2016) 719}
  [\href{https://arxiv.org/abs/1501.07421}{{\ttfamily 1501.07421}}].

\bibitem{Frenkel:2020iqq}
E.~Frenkel, P.~Koroteev, D.~S. Sage and A.~M. Zeitlin, \emph{{q-Opers,
  QQ-Systems, and Bethe Ansatz}},
  \href{https://arxiv.org/abs/2002.07344}{{\ttfamily 2002.07344}}.

\bibitem{Kazakov:2018ugh}
V.~Kazakov, \emph{{Quantum Spectral Curve of $\gamma$-twisted ${\cal N}=4$ SYM
  theory and fishnet CFT}}, vol.~30, pp.~293--342.
\newblock 2018.
\newblock \href{https://arxiv.org/abs/1802.02160}{{\ttfamily 1802.02160}}.

\bibitem{Marboe:2016yyn}
C.~Marboe and D.~Volin, \emph{{Fast analytic solver of rational Bethe
  equations}}, \href{https://doi.org/10.1088/1751-8121/aa6b88}{\emph{J. Phys.
  A} {\bfseries 50} (2017) 204002}
  [\href{https://arxiv.org/abs/1608.06504}{{\ttfamily 1608.06504}}].

\bibitem{bernstein1975differential}
I.~Bernstein, I.~M. Gelfand and S.~I. Gelfand, \emph{Differential operators on
  the base affine space and a study of g-modules}, {\emph{Lie groups and their
  representations (Proc. Summer School, Bolyai J{\'a}nos Math. Soc., Budapest,
  1971)} (1975) 21}.

\bibitem{Kuniba:1993cn}
A.~Kuniba, T.~Nakanishi and J.~Suzuki, \emph{{Functional relations in solvable
  lattice models. 1: Functional relations and representation theory}},
  \href{https://doi.org/10.1142/S0217751X94002119}{\emph{Int. J. Mod. Phys. A}
  {\bfseries 9} (1994) 5215}
  [\href{https://arxiv.org/abs/hep-th/9309137}{{\ttfamily hep-th/9309137}}].

\bibitem{Shankar:1978rb}
R.~Shankar and E.~Witten, \emph{{The S Matrix of the Kinks of the ($\psi^-$bar
  $\psi$)**2 Model}},
  \href{https://doi.org/10.1016/0550-3213(78)90031-7}{\emph{Nucl. Phys. B}
  {\bfseries 141} (1978) 349}.

\bibitem{Frassek:2018try}
R.~Frassek and V.~Pestun, \emph{{A Family of ${\rm GL}_r$ Multiplicative Higgs
  Bundles on Rational Base}},
  \href{https://doi.org/10.3842/SIGMA.2019.031}{\emph{SIGMA} {\bfseries 15}
  (2019) 031} [\href{https://arxiv.org/abs/1808.00799}{{\ttfamily
  1808.00799}}].

\bibitem{nakai2007paths}
W.~Nakai and T.~Nakanishi, \emph{{Paths and tableaux descriptions of
  Jacobi-Trudi determinant associated with quantum affine algebra of type
  $D_n$}}, {\emph{Journal of Algebraic Combinatorics} {\bfseries 26} (2007)
  253}.

\bibitem{Kedem:2007zz}
R.~Kedem, \emph{{Q-systems as cluster algebras}},
  \href{https://doi.org/10.1088/1751-8113/41/19/194011}{\emph{J. Phys. A}
  {\bfseries 41} (2008) 194011}
  [\href{https://arxiv.org/abs/0712.2695}{{\ttfamily 0712.2695}}].

\bibitem{DiFrancesco:2008mc}
P.~Di~Francesco and R.~Kedem, \emph{{Q-systems as cluster algebras II: Cartan
  matrix of finite type and the polynomial property}},
  \href{https://doi.org/10.1007/s11005-009-0354-z}{\emph{Lett. Math. Phys.}
  {\bfseries 89} (2009) 183} [\href{https://arxiv.org/abs/0803.0362}{{\ttfamily
  0803.0362}}].

\bibitem{Leurent:2012xc}
S.~Leurent, \emph{{Integrable systems and AdS/CFT duality}}, {\emph{PhD thesis}
  (2012) } [\href{https://arxiv.org/abs/1206.4061}{{\ttfamily 1206.4061}}].

\bibitem{Chervov:2006xk}
A.~Chervov and D.~Talalaev, \emph{{Quantum spectral curves, quantum integrable
  systems and the geometric Langlands correspondence}},
  \href{https://arxiv.org/abs/hep-th/0604128}{{\ttfamily hep-th/0604128}}.

\bibitem{Frassek:2015mra}
R.~Frassek and I.~M. Szécsényi, \emph{{Q-operators for the open Heisenberg
  spin chain}},
  \href{https://doi.org/10.1016/j.nuclphysb.2015.10.010}{\emph{Nucl. Phys. B}
  {\bfseries 901} (2015) 229}
  [\href{https://arxiv.org/abs/1509.04867}{{\ttfamily 1509.04867}}].

\bibitem{Baseilhac:2017hoz}
P.~Baseilhac and Z.~Tsuboi, \emph{{Asymptotic representations of augmented
  q-Onsager algebra and boundary K-operators related to Baxter Q-operators}},
  \href{https://doi.org/10.1016/j.nuclphysb.2018.02.017}{\emph{Nucl. Phys. B}
  {\bfseries 929} (2018) 397}
  [\href{https://arxiv.org/abs/1707.04574}{{\ttfamily 1707.04574}}].

\bibitem{Vlaar:2020jww}
B.~Vlaar and R.~Weston, \emph{{A Q-operator for open spin chains I: Baxter's TQ
  relation}}, \href{https://doi.org/10.1088/1751-8121/ab8854}{\emph{J. Phys. A}
  {\bfseries 53} (2020) 245205}
  [\href{https://arxiv.org/abs/2001.10760}{{\ttfamily 2001.10760}}].

\bibitem{Frassek:2017bfz}
R.~Frassek, C.~Marboe and D.~Meidinger, \emph{{Evaluation of the operatorial
  Q-system for non-compact super spin chains}},
  \href{https://doi.org/10.1007/JHEP09(2017)018}{\emph{JHEP} {\bfseries 09}
  (2017) 018} [\href{https://arxiv.org/abs/1706.02320}{{\ttfamily
  1706.02320}}].

\bibitem{Balog:2005yz}
J.~Balog and A.~Heged\H{u}s, \emph{{TBA equations for the mass gap in the O(2r)
  non-linear sigma-models}},
  \href{https://doi.org/10.1016/j.nuclphysb.2005.07.032}{\emph{Nucl. Phys. B}
  {\bfseries 725} (2005) 531}
  [\href{https://arxiv.org/abs/hep-th/0504186}{{\ttfamily hep-th/0504186}}].

\bibitem{Chicherin:2012yn}
D.~Chicherin, S.~Derkachov and A.~P. Isaev, \emph{{Conformal group: R-matrix
  and star-triangle relation}},
  \href{https://doi.org/10.1007/JHEP04(2013)020}{\emph{JHEP} {\bfseries 04}
  (2013) 020} [\href{https://arxiv.org/abs/1206.4150}{{\ttfamily 1206.4150}}].

\bibitem{Gurdogan:2015csr}
O.~Gürdo\u{g}an and V.~Kazakov, \emph{{New Integrable 4D Quantum Field
  Theories from Strongly Deformed Planar $\mathcal N = $ 4 Supersymmetric
  Yang-Mills Theory}}, \href{https://doi.org/10.1103/PhysRevLett.117.201602,
  10.1103/PhysRevLett.117.259903}{\emph{Phys. Rev. Lett.} {\bfseries 117}
  (2016) 201602} [\href{https://arxiv.org/abs/1512.06704}{{\ttfamily
  1512.06704}}].

\bibitem{Basso:2019xay}
B.~Basso, G.~Ferrando, V.~Kazakov and D.-l. Zhong, \emph{{Thermodynamic Bethe
  Ansatz for Fishnet CFT}},
  \href{https://doi.org/10.1103/PhysRevLett.125.091601}{\emph{Phys. Rev. Lett.}
  {\bfseries 125} (2020) 091601}
  [\href{https://arxiv.org/abs/1911.10213}{{\ttfamily 1911.10213}}].

\bibitem{Zamolodchikov:1980mb}
A.~B. Zamolodchikov, \emph{{'Fishnet' diagrams as a completely integrable
  system}}, \href{https://doi.org/10.1016/0370-2693(80)90547-X}{\emph{Phys.
  Lett.} {\bfseries 97B} (1980) 63}.

\bibitem{Gromov:2017blm}
N.~Gromov, \emph{{Introduction to the Spectrum of $N=4$ SYM and the Quantum
  Spectral Curve}},  \href{https://arxiv.org/abs/1708.03648}{{\ttfamily
  1708.03648}}.

\bibitem{Gromov:2019bsj}
N.~Gromov and A.~Sever, \emph{{Quantum fishchain in AdS$_{5}$}},
  \href{https://doi.org/10.1007/JHEP10(2019)085}{\emph{JHEP} {\bfseries 10}
  (2019) 085} [\href{https://arxiv.org/abs/1907.01001}{{\ttfamily
  1907.01001}}].

\bibitem{Gromov:2016rrp}
N.~Gromov and F.~Levkovich-Maslyuk, \emph{{Quark-anti-quark potential in $
  \mathcal{N} =$ 4 SYM}},
  \href{https://doi.org/10.1007/JHEP12(2016)122}{\emph{JHEP} {\bfseries 12}
  (2016) 122} [\href{https://arxiv.org/abs/1601.05679}{{\ttfamily
  1601.05679}}].

\bibitem{Cavaglia:2014exa}
A.~Cavagli\`a, D.~Fioravanti, N.~Gromov and R.~Tateo, \emph{{Quantum Spectral
  Curve of the $\mathcal N=$ 6 Supersymmetric Chern-Simons Theory}},
  \href{https://doi.org/10.1103/PhysRevLett.113.021601}{\emph{Phys. Rev. Lett.}
  {\bfseries 113} (2014) 021601}
  [\href{https://arxiv.org/abs/1403.1859}{{\ttfamily 1403.1859}}].

\bibitem{Bombardelli:2017vhk}
D.~Bombardelli, A.~Cavagli\`a, D.~Fioravanti, N.~Gromov and R.~Tateo,
  \emph{{The full Quantum Spectral Curve for $AdS_4/CFT_3$}},
  \href{https://doi.org/10.1007/JHEP09(2017)140}{\emph{JHEP} {\bfseries 09}
  (2017) 140} [\href{https://arxiv.org/abs/1701.00473}{{\ttfamily
  1701.00473}}].

\bibitem{Bombardelli:2018bqz}
D.~Bombardelli, A.~Cavagli\`a, R.~Conti and R.~Tateo, \emph{{Exploring the
  spectrum of planar AdS$_{4}$/CFT$_{3}$ at finite coupling}},
  \href{https://doi.org/10.1007/JHEP04(2018)117}{\emph{JHEP} {\bfseries 04}
  (2018) 117} [\href{https://arxiv.org/abs/1803.04748}{{\ttfamily
  1803.04748}}].

\bibitem{Braverman:2016pwk}
A.~Braverman, M.~Finkelberg and H.~Nakajima, \emph{{Coulomb branches of $3d$
  $\mathcal{N}=4$ quiver gauge theories and slices in the affine
  Grassmannian}}, \href{https://doi.org/10.4310/ATMP.2019.v23.n1.a3}{\emph{Adv.
  Theor. Math. Phys.} {\bfseries 23} (2019) 75}
  [\href{https://arxiv.org/abs/1604.03625}{{\ttfamily 1604.03625}}].

\bibitem{Nakajima:2019olw}
H.~Nakajima and A.~Weekes, \emph{{Coulomb branches of quiver gauge theories
  with symmetrizers}},  \href{https://arxiv.org/abs/1907.06552}{{\ttfamily
  1907.06552}}.

\bibitem{Frassek:2020lky}
R.~Frassek, V.~Pestun and A.~Tsymbaliuk, \emph{{Lax matrices from
  antidominantly shifted Yangians and quantum affine algebras}},
  \href{https://arxiv.org/abs/2001.04929}{{\ttfamily 2001.04929}}.

\end{thebibliography}\endgroup
\bibliographystyle{JHEP.bst}

\end{document}